\documentclass[twocolumn,apj,showpacs,showkeys]{emulatearxiv}
\usepackage{amsmath}
\usepackage{url}

\begin{document}

\title{Explosive Nucleosynthesis in Near-Chandrasekhar Mass White Dwarf 
Models for Type Ia supernovae: Dependence on Model Parameters}

\author{Shing-Chi Leung\thanks{Email address: shingchi.leung@ipmu.jp}, 
Ken'ichi Nomoto\thanks{Email address: nomoto@astron.s.u-tokyo.ac.jp}}

\affiliation{Kavli Institute for the Physics and 
Mathematics of the Universe (WPI), The University 
of Tokyo Institutes for Advanced Study \\ The 
University of Tokyo, Kashiwa, Chiba 277-8583, Japan}

\begin{abstract}

We present two-dimensional hydrodynamics simulations of near-Chandrasekhar mass white dwarf (WD) models for
Type Ia supernovae (SNe Ia) using the turbulent deflagration model with deflagration-detonation transition (DDT). 
We perform a parameter survey for 41 models to study the effects of the initial central density (i.e., WD mass), metallicity, flame shape, DDT criteria, and turbulent flame formula 
for a much wider parameter space than earlier studies. 
The final isotopic abundances of $^{11}$C to $^{91}$Tc in these simulations are obtained by 
post-process nucleosynthesis calculations. The survey includes SNe Ia models with the central 
density from $5 \times 10^8$ g cm$^{-3}$ to $5 \times 10^9$ g cm$^{-3}$ (WD masses of 1.30 - 1.38 $M_\odot$), 
metallicity from 0 to 5 $Z_{\odot}$, 
C/O mass ratio from 0.3 - 1.0
and ignition kernels including centered and 
off-centered ignition kernels. 
We present the yield tables of stable isotopes from $^{12}$C to 
$^{70}$Zn as well as the major radioactive isotopes for 33 models.
Observational abundances of $^{55}$Mn, $^{56}$Fe,
$^{57}$Fe and $^{58}$Ni obtained from the solar composition, 
well-observed SNe Ia and SN Ia remnants
are used to constrain the explosion models
and the supernova progenitor.
The connection between the pure turbulent deflagration model and the 
subluminous SNe Iax is discussed.  
We find that dependencies of the nucleosynthesis yields on the
metallicity and the central density (WD mass) are large.
To fit these observational abundances and also for the application of 
galactic chemical evolution modeling, these dependencies on the metallicity and WD mass
should be taken into account.

\end{abstract}




\section{Introduction}
\label{sec:intro}

Type Ia supernovae (SNe Ia) have been shown to be the major source of many
iron-peak elements in the galaxies 
\citep[e.g.,][]{Arnett1996, Matteucci2001, Matteucci2012, Nomoto1984, Nomoto1997, Nomoto2017, Nomoto2018}.
To understand how SNe Ia contribute to the metal enrichment
process in the galaxies, and to explain the growing diversities of 
the observational results, simulations of SN Ia models 
with much wider parameter ranges need to be done.

SNe Ia have been well-modeled by the
thermonuclear explosions of carbon-oxygen white 
dwarfs (WDs) \citep[e.g.,][]{Hillebrandt2000}.
Both the Chandrasekhar mass WD model and sub-Chandra 
mass WD model can be consistent with the 
similarity of SN Ia light curves \citep[e.g.,][]{Branch2017}.
The empirical similarity later leads to
the discovery of accelerating cosmological expansion 
and the existence of dark energy \citep{Riess1998, Perlmutter1999}. 

However, recent observations have suggested that
there exists a wide diversity in SNe Ia 
\citep[see, e.g., a review by][]{Taubenberger2017}.
In addition to previously known variations ranging
from luminous SNe Ia (super-Chandra and SN 1991T-like) 
to SN 1991bg-like faint SNe Ia, a new sub-class of 
SN 2002cx-like, or Type Iax SNe \citep[see, e.g., a
recent review by][]{Jha2017} and SN 2002es-like SNe
\citep[e.g.,][]{Taubenberger2017} have been reported.
Such brightness variations imply a large variation 
of the $^{56}$Ni mass ($\sim$ 0.1 $M_{\odot}$ 
to $\sim$ 1.4 $M_{\odot}$) produced in the explosions.

Studies of nucleosynthesis of one-dimensional models
have shown some important dependencies on the model parameters.
For example, Fe-peak elements synthesized in 
the Chandrasekhar mass model W7 \citep{Nomoto1984}
is shown to be consistent with the solar abundances,
except for a significant over-production of $^{58}$Ni,
where the rate of electron capture is important 
\citep[e.g.,][]{Thielemann1986,Iwamoto1999,Brachwitz2000,Mori2016}. The solar abundance pattern 
of Fe-peak elements can also be reproduced by 
the sub-Chandra models with solar metallcity 
\citep[e.g.,][]{Shigeyama1992,Nomoto1994}, 
although the Ni/Fe ratio (which depends on 
metallicity) tends to be under solar because 
of the low central density of exploding WDs.
As shown in [Ni/Fe] mentioned above, the 
central density of the WD is an important 
parameter because of the level of electron capture.
If the central density is high enough, synthesis of
certain neutron-rich isotopes, such as $^{48}$Ca, 
$^{50}$Ti, and $^{54}$Cr can be 
significant \citep{Woosley1997}. 

Another interesting example is [Mn/Fe], 
which increases with increasing [Fe/H] in 
the Galactic Halo \citep[e.g.,][]{Hinkel2014, Mishenina2015}.
[Mn/Fe] in metal-poor stars in dwarf-spheroidal 
galaxies \citep[e.g.,][]{Larsen2014, Sbordone2015} 
provides another information on metallicity dependence. 
In order to calculate the chemical evolution
of dwarf-spheroidal galaxies, metallicity-dependent 
SN Ia yields are necessary \citep[e.g.,][]{Kobayashi2015}.

Recent observations of SNe Ia remnants in the nebular phase
have provided important insights to the models of
SNe Ia. They include Tycho \citep{Yamaguchi2015}, 
Kepler \citep{Park2013}, and 3C 397 \citep{Yamaguchi2015}.
The relative X-ray flux of iron-peak 
elements can give promising constraints on
the explosion conditions. For example, these three remnants 
have been suggested to have progenitors with super-solar metallicity 
\citep{Yamaguchi2015}.

These would suggest the importance to obtain the SN Ia yields for
a wide range of environmental conditions, such as metallicity
and the mass (and thus the central density) of the WDs.

Such study will be important for the future use
of galactic chemical evolution for an accurate 
modeling of isotopes as a function of metallicity.
Nucleosynthesis of multi-dimensional hydrodynamics simulations was
made in \cite{Travaglio2004} with the use of 
tracer particle scheme \citep{Seitenzahl2010}. 
The effects of initial flame structure to the 
chemical yield was studied in \cite{Maeda2010, 
Fink2014, Seitenzahl2013} for different explosion
models.

A few more recent works have studied these objects. 
In \cite{Shen2017} the sub-Chandrasekhar 
SN Ia models are revisited and showed that the 
sub-Chandrasekhar SNe Ia can be connected to the 
remnant 3C 397 when appropriate amount of
reverse shock-heating is considered. Similar explorations 
were done in \cite{Dave2017}, where some representative 
models of pure turbulent deflagration, deflagration-detonation transition (DDT) and 
gravitationally confined detonation are explored. It is shown that 
for a sub-set of central densities, C/O ratio and high 
offset in the initial flame allow models to produce
super-solar [Mn/Fe] to match the observed data.
See, e.g. \cite{Nomoto2017} for a general review 
of nucleosynthesis pattern and its connection 
to explosion mechanism.  

Such results indicate that properties of these 
SNe Ia might have important metallicity effects.
Nucleosynthesis in SNe Ia has been studied extensively
but still only a small parameter space has been explored.

In view of this background, we make systematic 
modeling of SNe Ia for various explosion configuration and 
setting to see how the model parameters of SNe Ia affect the
WD explosions and their chemical yields for Chandrasekhar mass WD
for much more wide range of parameters 
(i.e., WD masses (central densities), metallicity, flame structure).
In the forthcoming paper, we will present our sub-Chandrasekhar 
mass models. The chemical yields of SNe Ia, which depend on the model
parameters and environmental conditions, can be constrained by the
observed abundance patterns of Fe-peak elements in various stars and 
systems.

We use our own 2D hydrodynamics code (see Appendix \ref{sec:methods}), because 
our 2D hydro code is suitable to calculate many models 
for a wide range of parameters than 3D hydro code.
The typical running time for one 2D model is $\sim$ 
days on a single machine;
while it takes weeks to months for a cluster 
to calculate the explosion phase of a model in 3D. 
Certainly, the 2D simulations have some qualitative
differences from the 3D simulations in two ways. 
First, the flame in 2D models tends to propagate faster than
in 3D models because of the larger surface area
for the same 2D-projection.
Second, the imposed symmetry may enhance the growth
of hydrodynamical instabilities owing to the 
imposed reflective boundaries, which stimulate
the growth of boundary flows. 

In Section \ref{sec:benchmark} we construct the benchmark to be a
typical SNe Ia model. In Section \ref{sec:survey}, we present the
nucleosynthesis yields of our models and show how large the 
effects of each model parameter are. We then discuss how our results
can be applied to observational data to constrain the model parameter.
In Section \ref{sec:discuss} we compare our results with earlier calculations. 
In Appendix, we summarize the numerical 
code which is specifically developed for modeling 
SNe Ia \citep{Leung2015a}. We describe the updates
and changes in the hydrodynamics and 
nucleosynthesis. Finally, in order to apply for the chemical evolution modeling
and comparison with observational data, we present tables of
the nucleosynthesis yields of our 24 models.

\section{Initial Models and Methods}

\subsection{Initial Models}

We construct the initial C+O WD models at the central carbon ignition
with the masses from $M = 1.30 M_\odot$ to 1.39 $M_\odot$ (and thus
the central densities from $5 \times 10^8$ g cm$^{-3}$ to $5 \times
10^9$ g cm$^{-3}$) for various metallicity and the carbon fraction
(see section \ref{sec:survey} for details).
The internal temperature is assumed to be $1 \times 10^8$ K.

To carry out the two-dimensional simulations, we set
the model parameters as follows. For each CO WD,  
we choose a given central density $\rho_c$, metallicity
$Z$ and CO ratio [C/O] with an isothermal profile. 
Then we follow the hydrodynamics simulation without 
further alternation. In the initial model, we 
solve the hydrostatic equilibrium of the WD 
assuming a constant composition and a constant
[C/O]. To model metallicity in the simplified 
network, we treat $^{22}$Ne as the proxy of 
metallicty.

The above initial model for the simulation of the explosion is a
simplified approximation of the realistic evolutionary model of an
accreting WD from its formation through the initiation of a
deflagration.  To clarify the simplification of our initial model, let
us compare with the evolutinary models calclated by
\cite{Nomoto1976,Nomoto1984}.

In \cite{Nomoto1976,Nomoto1984}, the initial mass of a C+O WD is 1.0
$M_\odot$ with uniform mass fractions of C, O, and $^{22}$Ne as $X$(C)
= 0.475, $X$(O) = 0.50, and $X(^{22}$Ne) = 0.025.  Here $^{22}$Ne is
converted from the initial CNO elements during H and He burning so
that $X(^{22}$Ne) is treated as the proxy of initial metallicity.  In
the present intial models, we also adopt a uniform abundance
distribution with $X$(O) = 0.50 and $X$(C) = 0.50 $-$ $X(^{22}$Ne) where
different $X(^{22}$Ne) is the proxy of different metallicity.
$X(^{22}$Ne) = 0.025 is regarded as the solar metallicity, although the
latest solar abundances correspond to $X(^{22}$Ne) = 0.0134 \citep{Asplund2009}.

In \cite{Nomoto1976,Nomoto1984}, the WD is cooled down to the central
temperature of $T_{\rm c}$ = 10$^8$ K and 10$^7$ K for two cases,
respectively, with almost isothermal distribution.  Afterwards, mass
accretion onto the WD starts with different accretion rates, which give
the rate of compressional heating of the WD interior.

In the SD scenario, the WD mass (and thus the central density)
increases by mass accretion from the companion star
\citep[e.g.,][]{Nomoto1994}.  The internal
temperature depends on the competition between the compressional
heating and radiative cooling \citep[e.g.,][]{Nomoto1982a,Nomoto1982b}. For a
higher accretion rate, the central temperature increases faster and
carbon is ignited at the center at a lower central density (and thus a
smaller WD mass).

For heating, heat inflow from the H/He shell burning is also important
\citep{Nomoto1984}.  Since the timescale of heat conduction in the WD
interior is shorter than the accretion timescale, the WD interior
is close to isothermal with the temperature of $\sim 10^8$ K
\citep{Nomoto1984}.

In this way, the adopted WD masses at the carbon
ignition correspond to different mass accretion rates from the
companion stars \citep{Nomoto1984} and/or the delay time in uniformly
rotating WDs \citep{Benvenuto2013}.

We should note that the highest accretion rate for the central carbon
ignition is limited to $\sim 7 \times 10^{-7} M_\odot$ y$^{-1}$ by the
rate of steady hydrogen burning above which a strong WD wind blows
\citep{Nomoto2007, Kato2014} or to $\sim 3 \times 10^{-6} M_\odot$
y$^{-1}$ by the off-center carbon ignition for the accretion of He
from a He star companion \citep{Nomoto1985}.  For these limitations,
the lowest WD mass at the carbon ignition is $\sim 1.35 M_\odot$
\citep{Nomoto1984}.

After C-ignition in the center due to strong screening effects, a
simmering phase starts with developing convective core, which was
calculted using the time-dependent mixing length theory
\citep{Unno1967,Nomoto1976,Nomoto1984}.
\citep[For recent works on simmering phase, see,
e.g.,][]{Piro2008,Jackson2010,Krueger2012,Ohlmann2014,MartinezRodriguez2016}.  
In these calculations,
so-called convective URCA processes were not included.  The extent of
the convective core might be limited by convective URCA process
although it is quite uncertain \citep[e.g.,][]{Arnett1996,Lesaffre2006}, In view
of the large uncertainty of convective URCA process, the exact
distributions of the temperature and abundances in the WD should be
regarded as highly uncertain and further study of simmering phase is
necessary \citep[See, e.g.,][for an analytic analysis]{Piro2008}.
We study the effects of the initial C/O                     
ratio as the origin of model diversity.

Near the end of simmering phase in the models by
\cite{Arnett1969,Nomoto1976,Nomoto1984}, it was found that a super-adiabatic
temperature gradient appears at the central temperature of $T_{\rm c}
> 8 \times 10^8$ K and increases sharply.  The timescale of the
temperature rise becomes shorter than the dynamical timescale around
$T_{\rm c} > 3 \times 10^9$ K.  At $T_{\rm c} > 5 \times 10^9$ K,
Nuclear reactions become rapid enough to realize nuclear statistical
equilibrium (NSE) and $T_{\rm c} \sim 10^{10}$ K is reached.  The
steep temperature jump, i.e., a deflagration front is formed.  Such an
evolution through NSE takes place in a timescale shorter than the
convective energy transport timescale, the convective core size and
the abundances in the bulk of the convective core are frozen, i.e.,
nuclear burning products in the center is not well-mixed with
the outer layers.

In the deflagrated region, NSE is realized so that the details of the
abundance change during rapid nuclear reaction is not important.
Decrease in $Y_{\rm e}$ due to electron capture during simmering phase is
also negligibly small compared with electron capture after NSE is
realized. (Here $Y_{\rm e} = \sum\limits_{i} Z_{\rm i}/A_{\rm i}$ 
with $Z_{\rm i}$ and $A_{\rm i}$ being respectively the atomic number and mass number of
the specie {\rm i}. For convenience in this paper, we use $Y_{\rm e} = Z_{\rm i}/A_{\rm i}$
for the individual species {\rm i} as well.)
Thus the neglection of convective region and the
composition change during the simmering phase does not much affect the
current results.

We also note that, owing to the degenerate electron gas, the mass and
radius is less sensitive to the choice of temperature profile. We
observe that the masses, radii and density profiles are still
comparable with those presented in the literature
\citep[See, e.g.,][for the WD model obtained from stellar evolutionary
  model]{Krueger2010}.


In the present study, we extend the WD mass down to the range of 1.30 - 1.35
$M_\odot$.  Such low WD masses may be called as the sub-Chandrasekhar
mass.  The central carbon ignition in such a sub-Chandrasekhar mass WD
would be possible by shock compression of the central region due to
the surface He detonation \citep[e.g.,][]{Arnett1996}.  The important
difference from those ``double detonation'' models is that, because of
the relatively large WD mass and thus the high central density,
the central carbon ignition does not necessary produce strong shock wave to induce the 
detonation but rather develops a carbon deflagration due to the large
electron-degeneracy pressure compared with the thermal pressure
released by nuclear burning \citep{Nomoto1976}.  Whether the surface
He detonation induces the carbon deflagration or direct detonation
will be studied in forth coming papers. 
%

\subsection{Input Physics}
\label{sec:input}

Here we briefly describe the new input physics 
used in the code. For the basic data structure 
of the code and the code test, we refer the 
readers to \cite{Leung2015a}.
We also refer the readers Appendix \ref{sec:methods} for
the numerical implementation of the SN Ia physics in our code.
Here we only list the parts relevant to SN Ia.
We use the most recent rates we have for 
describing the microphysics, including:
\par\noindent the Helmholtz Equation of state \citep{Timmes1999a};
\par\noindent nuclear reaction rates \citep{Rauscher2000};
\par\noindent strong screening factor \citep{Kitamura2000}; 
\par\noindent electron capture rates \citep{Langanke2001}.

\subsection{Methods}

In the present study, we assume that the central carbon flash develops
into the deflagration as follows.
The exact pattern of the initial flame is
not well constrained. 
To trigger the deflagration phase, therefore, we impose a flame by hand
in the star. The zone is assumed to be incinerated into 
NSE. We choose two different 
morphology. First, it is a centered flame with some
sinusoidal perturbations. This is similar to the $c3$
flame as used in \cite{Reinecke1999b}, which mimics that the
flame grows at center and then it is perturbed by 
Rayleigh-Taylor instabilities. Second, it is an off-center
"bubble" (in 2D the bubble corresponds to a ring in 3D),
similar to $b1$ in \cite{Reinecke1999b}. 
This pattern mimics the evolution that the convection in 
the star is rapid enough to bring the hot parcel from 
the center before it runaways. 
\footnote{Notice that for a fully self-consistent manner, one should
perform multi-D simulations of evolution of the WD from the simmering 
phase \citep{Woosley2004,Wunsch2006},
such that the convective pattern as well as the runaway
location can be captured naturally. However, this requires
the use of low-Mach number solver \citep[e.g.,][]{Zingale2011}
and high resolution 
due to the much slower physical process ($\sim$ hours)
and the typical runaway size of the flame ($\sim$ cm)
\citep{Zingale2007}.}

To determine the deflagration-detonation transition (DDT),
we compare the eddy turnover scale with the flame width,
i.e. the {\sl Karlovitz Number, Ka}, which is defined as \citep{Niemeyer1995a}
\begin{equation}
{\rm Ka} = \frac{l_{{\rm flame}}}{l_{{\rm turb}}}.
\end{equation}
Here $l_{{\rm flame}}$ and $l_{{\rm turb}}$ are the representative
length scale of the deflagration wave and the turbulent eddy motion
(see Appendix for details).
At the end of each time step we scan across the 
flame surface to see if the Karlovitz number, $Ka$, 
satisfies the DDT condition, which we pick $Ka = 1$ to be the 
required DDT condition. Once this condition is satisfied,
we put in the initial C-detonation in the form of 
2D bubble (a ring) at that point, and allow that
detonation to freely evolve. Extra detonation bubbles
are added as long as the flame surface is not 
yet swept by detonation wave. 
We follow the evolution until the whole star expands
sufficiently so that the whole star becomes 
sparse and cold that all electron capture and 
major nuclear reactions have stopped. 

We emphasize that there still exists theoretical
uncertainties in the DDT model, especially related
to the robustness of trigger detonation in 
an unconfined media (see Appendix \ref{sec:numerical}
for a comparison of how this certainty 
affects the nucleosynthesis).
 
\section{Benchmark Model of typical SNe Ia}
\label{sec:benchmark}

In doing the comparison, we first describe the parameters
for the benchmark model, its hydrodynamics behaviour and 
nucleosynthesis. 
The benchmark model is assumed to represent a statistical average of 
the SNe Ia which we observed, i.e. with solar metallicity, $\sim$ 0.6 $M_{\odot}$
of $^{56}$Ni and composition compatible with the solar abundance. 
This allows us to calibrate the validity of our models. For example, 
models which produce incompatible chemical abundances are regarded
as less frequent events in nature, thus casting constraints 
on the parameters space correspondingly.  

\begin{figure}
\centering
\includegraphics*[width=8cm,height=8cm]{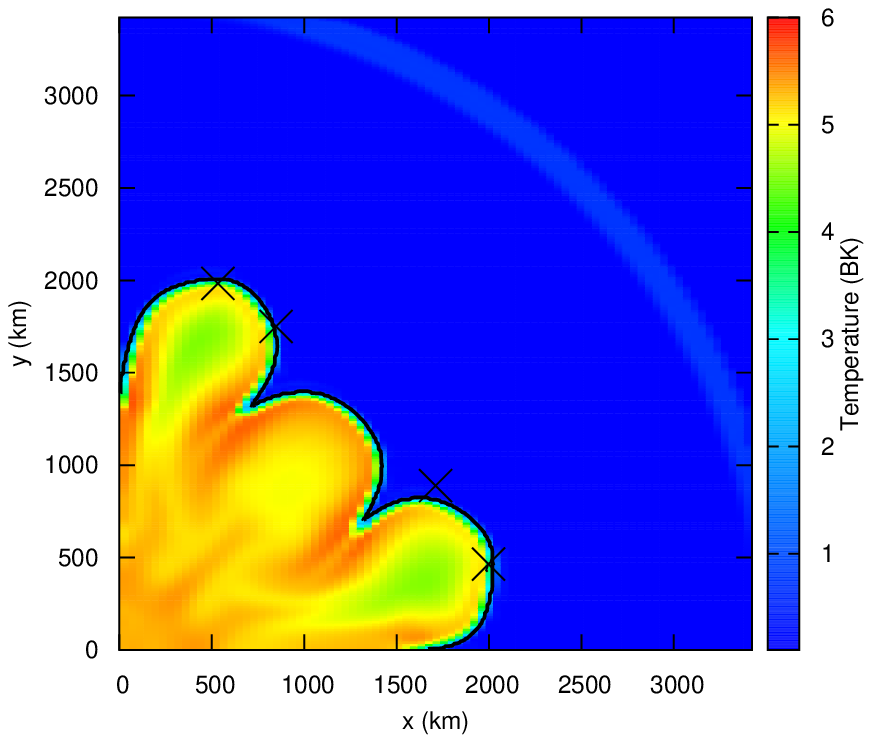}
\includegraphics*[width=8cm,height=8cm]{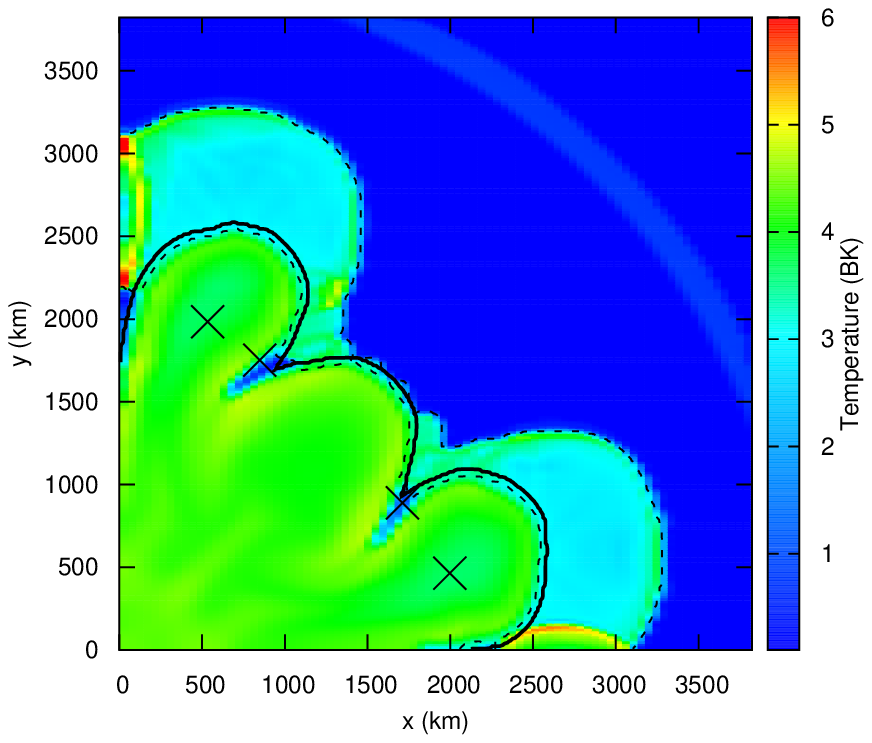}
\caption{(upper panel) The temperature colour plot together with the 
deflagration when the first detonation is triggered.
The solid line contour stands for the deflagration front and 
the crosses stand for the first four positions whose
DDT criteria is satisfied. (lower panel) Similar to the upper
panel, but at 0.1 s after the first DDT is triggered.
The solid (dashed) line stands for the deflagration (detonation)
front. The crosses are the same as above. We remind that
in the simulations, the detonation is triggered 
only along the deflagration front. In this figure the crosses
lie inside the deflagration because of fluid advection.}
\label{fig:flame_std}
\end{figure}

\begin{figure}
\centering
\includegraphics*[width=8cm,height=5.7cm]{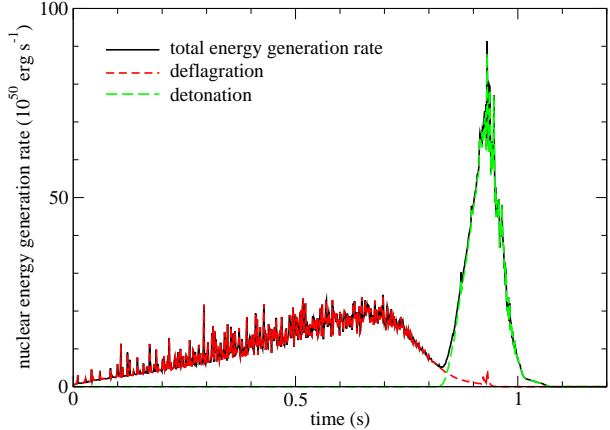}
\caption{The total energy generation rate and its components as a function
of time for the benchmark model. The solid line corresponds
to the total energy generation rate; the dotted, dashed, dot-dash
lines correspond to the luminosity by carbon deflagration, carbon 
detonation, NQSE + NSE burning respectively.}
\label{fig:lumin_std}
\end{figure}

\begin{figure}
\centering
\includegraphics*[width=8cm,height=5.7cm]{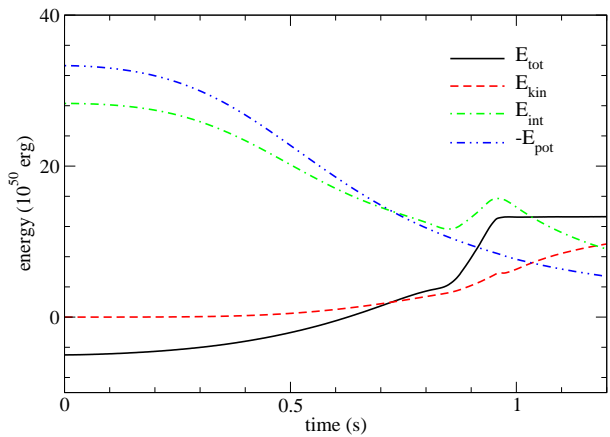}
\caption{The total energy and its components as a function
of time for the benchmark model. The solid line corresponds
to the total energy; the dotted, dashed, dot-dash lines
correspond to the total kinetic, internal and the absolute
value of gravitational potential energy.}
\label{fig:energy_std}
\end{figure}

In Table \ref{table:std} we tabulate the 
basic stellar parameters we found in producing the 
benchmark model. 
In Figure \ref{fig:flame_std} we plot the temperature 
in colour and the deflagration and detonation wave fronts
at the moment of the first DDT and at 0.1 s after the 
first DDT. At the moment of DDT (upper plot), the flame has developed 
from its initial size of $\sim 100$ km to a size of $\sim$ 
2000 km. We also mark the first four detonation spots by
crosses in both figures, where the transition density 
is $\sim 2 \times 10^7$ g cm$^{-3}$. It can be seen that the initial DDT
occurs on the "fingers" near the axis. We remind that DDT
can occur on the flame surface when the criteria 
is satisfied, and that flame surface is not yet
swept by detonation wave. The whole deflagration ash is
still hot at a temperature $\sim 5 - 6 \times 10^9$ K.
A thin ring of radius $\sim$ 3500 km can be seen due to 
the excitation of the initial flame which is put 
by hand. The sudden pulse creates a weak heating 
to that part up to $2 \times 10^9$ K. 
At 0.1 s after DDT (lower plot), the flame 
continued to grow to a size $\sim 2500$ km 
due to thermal expansion. The deflagration ash has 
drastically cooled down to a temperature 
$\sim 4 - 5 \times 10^9$ K. The detonation wave has 
quickly covered the deflagration front. Due to a
lower density, the detonation ash is in general 
cooler, about $3 \times 10^9$ K. Exception appears
when the detonation wave collides with the symmetry
boundary or another detonation wave. In these cases,
the shock compression can easily make the matter
to a temperature above $5 \times 10^9$ K.

In Figure \ref{fig:lumin_std} we plot the nuclear energy generation rate
and its components as a function of time for the benchmark model. 
We show separately the 
nuclear energy generation rate by deflagration and by detonation.
In Figure \ref{fig:energy_std} we plot the total energy and
its components as a function of time. 
For a more detailed discussion about the hydrodynamics evolution
of the benchmark model, we refer the readers to the Appendix \ref{sec:hydro}.

\begin{table*}

\begin{center}
\caption{Model setup for the benchmark model: 
central densities of NM $\rho_{ c{\rm (NM)} }$
are in units of $10^{9}$ g cm$^{-3}$.  
Metallicity is in units of solar metallicity.
The total mass $M$ and
and the final nickel-56 mass $M_{{\rm Ni}}$ 
are in units of solar mass. 
Ka is the critical Karlovitz number above which
C-detonation is assumed to occur in our simulations. 
See also Appendix for further details of the
deflagration-detonation transition criteria.
$R$ is the initial stellar radius in kilometers. 
$E_{\rm nuc}$ and $E_{\rm tot}$ are the energy released by nuclear reactions
and final total energy, respectively, both in units of $10^{50}$ erg. 
$Y_{{\rm e(min)}}$ is the minimum value of electron fraction within
the simulation box at the end of simulation. $t_{{\rm DDT}}$ is 
the first detonation transition time in units of second. 
$M_{{\rm Ni}}$ and $M_{{\rm Mn}}$ are the masses of $^{56}$Ni 
of $M_{{\rm Mn}}$ at the end of 
simulations, after all short-live radioactive isotopes 
have decayed.}
\label{table:std}
\begin{tabular}{|c|c|c|c|c|c|c|c|c|c|c|c|c|}
\hline

Model & $\rho_{c({\rm NM})}$ & Metallicity & flame shape & $Ka$ & $M$ &
$R$ & $Y_{{\rm e(min)}}$ & $E_{{\rm nuc}}$ & $E_{{\rm tot}}$ & 
$t_{{\rm DDT}}$ & $M_{{\rm Ni}}$ & $M_{{\rm Mn}}$ \\ \hline
300-1-c3-1 & 3 & 1 & $c3$ & 1 & 1.38 & 1900 & 
0.462 & 17.7 & 12.7 & 0.78 & 0.63 & $9.55 \times 10^{-3}$ \\ \hline

\end{tabular}
\end{center}
\end{table*}

\subsection{Pure Turbulent Deflagration Phase}

We show in Figure \ref{fig:flame_std} the temperature 
colour plot and the deflagration wave fronts. 
At early phase, the matter density is sufficiently high that
most matter is incinerated into NSE (including endothermic photo-disintegration
of $^{56}$Ni into $^{4}$He).
In the first 0.8 s, deflagration takes place, where the energy
release is slow. The deflagration wave, and its subsequent 
advanced burning releases about $10^{51}$ erg s$^{-1}$. 

In the pure turbulent
deflagration phase before the DDT, namely from $t$ = 0 - 1.12 s,
deflagration burns about 0.3 $M_{\odot}$ of matter.
As seen in Figure \ref{fig:lumin_std}, the deflagration
releases nuclear energy slowly, in the order of $10^{50}$ erg s$^{-1}$.
The nuclear energy production is slow so that
the total energy of the WD increases but remains negative.

During the deflagration phase, the star expands considerably.
As the flame front reaches the 
low density region ($\sim 10^7$ g cm$^{-3}$) beyond $t = 0.8$ s, 
the carbon deflagration release much less energy than what it 
original does at stellar core. The drop of luminosity 
near $t = 1$ s suggests that the matter has expanded and
cooled down so that the NSE timescale becomes comparable
or even longer than the hydrodynamics timescale.

\subsection{Detonation Phase}
When the density at the flame front decreases to $\approx 2.3 \times 10^7$ g cm$^{-3}$, 
the transition to the detonation takes place. 
We plot in Figure \ref{fig:flame_std} the temperature 
colour plot and the detonation wave fronts. 
The detonation starts from the tip of the finger shape,
around $r = 2000$ km. The detonation wave is almost 
unperturbed by the fluid motion that the flame structure 
appears to be almost spherical.
The temperature profile shows that most matter are no longer
in NSE. Due to the uneven surface of the flame at the 
moment of DDT, there is unburnt material left behind in the 
high density region. At the radius defined by the 
outermost radius reached by the deflagration wave, there always 
exists fuel inside. These matter is later burnt into NSE
by the detonation wave. This provides an additional source
of iron-peak elements. Notice that this feature does not
exist in one dimensional models because the spherical model
allows all matter to be burnt inside the same outermost radius 
reached by the flame. Therefore, the detonation can only burn
the low density matter and produce fewer iron-peak elements.
\footnote{Outside the flame front, the matter
is mildly heated up due to numerical effects. Notice that
even the flame propagation is slow compared to the speed
of sound, the injection of energy in a discrete manner
still creates sound wave which propagates outward.
Thus the nearly isobaric property of the flame cannot be
exactly preserved. This mildly heats up the matter
outside the flame front by compression. Notice that
the details of this compressional heating depends
on some model parameters, for example the minimum
temperature. In Appendix \ref{sec:tmin} 
we further discuss this aspect.}

The detonation wave quickly burns the remaining material, making 
the total energy positive. 
Then the WD expands rapidly and increases its kinetic energy.
In contrast to the slow deflagration wave, 
the detonation is a much efficient 
source for producing nuclear energy.
It burns the 1.0 $M_{\odot}$ matter within 
the next 0.2 second. 
The typical luminosity is of the order $10^{52}$ erg s$^{-1}$.

\subsection{Explosive Nucleosynthesis}

The chemical composition of the ejected matter is presented.
To obtain the nucleosynthesis yield, we use the tracer particle 
scheme to keep track the thermodynamics history. Then, we 
calculate nucleosynthesis by using a 495-isotope network,
which includes isotopes from $^{1}$H to $^{91}$Tc. Stable neutron-rich
isotopes, such as $^{48}$Ca, $^{50}$Ti, $^{54}$Cr and $^{60}$Fe
are included so that the nucleosythesis 
with electron capture can be consistently 
calculated for $Y_{\rm e} =$ 0.45 - 0.50.
For the numerical details, see Appendix \ref{sec:methods}.

\begin{figure}
\centering
\includegraphics*[width=8cm,height=5.7cm]{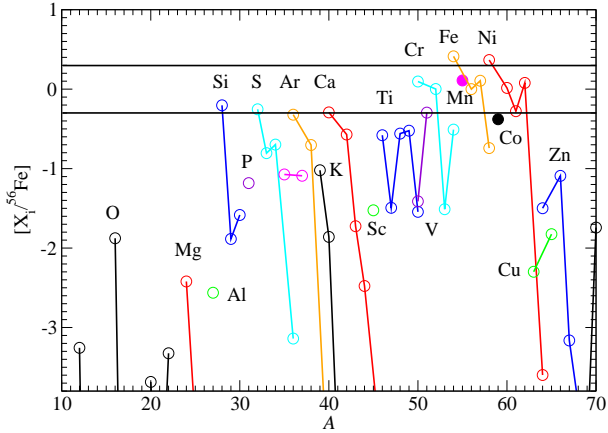}
\caption{$[{\rm X_i}/^{56}{\rm Fe}]$ of stable isotopes in the benchmark model 
after the short-lived radioactive isotopes have 
decayed. The ratio are scaled with solar value. The lines at 
$\pm 0.3$ (corresponding to 0.5 or 2.0 times of the solar value) 
are included. ($Remark$: The figure is replaced
due to the updated table.)}
\label{fig:final_std}
\end{figure}

In Figure \ref{fig:final_std} we plot $[{\rm X_i}/^{56}{\rm Fe}]$
of stable isotopes, after the decay of
short-lived radioactive elements are accounted. All
quantities are given by $[{\rm X_i}/^{56}{\rm Fe}] = 
{\rm log}_{10}((X_i/X(^{56}{\rm Fe}))/(X_i/X(^{56}{\rm Fe}))_\odot))$.
It can be seen that in general considerable
number of elements have $[{\rm X_i}/^{56}{\rm Fe}]$ between
-0.3 to 0.3 as marked in the figure. This shows that these
elements are consistent with the solar abundances.
Notice that many elements in the Sun come from
both Type Ia and Type II supernovae. For the case of
under-production, it is possible that such isotopes
may come from solely from Type II supernovae, 
such as the $\alpha$-chain isotopes. However, for 
the case of over-production, it will be a strong constraint
for that particular SN Ia model. It is because the typical 
rate of SNe Ia has the same order-of-magnitude as
Type II supernovae. Any severe over-production of such isotope, 
for instance 10 times above solar abundance, means that such 
explosion model is not a typical one since that isotope
cannot be "diluted" by the under-production (or null
production) of the other type of SN. 
Representative elements include $^{28}$Si,
$^{32}$S, $^{36}$Ar, $^{40}$Ca, $^{50-52}$Cr, $^{55}$Mn,
$^{54-57}$Fe, $^{58-62}$Ni. However, $^{50}$Ti and $^{66}$Zn
are under-produced. Furthermore, isotopes with 
odd-number atomic number, such as P, Cl and K, are mostly 
under-produced, with an exception of $^{51}$V. This is 
expected as the system is initially void of hydrogen for 
proton capture. In this benchmark model, we observed 
a production of $^{56}$Ni to be 0.67 $M_{\odot}$,
$3.34 \times 10^{-2} M_{\odot}$ neutron-rich 
species (such as $^{48-50}$Ti, $^{53-54}$Cr, 
$^{57-58}$Fe and $^{61-64}$Ni) and 
$0.385 M_{\odot}$ intermediate mass elements (IMEs). 
For the detailed velocity distribution of the 
products by pure turbulent deflagration, we refer
to Section \ref{sec:PTD} for a more
detailed discussion. 

\begin{figure}
\centering
\includegraphics*[width=8cm,height=5.7cm]{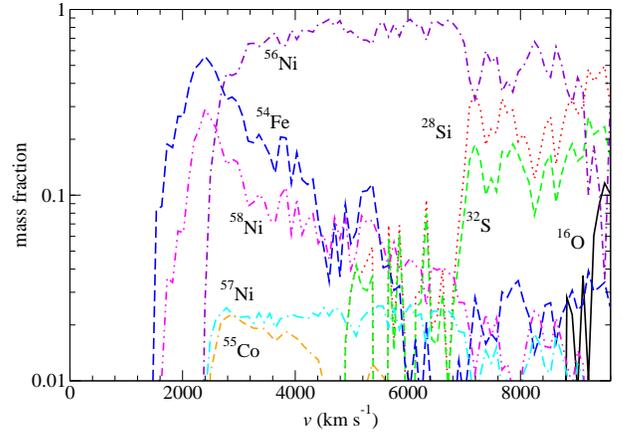}
\caption{The velocity profiles of major isotopes for 
benchmark model for the angle from $0 - 9$ degree.}
\label{fig:slice1_std}
\end{figure}

\begin{figure}
\centering
\includegraphics*[width=8cm,height=5.7cm]{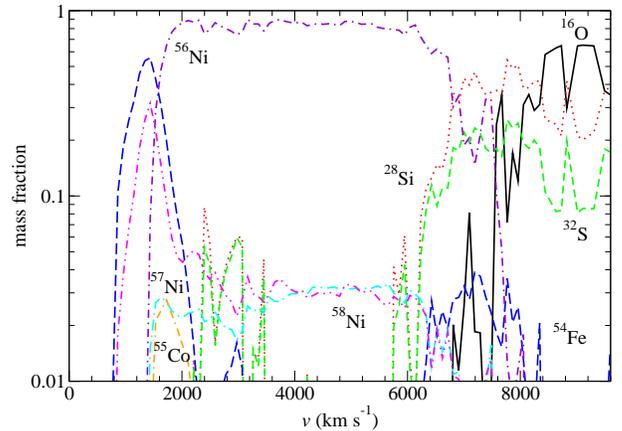}
\caption{The velocity profiles of major isotopes for 
benchmark model for the angle from $36 - 45$ degree.}
\label{fig:slice5_std}
\end{figure}

In Figure \ref{fig:slice1_std} we plot the velocity distribution
of major isotopes for the benchmark model at the end of simulations for 
the polar slide from $0 - 9$ degree (Slice 1). The chemical composition
is again obtained from the post-processing nucleosynthesis. 
The figure is consistent with the standard framework that 
the inner part, namely matter with low velocity, contains
mostly $^{56}$Ni. $^{54}$Fe and $^{58}$Ni located 
at the innermost part. In matter with $v \approx 7200 $ km s$^{-1}$, 
intermediate mass elements such as $^{28}$Si
and $^{32}$S becomes prominent. This corresponds to the 
flame entering the low density region. For $v \approx 6600$ km, 
there is a abrupt jump of $^{56}$Ni again, which 
results from detonation near the flame edge, where part of
the matter has a density $\sim 10^8$ g cm$^{-3}$. Close to 
$v \approx 9000$ km s$^{-1}$, intermediate mass elements (IMEs)
becomes prominent again. 
At the most outer part, the density is too low for nuclear 
reaction beyond carbon burning even for detonation. A trace of 
$^{16}$O is left behind. 

In Figure \ref{fig:slice5_std}, we make a plot similar to Figure
\ref{fig:slice1_std} but for the polar slide of
$36 - 45$ degrees (Slice 5). We choose this slide so as to 
make a contrast on the time difference between the 
quenching of deflagration and the arrival of 
detonation wave. As shown in Figure \ref{fig:flame_std},
detonation starts from the two opposite "fingers" of the far
end of the flame, but not the central "finger". This means,
before the detonation wave arrives the matter around the 
central "finger", the matter has certain time to
expand before being incinerated. Similar to the previous case,
isotopes with $Y_{\rm e} < 0.50$ are mostly found in 
the core, where $v < 3000$ km s$^{-1}$. The velocity space up to 
$v \approx 6000$ km s$^{-1}$ is filled with $^{56}$Ni. The IME
gap in this case is larger that of slice 1, that almost
no $^{56}$Ni is detected from $v = 5400 - 7200$ km s$^{-1}$. 
The second peak of $^{56}$Ni appears near $v= 7800$ km s$^{-1}$.
Close to $v \approx 9000$ km s$^{-1}$, the IMEs become prominent. 
Different from Slice 1, $^{16}$O appears in matter
with a velocity slightly less and also beyond than $8000$ km s$^{-1}$
for two distinctive reasons. For $v < 7800$ km s$^{-1}$, the 
remaining $^{16}$O comes from the tip of deflagration;
while for $v$ between 7800 - 9000 km s$^{-1}$, $^{16}$O appears 
due to the longer expansion time between the end of
deflagration and detonation. The amount of unburnt 
$^{16}$O is comparatively higher than that in Slice 1. 

In Figure \ref{fig:slice5_std}, we make a plot similar to Figure
\ref{fig:slice1_std} but for the polar slide of
$36 - 45$ degrees (Slice 5). We choose this slide so as to 
make a contrast on the time difference between the 
quenching of deflagration and the arrival of 
detonation wave. As shown in Figure \ref{fig:flame_std},
detonation starts from the two opposite "fingers" of the far
end of the flame, but not the central "finger". This means,
before the detonation wave arrives the matter around the 
central "finger", the matter has certain time to
expand before being incinerated. Similar to the previous case,
isotopes with $Y_{\rm e} < 0.50$ are mostly found in 
the core, where $v < 3000$ km s$^{-1}$. The velocity space up to 
$v \approx 6000$ km s$^{-1}$ is filled with $^{56}$Ni. The IME
gap in this case is larger that of slice 1, that almost
no $^{56}$Ni is detected from $v = 5400 - 7200$ km s$^{-1}$. 
The second peak of $^{56}$Ni appears near $v = 7800$ km s$^{-1}$.
Close to $v \approx 9000$ km s$^{-1}$, the IMEs become prominent. 
Different from Slice 1, $^{16}$O appears in matter
with a velocity slightly less and also beyond than 8000 km s$^{-1}$
for two distinctive reasons. For $v < 7800$ km s$^{-1}$, the 
remaining $^{16}$O comes from the tip of deflagration;
while for $v$ between 7800 - 9000 km s$^{-1}$, $^{16}$O appears 
due to the longer expansion time between the end of
deflagration and detonation. The amount of unburnt 
$^{16}$O is comparatively higher than that in Slice 1. 

\begin{figure}
\centering
\includegraphics*[width=8cm,height=5.7cm]{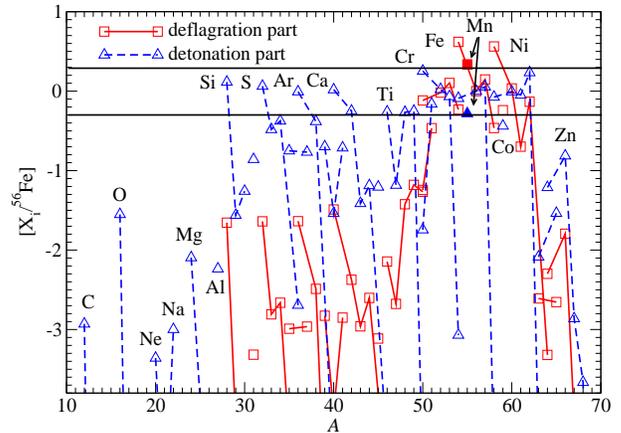}
\caption{The relative chemical abundance to $^{56}$Fe 
of benchmark model after the short-lived radioactive isotopes have 
decayed for the deflagration
and detonation regions in the benchmark model. 
The ratio are scaled with solar value.
($Remark$: The figure is replaced
due to the updated table.)}
\label{fig:final_defndet_std}
\end{figure}

We further classify the 
chemical yields of the tracer particles by checking whether
they reach their runaway by being swept by the deflagration
or detonation wave. Notice that it is possible that the 
tracer particles, first swept by the deflagration wave, 
are reheated by the shock collision from the detonation wave.
In these cases we still regard the chemical yield to 
be contributed by the deflagration wave. In Figure \ref{fig:final_defndet_std} 
we plot their corresponding chemical composition 
ratio to $^{56}$Fe scaled with solar abundance,
together with the total yield. 
It can be seen that the deflagration wave, similar to the 
one-dimensional model, contributes mostly to the 
formation of iron-peak elements, especially 
neutron-rich ones. For example, it has a higher $[{\rm X_i}/^{56}{\rm Fe}]$
fraction for $^{54}$Cr, $^{55}$Mn, $^{54}$Fe, $^{58}$Ni and $^{59}$Co. 
On the other hand, detonation, which swept mostly 
low-density region, produces less massive isotopes. 
IMEs such as $^{28}$Si, $^{32}$S, $^{36}$Ar and $^{40}$Ca
are mostly produced in detonation wave. 
Some lighter iron-rich elements, such as 
$^{46,48,49}$Ti and $^{54}$Cr are also produced by detonation.
As mentioned before, the unburned field surrounded 
by detonation wave is most of the time swept by the detonation
wave, which produces the necessary heating for producing
iron-peak elements. As a result, one can observe its contribution
to iron-peak elements including $^{62}$Ni and $^{66}$Zn. 

\section{Parameter Survey}
\label{sec:survey}

In this section, we study the
dependence on model parameters of carbon-oxygen WDs, by comparing
the results with the benchmark model. In Table 
\ref{table:Models}, we tabulate all model parameters
and their global results from hydrodynamics and 
nucleosynthesis. We follow the nomenclature in the 
literature \citep{Reinecke1999b} that $c3$ flame corresponds to the central
burning configuration, with a three-finger structure
to mimic the initial Rayleigh-Taylor instabilities. 
$b1a$ ($b1b$) is the one-ring configuration with at 
50 (100) km from the center. In Table \ref{table:Isotopes} - \ref{table:Isotopes3b}
we list the nucleosynthesis yield of the stable isotopes in
the representative models.

We remark that the nucleosynthesis results 
can be sensitive to the input physics, especially
to the microphysics. In particular, we expect that
the nucleosynthesis yield can change, when the 
nuclear reaction rate or electron capture rate
drastically change in the future. 
To show how the input physics affects the 
nucleosynthesis yield, in Appendix \ref{sec:revisit} 
we demonstrate by calculating the nucleosynthesis of the classical 
W7 and WDD2 models, but with our updated microphysics.

\begin{table*}

\begin{center}
\caption{Model setup for the benchmark model: 
central densities of NM $\rho_{ c{\rm (NM)} }$
are in units of $10^{9}$ g cm$^{-3}$.  
Metallicity is in units of solar metallicity.
The total mass $M$ and
and the final nickel-56 mass $M_{{\rm Ni}}$ 
are in units of solar mass. 
$R$ is the initial stellar radius. 
$E_{\rm nuc}$ and $E_{\rm tot}$ are the energy released by nuclear reactions
and final total energy, respectively, both in units of $10^{50}$ erg. 
$Y_{{\rm e(min)}}$ is the minimum value of electron fraction within
the simulation box at the end of simulation. $t_{{\rm DDT}}$ is 
the first detonation transition time in units of second. 
$M_{{\rm Ni}}$ is the mass of $^{56}$Ni at the end of 
simulations, in units of $M_{\odot}$. 
In the column Others, miscellaneous setting is described. 
'only def.' and 'only det.' stands for models which are
fixed to have either deflagration or detonation only.}
\label{table:Models}
\begin{tabular}{|c|c|c|c|c|c|c|c|c|c|c|c|c|}

\hline
Model & $\rho_{c({\rm NM})}$ & Metallicity & flame shape & $X_{12C}$ & $M$ & 
$R$ & $Y_{{\rm e(min)}}$ & $E_{\rm nuc}$ & $E_{\rm tot}$ & $t_{{\rm DDT}}$ & $M_{{\rm Ni}}$ & Others \\ \hline
050-0-c3-1    & 0.50 & 0 & $c3$   & 0.50  & 1.30 & 3060 & 0.488 & 18.5 & 14.6 & 1.34 & 0.97 & \\
050-1-c3-1    & 0.50 & 1 & $c3$   & 0.49  & 1.30 & 3060 & 0.488 & 18.3 & 14.4 & 1.34 & 0.89 & \\ 
050-1-c3-1P   & 0.50 & 1 & $c3$   & 0.49  & 1.30 & 3060 & 0.488 &  5.0 &  1.1 &  N/A & 0.21 & only def. \\ 
050-1-c3-1D   & 0.50 & 1 & $c3$   & 0.49  & 1.30 & 3060 & 0.488 & 19.2 & 15.3 &  N/A & 1.14 & only det. \\ 
050-3-c3-1    & 0.50 & 3 & $c3$   & 0.47  & 1.30 & 3060 & 0.488 & 17.6 & 13.7 & 1.35 & 0.74 & \\
050-5-c3-1    & 0.50 & 5 & $c3$   & 0.45  & 1.30 & 3060 & 0.488 & 17.3 & 13.3 & 1.35 & 0.63 & \\ \hline
075-0-c3-1    & 0.75 & 0 & $c3$   & 0.50  & 1.31 & 2600 & 0.482 & 18.4 & 14.2 & 1.19 & 0.90 & \\
075-1-c3-1    & 0.75 & 1 & $c3$   & 0.49  & 1.31 & 2600 & 0.482 & 18.1 & 13.9 & 1.19 & 0.81 & \\ 
075-1-c3-1P   & 0.75 & 1 & $c3$   & 0.49  & 1.31 & 2600 & 0.482 &  6.0 &  1.8 &  N/A & 0.24 & only def. \\ 
075-1-c3-1D   & 0.75 & 1 & $c3$   & 0.49  & 1.31 & 2600 & 0.482 & 19.8 & 15.6 &  N/A & 1.12 & only det. \\ 
075-3-c3-1    & 0.75 & 3 & $c3$   & 0.47  & 1.31 & 2600 & 0.482 & 17.8 & 13.6 & 1.19 & 0.71 & \\
075-5-c3-1    & 0.75 & 5 & $c3$   & 0.45  & 1.31 & 2600 & 0.482 & 17.4 & 13.2 & 1.20 & 0.60 & \\ \hline
100-0-c3-1    & 1.00 & 0 & $c3$   & 0.50  & 1.33 & 2600 & 0.479 & 18.1 & 13.7 & 1.10 & 0.87 & \\ 
100-1-c3-1    & 1.00 & 1 & $c3$   & 0.49  & 1.33 & 2600 & 0.479 & 18.0 & 13.5 & 1.10 & 0.75 & \\ 
100-1-c3-1P   & 1.00 & 1 & $c3$   & 0.49  & 1.33 & 2600 & 0.479 &  6.7 &  2.2 &  N/A & 0.26 & only def. \\ 
100-1-c3-1D   & 1.00 & 1 & $c3$   & 0.49  & 1.33 & 2600 & 0.479 & 20.3 & 15.8 &  N/A & 1.10 & only det. \\ 
100-3-c3-1    & 1.00 & 3 & $c3$   & 0.47  & 1.33 & 2600 & 0.479 & 17.5 & 13.1 & 1.10 & 0.66 & \\ 
100-5-c3-1    & 1.00 & 5 & $c3$   & 0.45  & 1.33 & 2600 & 0.479 & 16.9 & 12.5 & 1.11 & 0.50 & \\ \hline
300-0-c3-1    & 3.00 & 0 & $c3$   & 0.50  & 1.38 & 1900 & 0.462 & 18.4 & 13.4 & 0.78 & 0.70 & \\ 
300-1-c3-1    & 3.00 & 1 & $c3$   & 0.49  & 1.38 & 1900 & 0.462 & 17.7 & 12.7 & 0.78 & 0.63 & \\ 
300-1-c3-1P   & 3.00 & 1 & $c3$   & 0.49  & 1.38 & 1900 & 0.462 &  9.5 &  4.5 &  N/A & 0.31 & only def. \\ 
300-1-c3-1D   & 3.00 & 1 & $c3$   & 0.49  & 1.38 & 1900 & 0.462 & 21.0 & 16.0 &  N/A & 1.09 & only det. \\ 
300-3-c3-1    & 3.00 & 3 & $c3$   & 0.47  & 1.38 & 1900 & 0.462 & 17.6 & 12.6 & 0.78 & 0.55 & \\ 
300-5-c3-1    & 3.00 & 5 & $c3$   & 0.45  & 1.38 & 1900 & 0.462 & 17.4 & 12.4 & 0.79 & 0.44 & \\ \hline
500-0-c3-1    & 5.00 & 0 & $c3$   & 0.50  & 1.39 & 1600 & 0.453 & 19.0 & 13.9 & 0.66 & 0.67 & \\ 
500-1-c3-1    & 5.00 & 1 & $c3$   & 0.49  & 1.39 & 1600 & 0.453 & 18.6 & 13.4 & 0.66 & 0.59 & \\ 
500-1-c3-1P   & 5.00 & 1 & $c3$   & 0.49  & 1.39 & 1600 & 0.453 & 11.1 &  5.9 &  N/A & 0.32 & only def. \\ 
500-1-c3-1D   & 5.00 & 1 & $c3$   & 0.49  & 1.39 & 1600 & 0.453 & 21.2 & 16.0 &  N/A & 1.05 & only det. \\ 
500-3-c3-1    & 5.00 & 3 & $c3$   & 0.47  & 1.39 & 1600 & 0.453 & 18.1 & 13.0 & 0.67 & 0.50 & \\ 
500-5-c3-1    & 5.00 & 5 & $c3$   & 0.45  & 1.39 & 1600 & 0.453 & 17.9 & 12.8 & 0.67 & 0.40 & \\ \hline
300-1-b1a-1   & 3.00 & 1 & $b1a$  & 0.49  & 1.38 & 1900 & 0.455 & 18.5 & 13.4 & 0.95 & 0.68 & \\ 
300-1-b1a-1P  & 3.00 & 1 & $b1a$  & 0.49  & 1.38 & 1900 & 0.486 &  5.3 &  0.2 &  N/A & 0.22 & only def. \\ 
300-1-b1b-1   & 3.00 & 1 & $b1b$  & 0.49  & 1.38 & 1900 & 0.459 & 18.2 & 13.6 & 1.03 & 0.78 & \\ 
300-1-b1b-1P  & 3.00 & 1 & $b1b$  & 0.49  & 1.38 & 1900 & 0.491 &  5.5 &  5.5 &  N/A & 0.23 & only def. \\ 
300-1-b1b-1D  & 3.00 & 1 & $b1b$  & 0.49  & 1.38 & 1900 & 0.457 & 21.1 & 16.5 &  N/A & 1.03 & only det. \\ \hline
300-1-c3-0.6  & 3.00 & 1 & $c3$   & 0.37  & 1.38 & 1700 & 0.462 & 14.8 & 9.7  & 0.77 & 0.46 & C/O = 0.6 \\ 
300-1-c3-0.6P & 3.00 & 1 & $c3$   & 0.37  & 1.38 & 1700 & 0.462 &  9.1 & 4.0  &  N/A & 0.48 & C/O = 0.6 \\
			  &      &   &        &       &      &      &       &      &      &      &      & only def. \\
300-1-c3-0.6D & 3.00 & 1 & $c3$   & 0.37  & 1.38 & 1700 & 0.462 & 20.2 & 15.1 &  N/A & 1.07 & C/O = 0.6 \\
			  &      &   &        &       &      &      &       &      &      &      &      & only det. \\ \hline			
300-1-c3-0.3  & 3.00 & 1 & $c3$   & 0.23  & 1.38 & 1700 & 0.462 & 11.0 & 6.0  & 0.76 & 0.32 & C/O = 0.3 \\ 
300-1-c3-0.3P & 3.00 & 1 & $c3$   & 0.23  & 1.38 & 1700 & 0.462 &  5.3 & 2.2  &  N/A & 0.32 & C/O = 0.3 \\
 			  &      &   &        &       &      &      &       &      &      &      &      & only def. \\
300-1-c3-0.3D & 3.00 & 1 & $c3$   & 0.23  & 1.38 & 1700 & 0.462 & 19.6 & 14.6 &  N/A & 0.82 & C/O = 0.3 \\ 
			  &      &   &        &       &      &      &       &      &      &      &      & only det. \\ \hline

\end{tabular}
\end{center}
\end{table*}

\subsection{Effects of metallicity}

\begin{figure*}
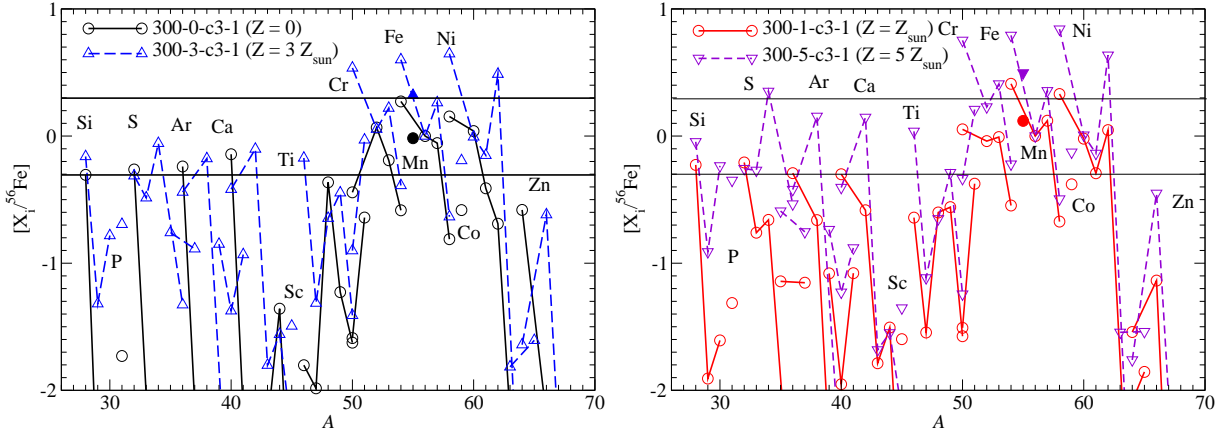

\centering
\includegraphics*[width=8cm,height=5.7cm]{fig8a_erratum.eps}
\includegraphics*[width=8cm,height=5.7cm]{fig8b_erratum.eps}
\caption{Dependence on metallicity (Z).
left: $Z = 0$ and 3,  300-0-c3-1 ($Z$=0), 300-3-c3-1 ($Z=3Z_\odot$);
right: $Z = 1$ and 5,  same as above (left panel), but for 
300-1-c3-1 ($Z = Z_{\odot}$), 300-5-c3-1 ($Z=5Z_\odot$). 
($Remark$: The figure is replaced
due to the updated table.)}
\label{fig:final_Z}
\end{figure*}

Models 300-0-c3-1, 300-1-c3-1, 300-3-c3-1 and 
300-5-c3-1 form a set to study the effect 
of metallicity on SNe Ia. Since all models start
from the same density and the same $^{12}$C/$^{16}$O ratio,
there is no observable difference in mass and radius.
The minimum electron fraction, which comes 
from matter burnt near the center, is insensitive
to metallicity and is about 0.460. But the energy release
and the final total energy decrease when metallicity
increases. This is because the $^{22}$Ne has a smaller
binding energy change when it is burnt compared to $^{12}$C. 
Also, the mixture with $^{22}$Ne lowers $Y_{\rm e}$,
which suppresses the $^{56}$Ni production at NSE. 
The detonation transition
time is also insensitive to metallicity. 

In Figure 
\ref{fig:final_Z} we plot the $[{\rm X_i}/^{56}{\rm Fe}]$ of stable isotopes in 
the three models. Metallicity can enhance strongly
certain isotopes, including $^{46}$Ti, $^{50-51}$V,
$^{50}$Cr, $^{55}$Mn, $^{54}$Fe, $^{57}$Fe, $^{58}$Ni and 
$^{62}$Ni. These isotopes are under-produced in the zero
metallicity limit, but are mostly overproduced for
$Z = 3 Z_{\odot}$ case. This suggests that in order to 
create the composition similar to the solar abundances, 
SN Ia itself has a metallicity close to the solar value.
From Tables \ref{table:Isotopes} and \ref{table:Isotopesb}, it can be seen that 
the presence of $^{22}$Ne strongly enhances the 
production of many isotopes, but suppresses the production 
of isotopes closely related to the alpha-chain, 
such as $^{32}$S, $^{40}$Ca, $^{52}$Cr, $^{56}$Fe and so on.

\subsection{Effects of central density}

\begin{figure*}
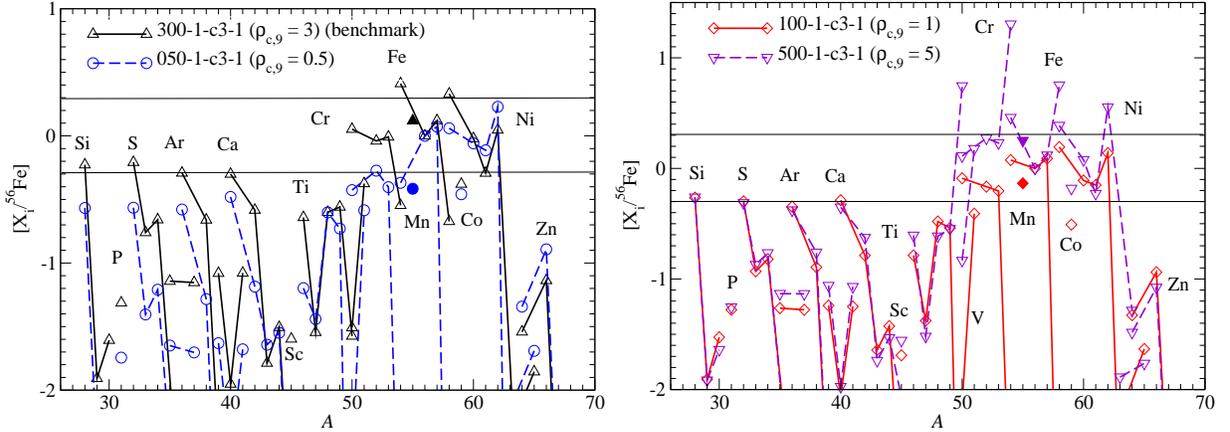

\centering
\includegraphics*[width=8cm,height=5.7cm]{fig9a_erratum.eps}
\includegraphics*[width=8cm,height=5.7cm]{fig9b_erratum.eps}
\caption{Dependence on the initial 
central density ($\rho_{c,9} = \rho_c / 10^9$ g cm$^{-3}$).
Left: 050 and 300, 050-1-c3-1 ($\rho_{c,9}$ = 0.5), 300-1-c3-1 ($\rho_{c,9}$ = 3); 
right: 100 and 500,  same as above (left panel), but for 
100-1-c3-1 ($\rho_{c,9}$ = 0.5), 500-1-c3-1 ($\rho_{c,9}$ = 3).
($Remark$: The figure is replaced
due to the updated table.)}
\label{fig:final_rho}
\end{figure*}

Models 050-1-c3-1, 100-1-c3-1, 300-1-c3-1 and 500-1-c3-1 form another series of models which 
studies the effects of initial central density. In this case, 
the initial mass increases with density while
the radius decreases, meaning that a more compact WD 
at the beginning for a higher central densities. Also, the higher
density can create a much hotter core, which makes the
electron capture rates higher and the 
minimum electron fraction lower. On the other hand, because
of the higher density, more energy is lost by neutrino
and electron capture, which means the total energy 
production decreases. But the higher density provides
a faster laminar flame at the beginning, which triggers
faster production of turbulence and leads to earlier
detonation transition. The lower electron
fraction decreases the $^{56}$Ni 
in NSE, so that the $^{56}$Fe
mass fraction decreases when central density increases. 

In Figure \ref{fig:final_rho} we plot
the $[{\rm X_i}/^{56}{\rm Fe}]$ of the stable isotopes 
for the five models to show
the effects of central density.
All models show an underproduction of IMEs (SI, S, Ar and
Ca). Their abundances increase slightly with $M$ and
become saturated at $\rho_c = 3 \times 10^9$ g cm$^{-3}$. 
Certain isotopes which are under-produced at
low density are significantly enhanced at high density.
They include $^{46}$Ti, $^{50-54}$Cr, $^{55}$Mn, 
$^{54}$Fe, $^{58}$Fe, $^{58}$Ni and $^{62}$Ni. 
It can be observed easily that the over-production
of low-$Y_{\rm e}$ isotopes including 
$^{50}$Ti, $^{54}$Cr, $^{58}$Fe and $^{62}$Ni
occur at high density. At $\rho_c = 5 \times 10^9$ g cm$^{-3}$,
[${\rm X_i}/^{56}$Fe] of these isotopes are
10 times higher than solar abundance ratios.
These isotopes are so much over-produced that we expect
that SNe Ia with this density should be less frequently to
occur. (See, however, Section \ref{sec:ye_mix} 
and Figure \ref{fig:final_mixing_std} for the effects
on $Y_{\rm e}$ mixing.) 

In terms of isotope masses in Tables \ref{table:Isotopes} and \ref{table:Isotopesb},
at low densities, most of the isotopes masses are smaller,
with representative exceptions of $^{50-51}$V 
and $^{56}$Fe. This is contributed to the more massive
zone being incinerated by detonation instead of deflagration.
On the other hand, at high density, in general most 
isotope masses increase, especially the low-$Y_{\rm e}$ isotopes,
for instance $^{46-48}$Ca, $^{54}$Cr, $^{60}$Fe, show
order-of-magnitude jump when density reaches $5 \times 10^9$
and $7 \times 10^9$ g cm$^{-3}$. This part reveals that 
in order to match the solar abundance for most isotopes, 
the suitable density is about $2 - 4 \times 10^9$ g cm$^{-3}$. 
For lower densities, there is a suppression in low-$Y_{\rm e}$
isotopes and $^{55}$Mn. On the other hand, these isotopes
are severely overproduced when density exceeds this range.

\subsection{Effects of $Y_{\rm e}$ mixing}
\label{sec:ye_mix}

In our calculations we have applied the tracer particle
algorithm to do the post-process nucleosythesis
calculation. The tracer particles record the local
density and temperature from the projected
Eulerian grid while they are advected by the
fluid motion. The nuclear reactions and the 
corresponding electron capture are calculated
based on the thermodynamics trajectories. 

However, subtlety appears in this scheme. As the star 
has finished its carbon deflagration and detonation, 
the star expands. Simultaneously, the density and 
temperature drops because locally the matter adiabatic
expands. There exists a period of time that the matter
remains sufficiently hot $(> 10^9$ K) while the turbulent
motion remains significant. The matter with different
density and temperature may mix during expansion 
before it reaches a real homologous expansion. 
The temperature and density after mixing can 
be naturally captured by the tracer particles. 
But it does not carry information if 
the mixing of electron fraction since it is a quantity
later derived from post-processing. Notice that
we have included electron capture in the NSE 
as done in \cite{Seitenzahl2009}. Notice that, this electron 
fraction can be different from the post-processed 
ones when strong mixing occurs. Effectively, 
the "real" $Y_{\rm e}$ in the fluid parcel
can be higher as the matter mixes
with the surrounding of lower densities. This effect
will be important if such mixing begins before the
tracer particles leave NSE. 

\begin{figure}
\centering
\includegraphics*[width=8cm,height=5.7cm]{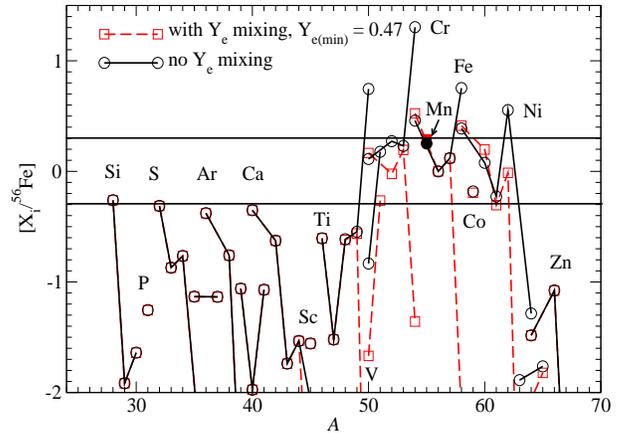}
\caption{The relative chemical abundance to $^{56}$Fe 
of Model 500-1-c3-1 ($\rho_{{\rm c, ini}} = 5 \times 10^9$ g cm$^{-3}$)
after the short-lived radioactive isotopes have 
decayed for the benchmark model with or without electron 
fraction mixing. The ratio are scaled with solar value.
($Remark$: The figure is replaced
due to the updated table.)}
\label{fig:final_mixing_std}
\end{figure}

To mimic this effect, we assume there exists some 
lower limit of electron fraction. This imitates the 
mixing of the lower $Y_{\rm e}$ matter with the 
surrounding high $Y_{\rm e}$ matter. In the Model
500-1-c3-1 ($\rho_c = 5 \times 10^9$ g cm$^{-3}$), 
the lower $Y_{\rm e}$ reaches by the star is 
$\approx 0.453$, while the typical $Y_{\rm e}$ in ash 
is $\approx 0.47$. In the post-processing, we stop
the electron capture as long as the $Y_{\rm e}$ of
the tracer particles reach this lower limit. 
In Figure \ref{fig:final_mixing_std} we plot the 
corresponding $[{\rm X_i}/^{56}{\rm Fe}]$ of the stable isotopes. The original one, 
which does not take $Y_{\rm e}$ mixing into account,
is included. It can be seen that the $Y_{\rm e}$-mixing
has a smaller effect to IMEs but stronger 
effect on iron-peak elements. 
Since $Y_{\rm e}$ influences mostly neutron-rich
isotopes of iron-peak elements, there is no
observable change to the mass fraction of IMEs. 
$^{28}$Si to $^{49}$Ti are equal in both models. 
Neutron rich isotopes, including $^{50}$Ti, $^{50-51}$V,
$^{52-54}$Cr, $^{58}$Fe, $^{59}$Co and $^{62}$Ni are
strongly dependent on the mixing process. 
A difference of an order-of-magnitude can be observed. 

\subsection{Effects of initial flame structure}

\begin{figure}
\centering
\includegraphics*[width=8cm,height=5.7cm]{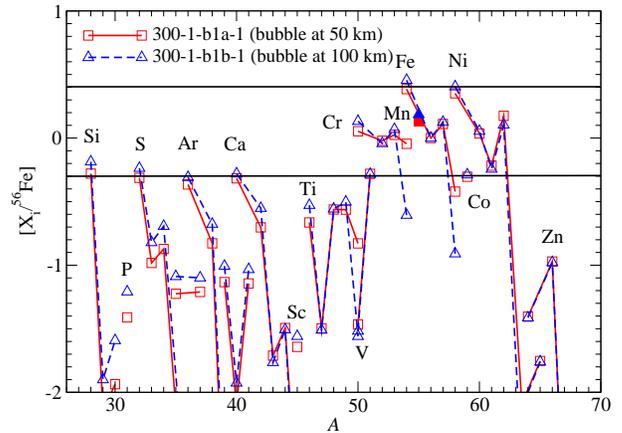}
\caption{Dependence on flame shape:
Similar to Figure \ref{fig:final_Z} but for Models
300-1-b1a-1 (one bubble at 50 km from origin) and 
300-1-b1b-1 (one bubble at 100 km from origin).
($Remark$: The figure is replaced
due to the updated table.)}
\label{fig:final_flame}
\end{figure}

Models 300-1-c3-1, 300-1-b1a-1 and 300-1-b1b-1 study the effects of initial 
flame shape. Model 300-1-b1a-1 (300-1-b1b-1) assumes the flame starts 
from a ring at around 50 (100) km from the origin with 
a radius of 15 km. The three cases have similar explosion
energy and nuclear energy release. But their minimum 
electron fraction is very different, where flame starts
from the center has the lowest electron fraction, which 
is expected as the matter at high density has sufficient 
time to burn and then to carry out electron capture when
matter is in NSE. On the contrary, off-center burning
cannot provide such condition for electron capture at
early time. In terms of detonation transition, off-center
burning tends to have detonation at later time, which is 
because the initial bubble is much weaker to create
expansion of the WD and also the turbulent flow. But 
rings located further out can start the explosion sooner
since the flame front can reach the low-density regime, 
one of the keys for distributed burning, at earlier time.
 
In Figure \ref{fig:final_flame} we plot the final nucleosynthesis
yield for the two models. In contrast to previous tests,
the flame structure, which alters significantly the explosion
dynamics, does not influence the qualitative pattern of 
chemical abundance. When the initial incinerated 
zone is farther from the center, the lower production of low Ye
isotopes, such as $^{48-49}$Ti, $^{52}$Cr, $^{60-62}$Ni and
so on, become more abundant. 
On the contrary, high $Y_{\rm e}$ isotopes are enhanced, such 
as $^{46}$Ti, $^{50}$Cr, $^{54}$Fe and $^{55}$Mn. However, in
general their production is lower than the centered
burning cases. It shows that the flame structure in
two-dimensional model is less important as long as the flame
front can reach the center at early time. It has more
influences on the production of IMEs.

\subsection{Effects of initial carbon-oxygen ratio}

\begin{figure}
\centering
\includegraphics*[width=8cm,height=5.7cm]{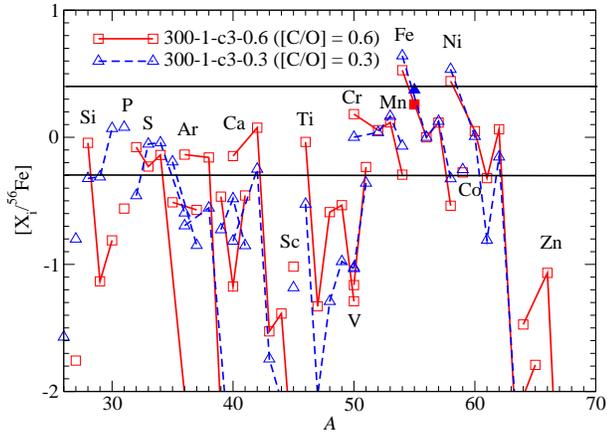}
\caption{Dependence on [C/O] ratio:
Similar to Figure \ref{fig:final_Z} but for Models
300-1-c3-0.6 ([C/O] = 0.6) and
300-1-c3-0.3 ([C/O] = 0.3). ($Remark$: The figure is replaced
due to the updated table.)}
\label{fig:final_coratio}
\end{figure}

Models 300-1-c3-1, 300-1-c3-0.6 and 300-1-c3-0.3 study the effects of initial 
carbon-oxygen ratio. In most works about the explosion phase
a $^{12}$C/$^{16}$O ratio is assumed to be unity.
The exact C/O ratio should depend on the stellar 
evolution, in particular whether there is carbon burning 
in the carbon-oxygen core. In \cite{Umeda1999}, it is shown that 
the C/O value can reach as low as 0.3 depending on 
the initial carbon-oxygen core mass and the 
metallicity, which is much lower than the value assumed 
in the literature. In these three models, we study the 
role of this value by choosing three contrasting values
from 0.3 to 1. In terms of explosion energetic, when 
C/O ratio decreases, the minimum $Y_{\rm e}$ increases. This is because
for a lower C/O ratio, the energy release by the carbon deflagration
is lower. This causes a lower final temperature of the 
ash temperature, which corresponds to a lower electron 
capture rate. The transition time does not show a
significant change, because the deflagration phase of the
three models are roughly similarly. Most fuel
is burnt to NSE. The explosion energy and 
the final total energy are also lower when the C/O 
ratio increases. Also, the global lower energy releases
due to the lower energy production in the detonation, 
At last $^{56}$Ni produced decreases as well owing to 
the weaker detonation. 

We plot the mass ratio to 
$^{56}$Fe relative to the solar value in Figure
\ref{fig:final_coratio} and their values in 
Tables \ref{table:Isotopes2} and \ref{table:Isotopes2b}. The chemical abundance 
shows contrasting resulting in the values and 
the mass ratio. Due to a weaker explosion, the 
lowered $^{56}$Ni production may boost the 
mass ratio. On the other hand, the lower energy
input also suppresses the burning in the later 
stage. When C/O ratio decreases,
the masses of lower $Y_{\rm e}$ isotopes in the iron-peaked elements
decreases, such as $^{50}$Cr, $^{52}$Cr, $^{54}$Fe, 
$^{56-57}$Fe; while those of higher $Y_{\rm e}$ increase, 
such as $^{53-54}$Cr, $^{58}$Fe, $^{60}$Fe.
In contrast, the mass ratio does not have a uniform
trend for these elements. For example, $^{54}$Fe and 
$^{58}$Ni show an increasing mass ratio when 
C/O ratio decreases, but no similar tendency for 
$^{61}$Ni and $^{62}$Ni. Similar feature appears
in intermediate-mass isotopes, such as $^{36}$Ar, 
$^{38}$Ar, $^{40}$Ca and $^{42}$Ca. In terms of 
total mass, there is a mild increase in these isotopes
when C/O ratio decreases from 1.0 to 0.6, but 
a significant drop when that further decreases from
0.6 to 0.3. This suggests that for low C/O ratio 
star, the reduction of explosion energy becomes
dominant in the nucleosynthesis process. Such
feature also suggests that a thorough knowledge 
in the progenitor C/O ratio is critical in determining
the correct global population of chemical species.

\subsection{Effects of the detonation trigger}

\begin{figure}
\centering
\includegraphics*[width=8cm,height=5.7cm]{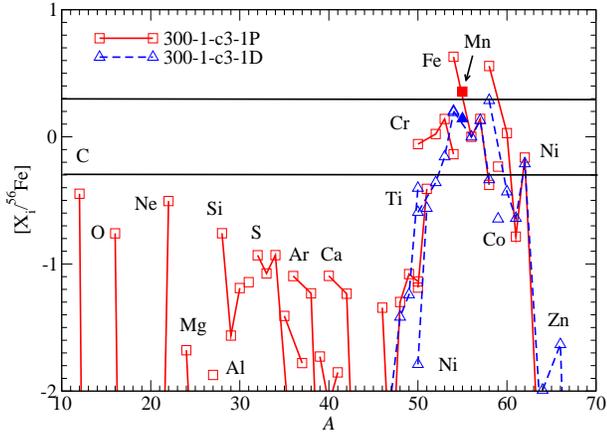}
\caption{Similar to Figure \ref{fig:final_Z} but for Models
300-1-c3-1P, 300-1-c3-1D. The nucleosynthesis of the whole star
is included. ($Remark$: The figure is replaced
due to the updated table.)}
\label{fig:final_DD}
\end{figure}

In modeling SNe Ia, deflagration-Detonation
transition (DDT) is important in order
to explain the observed brightness. 
The nature of DDT has been demonstrated in terrestrial
experiments, such as the air-H$_2$ experiments \citep{Poludenko2011}.
However, the counterpart in SNe Ia is unclear. 
Besides, in numerical estimations the turbulence required
to trigger the DDT is stronger than what is shown in 
the numerical experiments. 
Recent discoveries of SNe Iax hint on 
possibilities that no DDT or failed DDT occurs
in this scenario. 
This points to the needs for the pure turbulent deflagration models in our
model collection.

\subsubsection{Pure Turbulent Deflagration Models}

In Table \ref{table:std}, models name with an ending "P" corresponds
to the model with the DDT trigger switched off that
simulates the case of very large $Ka$ for the DDT criterion.  
Our treatment can also be
regarded as the approximation to the case of a failed DDT, 
caused by some external reasons,
such as the very small carbon fraction 
(and large fractions of O and Ne) in the progenitor WD.

In some pure turbulent deflagration models, a significant portion 
of $^{12}$C and $^{16}$O remains unburnt.  As a result, 
the WD has a
much lower final total energy after all deflagration 
wave has quenched, compared to the corresponding DDT
or pure detonation model. 
For example in Model 050-1-c3-1P the final total 
energy is $1.1 \times 10^{50}$ erg. In these cases, 
the nuclear energy is unlikely to make the whole
star explode. Instead, the hot ash floats 
upward and transfers its momentum to the 
outer lower density layers. This causes partial ejection
of the outer layers with some mixture from the deflagration ash
by convective mixing.
A WD remnant is left behind with the materials 
the original WD (C and O) and 
a range of isotopes from the deflagration.  
The failure of unbinding the star is also connected
to the missing of nebular spectra.
In Tables 
\ref{table:IsotopesPTD}, \ref{table:IsotopesPTDb} and \ref{table:DecayPTD}, 
we list the mass distributions of the stable isotopes
and some long-live radioactive isotopes
from some representative pure turbulent deflagration 
models. 

In Figure \ref{fig:final_DD} we plot the scaled
mass fraction similar to Figure \ref{fig:final_Z}
but for Models 300-1-c3-1P and 300-1-c3-1D. 
(Compare the benchmark model 300-1-c3-1 in Figure 4.)
The nucleosynthesis of the whole SN Ia is included. 
In the pure turbulent deflagration model, 
$^{12}$C and $^{16}$O are significant.
The deflagraation ash includes iron-peak elements, where iron-group elements 
(especially $^{54}$Fe and $^{58}$Ni) tend to be overproduced 
because of a lower $^{56}$Fe mass.
The ash also includes relatively small amount of intermediate mass
elements (IME) such as $^{24}$Mg, $^{28}$Si, $^{32}$S and $^{36}$Ar.

\subsubsection{Pure Detonation Models}

The other limiting case corresponds to the models exploded
by pure detonation. 
To reduce the uncertainties, 
models with their names ended with "D" are not necessary
the detonation is triggered by deflagration, as
in high density the flame size is always smaller than
typical eddy size, making the heat diffusion of the
ash to the fuel slow. In fact, another possible
scenario is similar to the double detonation model. 
Assuming a sufficiently slow helium accretion, 
the helium can be accumulated thick enough to trigger
helium detonation rather than helium deflagration. 
The shock wave created by the helium detonation can
trigger the consequent carbon detonation in the core, 
when the helium detonation possesses high degree of symmetry.

Nucleosynthesis yields in the pure detonation model are seen in Figure
\ref{fig:final_DD} for 300-1-c3-1D.
In the pure detonation models, most of materials are burnt into
iron-peak elements due to the strong detonation.  
Therefore, production of C+O, IMEs, Ti and Cr are suppressed.
The pattern of iron-group elements for the these models is similar. 
In pure detonation model, no significant
overproduction of iron-group elements is seen.

\subsection{Connections between Pure Turbulent Deflagration and Type Iax Supernovae}
\label{sec:PTD}

The pure turbulent deflagration model has been suggested as a possible
model for peculiar subluminous SNe Ia, i.e., SNe Iax (e.g., \cite{Jha2017}).
If the DDT is triggered, 
the detonation produces too much $^{56}$Ni to match with observations.
Also, the detonation tends to produce stratified composition in 
its ash, which conflicts with the strong mixing as shown in SNe Iax
spectra.  Furthermore, the pure turbulent deflagration can 
leave a WD remnant, which is consistent with the late time 
spectra of SNe Iax (e.g., \cite{Jha2017}) 

In view of that, we further discuss the hydrodynamics
and the nucleosynthesis of the pure turbulent deflagration models. Some of the models, 
such as Model 300-1-c3-1P, can be compared with some
models in the literature (see for example \cite{Fink2014} 
for the pure turbulent deflagration models with mainly different flame structure). 

In Table \ref{table:PTD} we tabulate the explosion energetic results
and some global quantities of nucleosythesis. (We intend to 
repeat some quantities as listed in \ref{table:Models} so
as to make the table comparable with Table 1 in \cite{Fink2014}.)
It can be seen that in general when the central density 
increases, corresponding to a more massive CO WD progenitor, 
the explosion becomes stronger. This is because the faster 
burning rate and faster flame propagation rate at high density.
Also, the star is more compact so that the star expands
only after more material is burnt to supply the first 
expansion. As a result, in a more massive CO WD, the 
pure turbulent deflagration model gives more massive ejecta, which spanned from 0.21
$M_{\odot}$ to 0.32 $M_{\odot}$, while the ejecta
mass has a range from 0.22 $M_{\odot}$ to 
1.10 $M_{\odot}$. 

\begin{figure}
\centering
\includegraphics*[width=8cm,height=5.7cm]{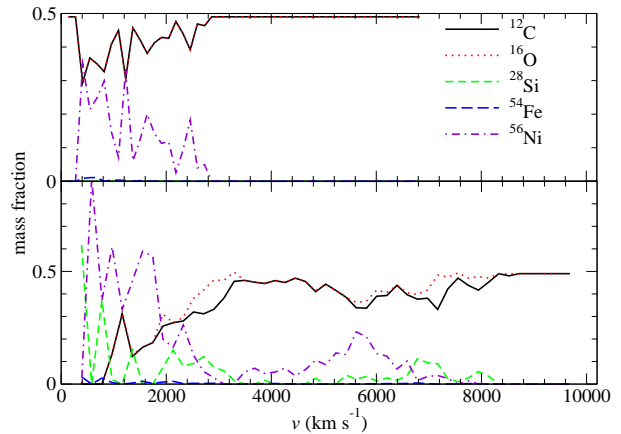}
\caption{The mass fraction of the ejecta against
asymptotic ejecta velocity for Models 050-1-c3-1P and 
500-1-c3-1P. Only the tracer particles with its
velocity exceeding the escape velocity is accounted. 
The velocity is derived by removing the gravitational
energy component. }
\label{fig:velmap_PTD}
\end{figure}

In Figure \ref{fig:velmap_PTD} we plot the chemical abundance
distribution in the asymptotic ejecta velocity space. 
The asymptotic ejecta velocity $v_{{\rm asy}}$ is derived from the 
tracer particle local gravitational potential $\phi$ and 
final velocity $v_{{\rm end}}$ by $v_{{\rm asy}} = \sqrt{v_{{\rm end}}^2 + \phi}$.
Particles with a velocity below the escape velocity is
ignored because they are bounded after the expansion. 
We plot the velocity map for two contrasting Models 
050-1-c3-1 and 500-1-c3-1. The model with a lower initial
mass has a lower maximum ejecta speed about 7000 km s$^{-1}$, 
compared to the high mass model of 9000 km s$^{-1}$. 
$^{56}$Ni can be found in the low velocity region from 
$0 - 3000$ km s$^{-1}$. Beyond that, only the shock compressed 
carbon and oxygen are found. On the other hand in the
high mass model, in the low velocity field, significant
amount of $^{28}$Si and $^{56}$Ni are observed. 
Then there are mostly $^{12}$C and $^{16}$O around
3000 - 6000 km s$^{-1}$, coming from the exciting atmosphere. 
At last, at high velocity region, a non-zero amount of 
$^{28}$Si and $^{56}$Ni are found again. This shows signs 
of mixing by the Rayleigh-Taylor instabilities.

\begin{table*}

\begin{center}
\caption{The explosion energetic and global nucleosynthesis 
quantities of the pure turbulent deflagration models. $M_{{\rm ej}}$ and $M_{{\rm b}}$
are the ejecta and remnant mass of the SNe Ia in these models.
$M_{{\rm ej}}$ ($M_{{\rm b}}$) is the total mass where the fluid elements have
a velocity above (below) the escape velocity.
$M_{{\rm burn}}$ is the ash mass at the end of simulations.
$M$(IGE) and $M$(IME) are the iron-peak elements and intermediate
mass elements derived after all short-life radioactive isotopes
have decayed. $M_{{\rm ej,^{56}Ni}}$,  $M_{{\rm ej}}({\rm IGE})$ and $M_{{\rm ej}}({\rm IME})$
are the masses of $^{56}$Ni, iron-peak elements and intermediate
mass elements derived after all short-life radioactive isotopes
have decayed in the ejecta. 
All masses are in united of $M_{\odot}$.
$E_{{\rm nuc}}$ and $E_{{\rm tot}}$ are the energy released by nuclear reaction
and the asymptotic energy at the end of simulations
in units of $10^{50}$ erg. }
\label{table:PTD}
\begin{tabular}{|c| c c c c c c c c | c c c|}

\hline
Model & $E_{{\rm nuc}}$ & $E_{{\rm tot}}$ & $M_{{\rm ej}}$ & $M_{{\rm b}}$ & $M_{{\rm burn}}$ & $M_{{\rm ^{56}Ni}}$ & 
$M$(IGE) & $M$(IME) & $M_{{\rm ej,^{56}Ni}}$ & $M_{{\rm ej}}({\rm IGE})$ & $M_{{\rm ej}}({\rm IME})$ \\ \hline
050-1-c3-1P   & 5.0  & 1.1 & 0.40 & 0.90 & 0.40 & 0.21 & 0.33 & 0.07 & $2.20 \times 10^{-2}$ & $2.40 \times 10^{-2}$ & $1.00 \times 10^{-2}$  \\
075-1-c3-1P   & 6.0  & 1.8 & 0.22 & 1.09 & 0.46 & 0.24 & 0.30 & 0.16 & $8.14 \times 10^{-3}$ & $8.89 \times 10^{-3}$ & $5.21 \times 10^{-3}$  \\
100-1-c3-1P   & 6.7  & 2.2 & 0.37 & 0.96 & 0.51 & 0.26 & 0.35 & 0.16 & $3.08 \times 10^{-2}$ & $3.36 \times 10^{-2}$ & $1.04 \times 10^{-2}$ \\
300-1-c3-1P   & 9.5  & 4.5 & 0.55 & 0.82 & 0.69 & 0.31 & 0.39 & 0.30 & 0.10 & 0.14 & $4.28 \times 10^{-2}$ \\
500-1-c3-1P   & 11.1 & 5.9 & 1.10 & 0.28 & 0.79 & 0.32 & 0.64 & 0.15 & 0.29 & 0.37 & $7.12 \times 10^{-2}$ \\ \hline

\end{tabular}
\end{center}
\end{table*}

\begin{figure*}
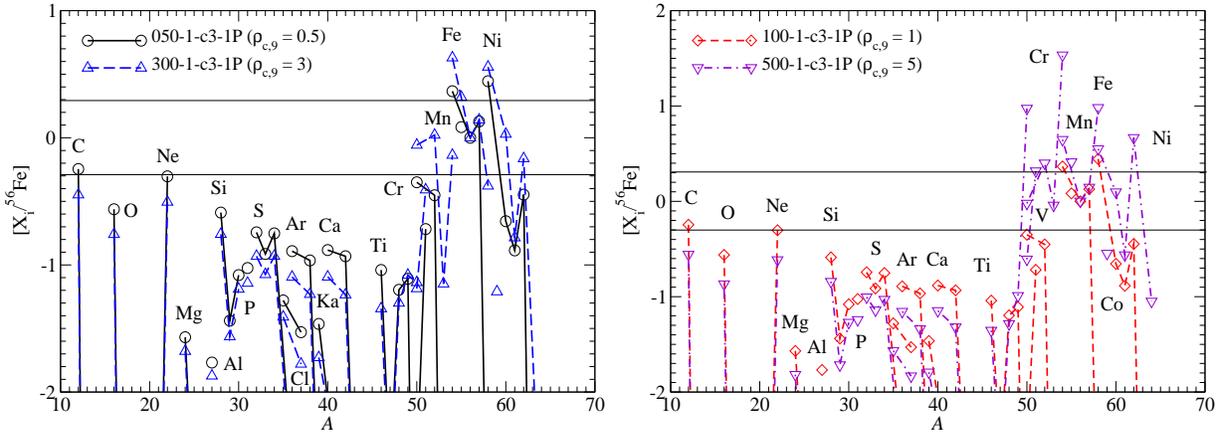

\centering
\includegraphics*[width=8cm,height=5.7cm]{fig15a.eps}
\includegraphics*[width=8cm,height=5.7cm]{fig15b.eps}
\caption{Dependence on the initial 
central density ($\rho_{c,9} = \rho_c / 10^9$ g cm$^{-3}$).
The nucleosynthesis of the whole SN Ia is included. 
Left: 050 and 300, 050-1-c3-1P ($\rho_{c,9}$ = 0.5), 300-1-c3-1P ($\rho_{c,9}$ = 3); 
right: 100 and 500,  same as above (left panel), but for 
100-1-c3-1P ($\rho_{c,9}$ = 0.5), 500-1-c3-1P ($\rho_{c,9}$ = 3).}
\label{fig:final_PTD}
\end{figure*}

In Figure \ref{fig:final_PTD} we plot $[{\rm X_i}/^{56}{\rm Fe}]$ of the stable isotopes
similar to Figure \ref{fig:final_DD} but for the pure turbulent deflagration
models. The nucleosynthesis of the whole SN Ia is included. 
The effects of central density are very similar
to the DDT model except the set of isotopes concerned
are different. There are non-negligible amounts of 
$^{12}$C, $^{16}$O and $^{20}$Ne. Their amounts decrease
when initial central density increases.  
Under-produced IMEs include Si, S, Ar, Ca and Ti. 
Similar effects of central density are observed. 
The iron-group elements are
in general well-produced already in the deflagration
phase except some neutron-rich isotopes. Their amounts
increase with the progetnior mass. At the Chandrasekhar mass
limit, the neutron-rich isotopes become very sensitive
to the density because of the electron capture. Including 
$^{50}$Ti, $^{54}$Cr, $^{58}$Fe and $^{62}$Ni, they
are severely over-produced by a factor from 3 to 30.

\subsection{Metallicity and Central Density Dependencies of Iron-Peak Elements}

It is widely believed that SNe Ia are the major source of the
iron-peak elements.
In the galactic chemical evolution, the metal produced in each
generation of stars increases the metal content of the stars in later
generations.  For supernovae, the increasing metallicity of the
progenitors affects the supernova nucleosynthesis.  Thus in modeling
such a chemical evolution including the time-delay of SN Ia
enrichment, one needs to apply the metallicity-dependent supernova
yields for both SNe II and SNe Ia.  As SNe Ia are the major source of
iron-peak elements, we summarize the metallicity-dependent yields
of $^{55}$Mn and $^{56-58}$Ni.

Generally, iron peak elements are synthesized by both deflagration and
detonation.  In the NSE region produced by deflagration, the density
is high enough for electron capture to reduce $Y_{\rm e}$.  Thus the
isotopic ratio is sensitive to the initial central density (i.e., the
C+O WD mass).  In the delayed-detonation phase, the density is too low for
electron capture to take place.  Instead, the isotopic ratios are
affected by the initial $Y_{\rm e}$, which is lower for higher metallicity
because a larger amount of $^{22}$Ne has been synthesized from the
initial CNO elements by H and He burning in the progenitor star of the
C+O WD.

The ratio between the deflagration yields and the detonation yields is
affected by the central density.  Generally, the lower central density
model has a larger detonation region, thus being more sensitive to
metallicity.  More specific dependencies are discussed below.

\begin{figure}
\centering
\includegraphics*[width=8cm,height=5.7cm]{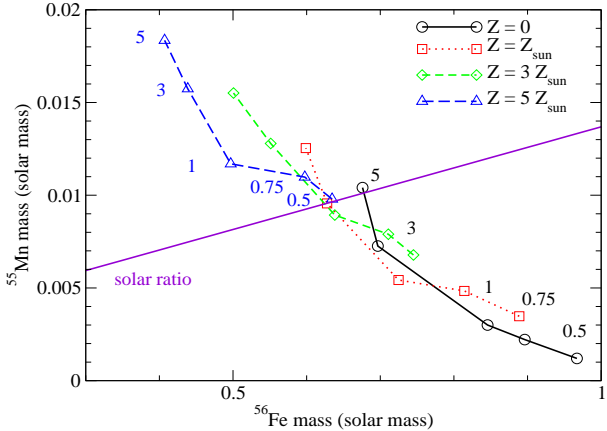}
\caption{The mass of $^{55}$Mn against $^{56}$Fe
in solar mass for SNe Ia models with a
metallicity from 0 - 5 $Z_{\odot}$ and a 
central density from $5 \times 10^8$ g cm$^{-3}$
to $5 \times 10^9$ g cm$^{-3}$. The numbers
in the figure denotes the central density of that
model in units of $10^9$ g cm$^{-3}$. The purple line 
(see online version for colour) corresponds to 
the solar value. Notice
that the mass of $^{56}$Fe corresponds to the 
mass of the decay product of $^{56}$Ni.}
\label{fig:55Mn_56Ni_plot}
\end{figure}

\subsubsection{$^{55}$Mn vs. $^{56}$Fe}

In Figure \ref{fig:55Mn_56Ni_plot} we plot the mass of
$^{55}$Mn against $^{56}$Ni for our SN Ia models 
using the c3 flame with metallicity of $Z = 0 - 5 Z_{\odot}$
and a central density from $5 \times 10^8$ g cm$^{-3}$
to $5 \times 10^9$ g cm$^{-3}$. $^{55}$Mn is an
interesting element as this element is not abundantly
produced in other types of SNe.  
SNe Ia might be the major source of 
$^{55}$Mn in realizing the observed solar abundance. 
$^{55}$Mn is in general produced by deflagration
as well as alpha freezeout at high metallicity region. 
It can be seen that for models with a constant $Z$,
increasing the central density leads to a lower
$^{56}$Ni and higher $^{55}$Mn production. The range
of $^{56}$Ni production drops from $(0.7 - 1) M_{\odot}$
(being larger for higher $\rho_{\rm c}$) at $Z = 0$ to $(0.4 - 0.6) M_{\odot}$
at $Z = 5 Z_{\odot}$.  On the
other hand, the $^{55}$Mn production increases from
$(0.001, 0.01) M_{\odot}$ (being larger for higher $\rho_{\rm c}$) at $Z = 0$ to 
$(0.01, 0.018) M_{\odot}$ at 5 $Z_{\odot}$. 
Along the same metallicity line, 
a higher central density model has more extended 
deflagration phase. As a result, more 
matter are incinerated into NSE and has more time for 
electron capture to take place. Since 
electron capture lowers $Y_{\rm e}$, 
it enhances the production of $^{55}$Mn ($Y_{\rm e} =$ 0.455),
but decreases the fraction of $^{56}$Ni ($Y_{\rm e} =$ 0.5) in NSE.

\begin{figure}
\centering
\includegraphics*[width=8cm,height=5.7cm]{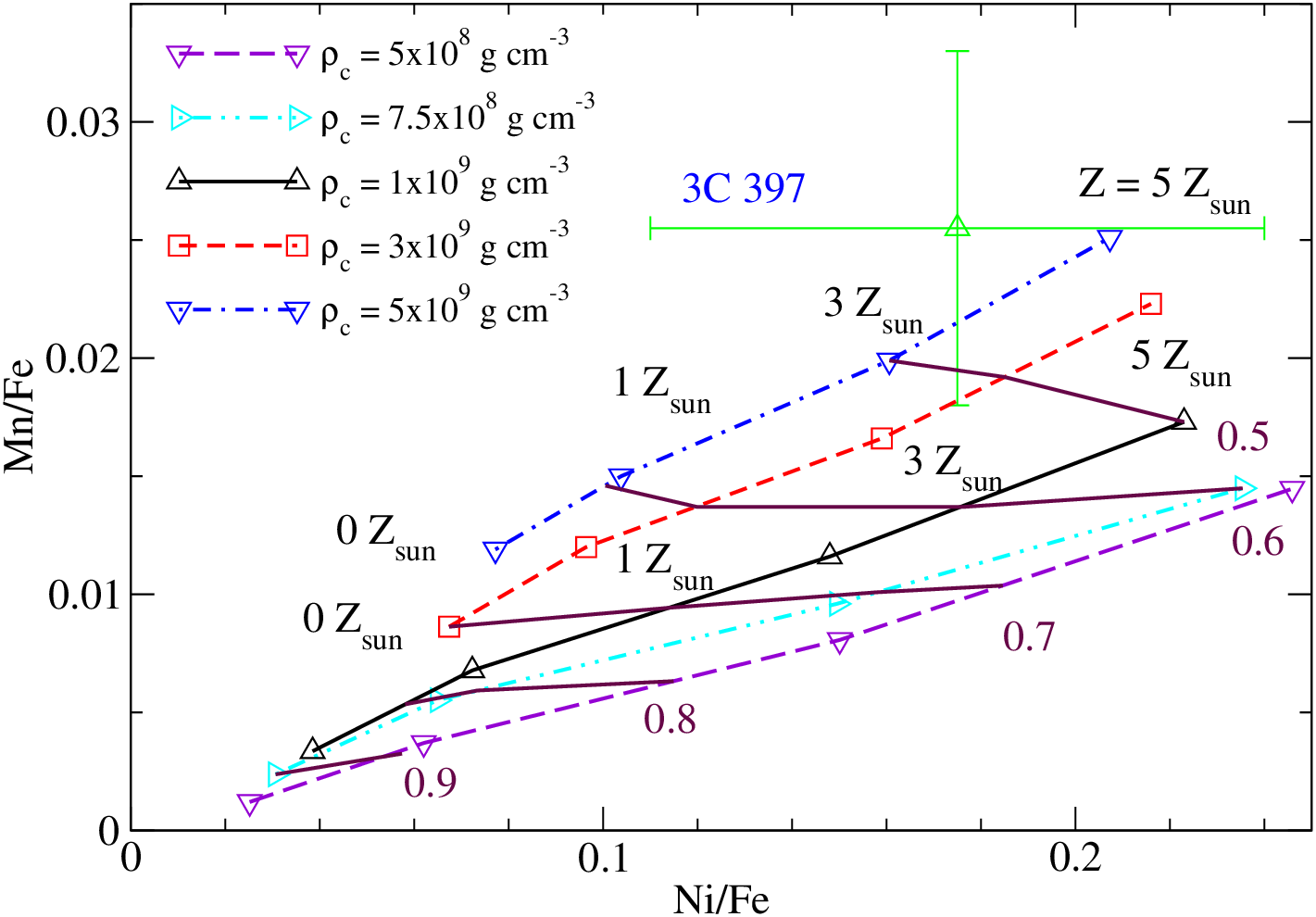}
\caption{The mass ratio of stable isotopes 
Mn/Fe against Ni/Fe.
The observational abundances obtained from SN remnants
3C 397 \citep{Yamaguchi2015}, Tycho \citep{Yamaguchi2014}
and Kepler \citep{Park2013} are included for comparison.
The sequence goes from
zero metallicity on the left to $Z = 5 Z_{{\rm sun}}$
on the right. 
The magenta lines (see the online version
for the colour plot) stand for the iso-$^{56}$Ni mass
models from 0.5 - 0.9 $M_{\odot}$
in an interval of 0.1 $M_{\odot}$. Notice
that the mass of $^{56}$Fe corresponds to the 
mass of the decay product of $^{56}$Ni.}
\label{fig:Mn_Ni}
\end{figure}

However, merely comparing the solar abundance cannot 
provide a comprehensive picture since the parameter
space, owing to the high dimensional parameter space,
could be degenerate. Qualitatively different models might 
provide the mass fraction distribution similar to the solar
abundance. 

To test whether the SNe model is compatible with
observational results, especially from nearby SNe Ia. 
One test is to compare the Mn/Fe mass ratio against 
Ni/Fe after the radioactive decay.
Mn is known to be an important indicator of metallicity
through its decay from $^{55}$Co $\rightarrow ^{55}$Fe  
$\rightarrow ^{55}$Mn. The parent isotope $^{55}$Co
is sensitive to the metallicity, in particular the 
amount of $^{22}$Ne. In the previous section we have described
the results that increasing the initial metallicity of 
the WD progenitor can drastically increase the $^{55}$Mn
abundance. In view of this, measuring Mn/Fe can 
point out accurately what metallicity the SNe Ia is,
at the time it was exploding. 

One example is given in \cite{Yamaguchi2015}.
The SNe Ia remnant 3C 397 is measured. 
They find the Mn/Fe mass ratio  
of $0.025^{+0.008}_{-0.007}$ and the Ni/Fe mass ratio 
of $0.17^{+0.07}_{-0.05}$. in this SN Ia remnant. 
Based on one-dimensional
models, they find that this SN Ia remnant is related
to an SN Ia with a high metallicity above 5 $Z_{\odot}$. 
In Figure \ref{fig:Mn_Ni} we plot the mass ratio Mn/Fe
against Ni/Fe for our models. 

In \cite{Shen2017} the sub-Chandrasekhar
SNe Ia are revisited as to supplement the lack of sub-Chandrasekhar
mass models with a very thin He envelope, which can produce effectively a direct
detonation of CO core. The one-dimensional
hydrodynamics with nucleosynthesis of such models are calculated. 
It is shown that the global nucleosynthesis pattern is still 
incapable of explaining the high Mn/Fe mass ratio 
unless one picks a subset of the ejecta 
by assuming reverse shock-heating effects.

In \cite{Dave2017}
the scenario is examined in the context of gravitationally confined detonation with some 
supplementary models from pure turbulent deflagration with or without DDT. This model is found to be
producing an incompatible pattern of [Ni/Fe] vs. [Mn/Fe] in 
low metallicity model such as $0 - 3 Z_{\odot}$. The pure turbulent deflagration model
and DDT model with the low [C/O] ratio and higher central density 
produces a more compatible chemical abundance. Our results are consistent
with theirs in our analysis of model parameters. 
As discussed in the main text, the lower C/O ratio
can enhance the [Mn/Fe] ratio owing to a weaker explosion.
The high density is also contributing to enhance Mn production
by the faster electron capture. The offset of initial flame,
as shown in our model 300-1-b1b-1, is also helpful in 
boosting the Mn production. 

It can be seen that the central density, metallicity and
detonation criteria can enhance both the production 
of manganese and nickel group isotopes. In contrast, 
the variation in the initial flame structure either 
suppresses Ni production and enhances Mn
production, or vice versa. To explain this
unusual object, similar to the one-dimensional
results as presented in \cite{Yamaguchi2015}, 
the $5 Z_{\odot}$ is necessary for explaining
the high Mn/Fe mass ratio. In particular, we need rather
higher central density at $5 \times 10^9$ g cm$^{-3}$
for the progenitor, with a metallicity from $3 - 5 Z_{\odot}$. 

\begin{figure}
\centering
\includegraphics*[width=8cm,height=5.7cm]{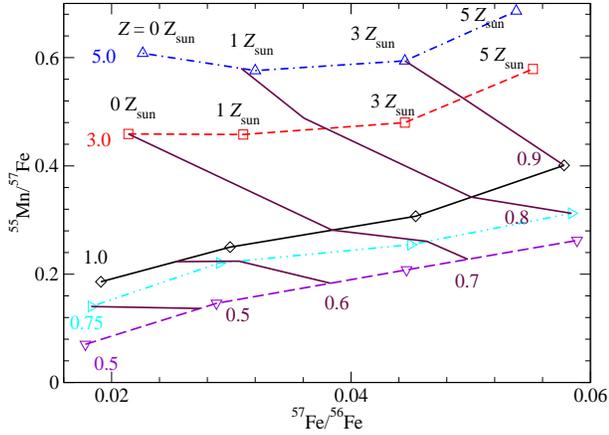}
\caption{The mass ratio of stable isotopes 
$^{55}$Mn/$^{57}$Fe against $^{57}$Fe/$^{56}$Fe. 
The sequence goes from
zero metallicity on the left to $Z = 5 Z_{\odot}$
on the right. 
The magenta lines (see the online version
for the colour plot) stand for the iso-$^{56}$Ni mass
models from 0.5 - 0.9 $M_{\odot}$
in an interval of 0.1 $M_{\odot}$.
Notice that the masses of $^{56}$Fe and $^{57}$Fe corresponds to the 
masses of the decay product of $^{56}$Ni and $^{57}$Ni
respectively.}
\label{fig:Mn_Fe}
\end{figure}

Another measurement that can be made in SN Ia remnants         
is the mass ratio $^{55}$Fe/$^{57}$Fe, which 
contains the intermediate isotopes of the decay chains
$^{57}$Ni $\rightarrow ^{57}$Co $\rightarrow ^{57}$Fe
and $^{55}$Co $\rightarrow ^{55}$Fe $\rightarrow ^{55}$Mn.
This mass ratio is used to analyze the 
density of the progenitor and to determine the progenitor
scenario.  \cite{Roepke2012} obtained the mass ratio
$^{55}$Fe/$^{57}$Fe $= 0.27$ for sub-Chandrasekhar mass model  
and $^{55}$Fe/$^{57}$Fe $= 0.68$ for Chandrasekhar mass model.

Here, we perform a similar analysis
based on our arrays of model with a central ignition kernel, 
and plot our results in Figure \ref{fig:Mn_Fe}. 
It can be seen that the
$^{55}$Fe/$^{57}$Co mass ratio is an increasing function of both density and metallicity. 
The comparison with the observations needs careful observations and
modeling of the light curve is necessary \citep{Roepke2012,Shappee2016}.

\begin{figure}
\centering
\includegraphics*[width=8cm,height=5.7cm]{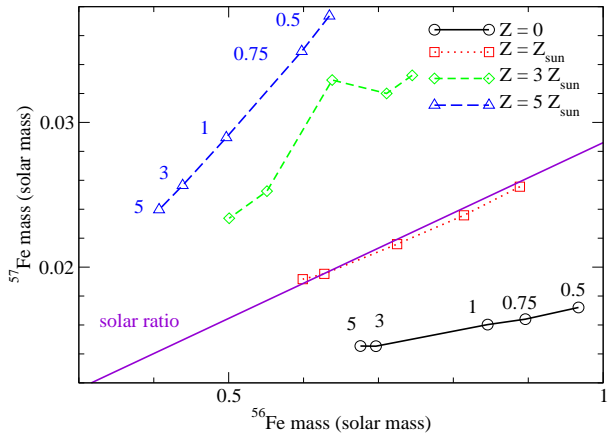}
\caption{Similar to Figure \ref{fig:55Mn_56Ni_plot}, 
but for the mass of $^{57}$Fe against $^{56}$Fe.
Notice that $^{56}$Fe and $^{57}$Fe are the 
decay products of $^{56}$Ni and $^{57}$Ni
respectively.}
\label{fig:57Ni_56Ni_plot}
\end{figure}

\subsubsection{$^{57}$Fe vs. $^{56}$Fe}

In Figure \ref{fig:57Ni_56Ni_plot} we plot the masses of 
$^{57}$Fe against $^{56}$Fe similar to Figure \ref{fig:55Mn_56Ni_plot}.
$^{57}$Fe is the decay product of the radioactive isotope $^{57}$Ni
by the chain $^{57}$Ni $\rightarrow ^{57}$Co $\rightarrow ^{57}$Fe,
which has a decay half life of 35.6 hours and 271.8 days respectively. 
The $^{57}$Ni is produced in both deflagration and detonation
zones. 
Along models of a constant metallicity, the increase in the 
central density moves the models 
towards a lower $^{56}$Fe and lower $^{57}$Fe.
The range of $^{57}$Fe varies from $(0.014, 0.018) M_{\odot}$
(being smaller for higher $\rho_{\rm c}$) at $Z = 0$ to $(0.024, 0.038) M_{\odot}$ 
at $Z = 5 Z_{\odot}$. The dependence of $^{57}$Fe on $\rho_{\rm c}$  also
stems from the electron capture rate 
being faster at higher densities.  In NSE, 
less amount of matter can have sufficient low 
$Y_{\rm e}$ for producing the parent isotope $^{57}$Ni ($Y_{\rm e} = 0.491$). 
On the other hand, the increase of metallicity
strongly enhances the production of $^{57}$Ni. 
 
\begin{figure}
\centering
\includegraphics*[width=8cm,height=5.7cm]{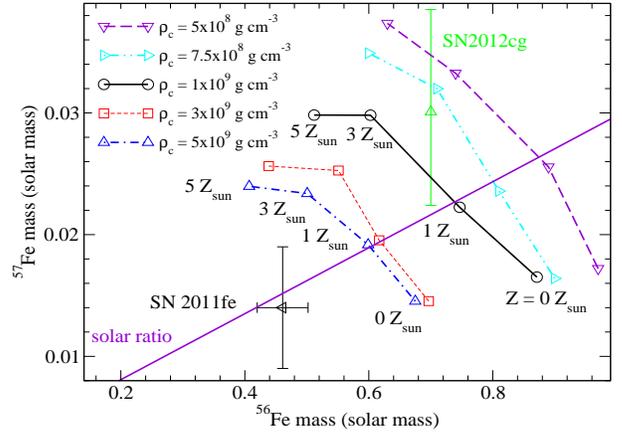}
\caption{The mass ratio of $^{57}$Fe and $^{56}$Fe
for the models presented in the article. The two data
points correspond to the observational constraints
of two SNe Ia SN 2012cg and SN 2011fe. The sequences go from 
zero metallicity (bottom) to $Z = 5 Z_{\odot}$ (top).
The solar ratio of $^{57}$Fe/$^{56}$Fe is given as the 
purple line (see online version for the colour plot).
Notice that $^{56}$Fe and $^{57}$Fe are the 
decay products of $^{56}$Ni and $^{57}$Ni
respectively.}
\label{fig:Ni56_Ni57}
\end{figure}

Similar to previous section, we carry out
an observational test for the ratio $^{57}$Fe/$^{56}$Fe
after the explosion. The parent isotope, $^{56}$Ni ($Y_{\rm e} = 0.5$),
is a direct end product from the $\alpha$-chain reaction of $^{12}$C
burning. This is a mostly produced in the detonation
after the transition. On the other hand, 
the parent isotope of $^{57}$Fe, $^{57}$Ni ($Y_{\rm e} = 0.491$),
is mostly a product of carbon deflagration in the intermediate 
regime due to its slightly lower neutron ratio. We emphasize that 
the presence of $^{57}$Fe varies case by case especially in 
the case of strong detonation. The detonation can also
produce zones with sufficiently high temperature so that 
electron capture can occur for a certain period of time. 
In that case $^{57}$Fe can also be found in high
density detonation zone. In our case, we find that most
$^{57}$Fe is still produced in the deflagration zone. 

Measurement of recently exploded supernova SN 2012cg
is made in \cite{Graur2016}. This supernova was located 
in nearby spiral galaxy NGC 4424 at a distance of 
15.2 $\pm$ 1.9 Mpc where the observations were made till 
1055 days after the maximum luminosity has reached. 
They find that the observed ratio is $\sim$ 0.043 $\pm^{+0.012}_{-0.011}$
by using analytic fit of theoretical models.
Here we carry out a similar 
analysis by using our arrays of models. 
In Figure \ref{fig:Ni56_Ni57}
we plot this relation of our models. 
Metallicity reduces the synthesis of $^{56}$Fe
but has not much impact on $^{57}$Fe, while the initial 
central density and detonation criteria increase the production of 
both isotopes. Also, it can be seen that most 
data lie within the range derived in \cite{Graur2016}. 
The relation is almost insensitive to the flame structure.
From the figure, we can conclude that in order to explain 
the observed ratio of SN 2012cg, we need SNe Ia 
models with log $\rho_c =$ $5 \times 10^8$ g cm$^{-3}$ - $1 \times 10^9$
g cm$^{-3}$. The lower the central density is, the 
higher metallicity we need. This is because the 
low density can suppress the electron capture
and delay the DDT time. The presence of $^{22}$Ne
can compensate this change. We also show another
example SN 2011fe \citep{Nugent2011}. The late time light curve of 
this SN Ia is also analyzed in \cite{Dimitriadis2017}
for extracting the $^{56}$Ni and $^{57}$Ni, which they
obtain $M(^{56}$Ni) $= 0.461 \pm 0.041 M_{\odot}$ and 
$M(^{57}$Ni) $= 0.014 \pm 0.005 M_{\odot}$ (Case 1). Our models
suggest that this SN Ia is has metallcity being slightly 
sub-solar, with its central density close to $5 \times 10^9$ g cm$^{-3}$.

\begin{figure}
\centering
\includegraphics*[width=8cm,height=5.7cm]{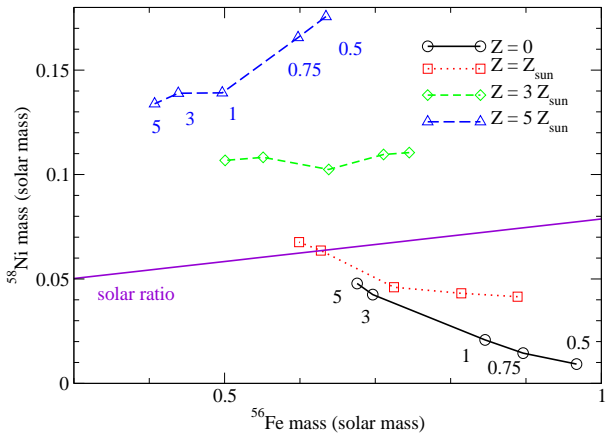}
\caption{Similar to Figure \ref{fig:55Mn_56Ni_plot}, 
but for the mass of $^{58}$Ni against $^{56}$Fe.
Notice that $^{56}$Fe is the 
decay product of $^{56}$Ni.}
\label{fig:58Ni_56Ni_plot}
\end{figure}

\subsubsection{$^{58}$Ni vs. $^{56}$Fe}

In Figure \ref{fig:58Ni_56Ni_plot} we plot similar
to Figure \ref{fig:55Mn_56Ni_plot} but for $^{58}$Ni against
$^{56}$Fe. It is the lightest isotope among all stable
Ni isotopes which is also the most abundant one. 
It is produced mostly by deflagration. 
Along models of constant density, there are two 
contrasting trends. At low metallicity, 0 to 1 $Z_{\odot}$,
the increasing central density leads to an increasing
production of $^{58}$Ni. On the other hand,
at high metallciity the increasing density suppress the 
production of $^{58}$Ni. The range varies from $(0.01,0.04) M_{\odot}$
at zero metallicity to $(0.013-0.018) M_{\odot}$
at $Z = 5 Z_{\odot}$. This is also 
related to the competition between the electron capture
and enhancement of $^{22}$Ne. $^{58}$Ni has a neutron 
ratio of 0.517. At low metallcity, the low abundance
of $^{22}$Ne suppresses the production of $^{58}$Ni
directly. Thus, the matter need to rely on electron capture
to increase the matter neutron ratio to produce 
$^{58}$Ni. However, as $^{22}$Ne abundance increases, 
as $^{22}$Ne is directly linked to $^{58}$Ni
by an $\alpha$-chain. An increasing metallicity
strongly favours the production of $^{58}$Ni. At this
point, the electron capture hinders the production
of this isotope because any electron capture can shift 
the neutron fraction of matter away from this 
$\alpha$-chain.  

\section{Discussion}
\label{sec:discuss}

\subsection{Comparison with previous models}

In the literature, multi-dimensional SNe Ia 
simulations have been done. A
trace back on multi-dimensional simulations can be 
as early as \cite{Mueller1982}. At first sight, our 
work might appear to have overlap with the 
previous works. This is not the case for several reasons. 
\par\noindent
(1) First, observations of 
the variety of SNe Ia light curves indicate that the progenitor
parameter space can be much broader than we have expected.
SNe Ia and SN remnants with unusual isotope ratios
are discovered consecutively \citep{Graur2016, Yamaguchi2015}.
\par\noindent
(2) Second, in terms of galactic chemical evolution, the diversified 
evolution of elements as a function of metallicity 
\citep{Sobeck2006, Reddy2003}
also implies the necessity of a wide parameter survey 
for SNe Ia. 
\par\noindent
(3) Third, there is not yet any systematic 
study of multi-dimensional SNe Ia in the literature, 
which spans the model parameter space while coupling
with the updated microphysics. 
A revised and consistent study is therefore important to 
update the SNe Ia modeling to be compatible with the 
rapidly growing observational data of SNe Ia.
\par\noindent
(4) Fourth, 
important updates in the microphysics have 
been found in the last decades and there can be implications
of these new updates to SN Ia simulations 
\citep{Langanke2001, Seitenzahl2009}. 
The changes of reaction rates including
the strong screening factor for $^{12}$C + $^{12}$C,
can influence the explosion dynamics 
\citep[e.g.,][]{Kitamura2000} through,
for example, the speed of deflagration wave.

\subsubsection{Travaglio et al.}

Here we compare our results with some of the representative
works in the literature which studied SNe Ia nucleosynthesis. 
In \cite{Travaglio2004}, the first multi-dimensional 
simulation with nucleosynthesis is done using the tracer particle
scheme. In their work, the pure turbulent deflagration with 
some initial flame bubbles are considered. 
The lack of detonation transition in this work has led to 
an overproduction of $^{54}$Fe and $^{58}$Ni
and underproduction of $\alpha$-chain isotopes. 
Their $Y_{\rm e}$ is similar
to ours from 0.462 to 0.500. But they observed an
overproduction of Fe and Ni as persisted in the 
classical W7 model \citep{Nomoto1984} as a result of 
the old weak interaction rates. Notice that the 
inclusion of DDT more likely further increases the production
of $^{56}$Fe in the group of iron-peak elements. 

\subsubsection{Maeda et al.}

In \cite{Maeda2010} the first large-scale
study in nucleosynthesis is presented which is based
on three hydrodynamics models. Their methodology is 
comparable to ours by including, for instance,  
NSE and electron capture.
They have used turbulent flame with similar prescriptions. 
They also have updated the weak interaction rate 
accordingly but the NSE does not take nuclear screening
into account. 

In their work, three models are
presented which include one pure turbulent deflagration
model and two DDT models from a centered- 
(C-DDT) and off-center (O-DDT) ignition kernel. 
Our model is the closest to their 
C-DDT model in terms of configuration and initial flame
structure. Their (our) model show a final kinetic energy and 
nuclear energy release at $9.6 \times 10^{50}$ ($1.27 \times 10^{51}$)
and $1.46 \times 10^{51}$ ($1.77 \times 10^{51}$) erg;
while the energy released by nuclear reaction 
at DDT is $7.67 \times 10^{50}$ ($8.10 \times 10^{50}$) in their (our) model.
The energy difference is most likely contributed by 
our three-step schemes that the low density matter
$1 - 5 \times 10^7$ g cm$^{-3}$ can still contribute 
to the energy budget as long as they can sustain the nuclear reactions.
In terms of isotope distributions, 
more differences can be spotted. 

They find IME such as $^{28}$Si at such as low velocity
$\approx 4 \times 10^3$ km s$^{-1}$, which is $\sim$ 30 \% lower
than ours. They also report the presence of $^{16}$O with a mass
fraction above 0.1 at $6 \times 10^3$ km s$^{-1}$,
which is also 10 \% lower than our model. Although the $^{56}$Ni
distribution is similar in both cases, They show a
drop of $^{56}$Ni around $8 \times 10^3$ km s$^{-1}$ in their model, 
while for our case, depending on the viewing angle, 
$^{56}$Ni starts to drastically drop at $6 - 7 \times 10^3$
km s$^{-1}$. 

In terms of isotope abundances after
decay, qualitative features of both models agree with each other.
For example, we all have a well-produced $\alpha$-chain
elements and iron-peak elements which increase with
atomic number. The non-$0.5$ $Y_{\rm e}$ isotope abundances
are also similar that, for instance, the mass 
fractions of $^{51}$Cr, 
$^{59}$Cr and $^{62}$Ni are higher than 
$^{50}$Cr, $^{58}$Fe and $^{61}$Ni. 
But there exist some differences. For example, 
they observe a higher production of $Y_{\rm e} < 0.5$
for also IME. They have a higher $^{38}$Ar and
$^{42}$Ca, with both of them well-produced. Our models
show both of them underproduced.

\begin{figure}
\centering
\includegraphics*[width=8cm,height=5.7cm]{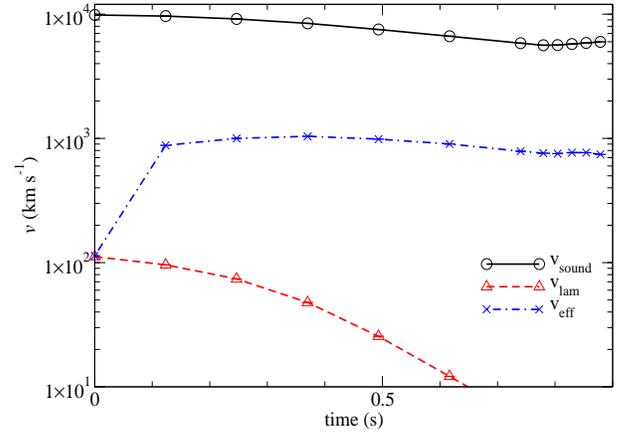}
\caption{The speed of sound, the laminar flame speed and
the effective turbulent flame speed of the benchmark model.
The error bar stands for the maximum and minimum velocities
found in the simulation and the data point is the mass-averaged
results collected from meshes which are locally swept by
the carbon deflagration.}
\label{fig:flamespeed}
\end{figure}

To make a further comparison of our results with their work, 
we plot in Figure \ref{fig:flamespeed} the 
speed of sound, the effective turbulent flame speed
and the laminar flame speed of the benchmark model
as a function of time. The data points correspond to the 
mass-averaged value of the corresponding quantities from
the mesh points which are undergoing carbon deflagration,
as indicated by the level-set function.
Since detonation always wraps the deflagration front, 
which impedes the further propagation of flame at
late time. The flame speed survey stops when the 
flame is surrounding by detonation ash. In the figure
we can see three quantities occupy characteristic
velocity range. The sound speed, which is the fastest among
all, has a typical velocity $\sim 10^{4}$ km s$^{-1}$. 
The effective turbulent flame is about one order of 
magnitude lower in the speed, $\sim 10^{3}$ km s$^{-1}$. The laminar
flame speed is the slowest that at the beginning 
it is $\sim 10^{2}$ km s$^{-1}$, but it gradually drops as 
the star expands, to $\sim 1$ km s$^{-1}$ or lower. On the contrary,
turbulence plays an important role to support the flame
propagation at an almost constant subsonic speed.

\begin{figure}
\centering
\includegraphics*[width=8cm,height=5.7cm]{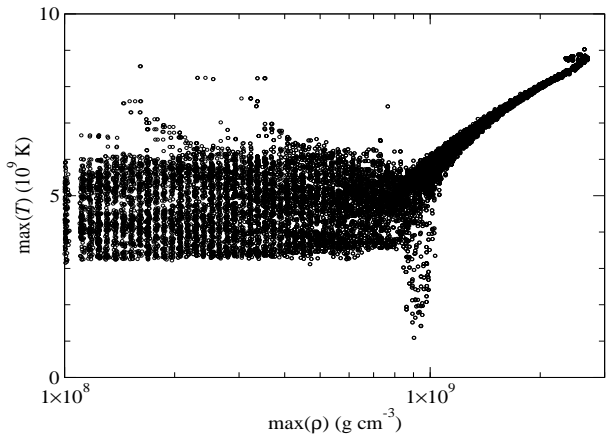}
\caption{The maximum temperature against maximum density
of the thermodynamics histories collected from
the tracer particles from our benchmark model.}
\label{fig:traj_sum}
\end{figure}

In Figure \ref{fig:traj_sum} we plot the maximum temperature
against maximum density of the thermodynamics history
obtained from the tracer particles in the benchmark model. 
This figure can be compared with Figure 6 in
their work.
The particles can be separated into two groups, the group
with $\rho_{{\rm max}} > 10^9$ 
g cm$^{-3}$ and the group with $\rho_{{\rm max}} \leq 10^9$
g cm$^{-3}$. For the first group, it has an exponential 
relation between density and temperature where 
$T_{{\rm max}} > 7 \times 10^9$ K. This group corresponds
to the particles burnt by carbon deflagration. 
For the second group, it can be seen
the particles have a range of $T_{{\rm max}}$ from $3 \times 10^9$ 
to $6 \times 10^9$ K for a wide range of density. 
This corresponds to the particles being burnt by carbon detonation.
The shock wave interaction due to multiple detonation
ignition creates a wide spectrum in the $\rho_{{\rm max}}-T_{{\rm max}}$
relation. Also, the matter density has dropped due to expansion, where 
incomplete burning makes $T_{{\rm max}}$ lower.  

\begin{figure}
\centering
\includegraphics*[width=8cm,height=5.7cm]{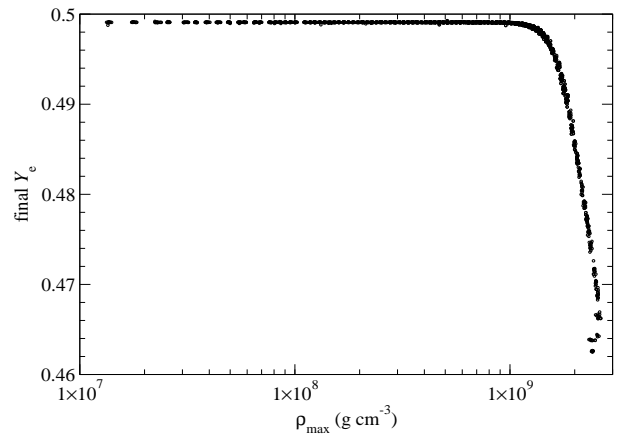}
\caption{Similar to Figure \ref{fig:traj_sum}
but for the final $Y_{\rm e}$ against $\rho_{{\rm max}}$.}
\label{fig:traj_ye}
\end{figure}

In Figure \ref{fig:traj_ye} we plot the final electron
fraction against maximum density of the tracer particle
thermodynamics histories. This plot is similar to 
Figure 9 in their work. For particles with a maximum
density $> 10^9$ g cm$^{-3}$, it has a final $Y_{\rm e}$ from
0.46 to 0.50. This corresponds to the particles which
experienced carbon deflagration. For particles with a 
maximum density lower than $10^9$ g cm$^{-3}$, it has a final 
$Y_{\rm e} = 0.5$. These are the particles which experienced
carbon detonation or incomplete burning, so that 
the particle does not have enough time to carry out 
electron capture before it cools down due to star
expansion. This figure can be compared with Figure 9
of \cite{Maeda2010}. In their work, they have a wider 
distribution of $\rho_{{\rm max}}$ for the same $Y_{\rm e}$.
During deflagration
there is always a clear discontinuity between unburnt 
matter and burnt matter along the density contour. This 
creates a spectrum of time difference of each tracer particle
to carry out electron capture. As a results, they have a wider
range of $\rho_{{\rm max}}$ for the same final $Y_{\rm e}$.

\subsubsection{Krueger et al.}

In \cite{Krueger2012}, the effects of the central
density are also studied in the range $1 \times 10^9$ to 
$5 \times 10^9$ g cm$^{-3}$ for WD models with solar metallicity.
The model configuration is very similar to ours except for
three points. First, in their 2D simulations, the turbulence-flame 
interaction is not implemented. The flame acceleration
before DDT is assumed to be attributed solely by
buoyancy stretching instead of turbulence stretching.
Without the notation of sub-grid scale turbulence 
strength, they parametrized the DDT criterion by the 
threshold density. Second, 
an adiabatic convective region is assumed for the initial WD 
as discussed in \S2 for the initial model of \cite{Nomoto1984}.
a non-isothermal WD is used as the initial condition.
Third, they assume the
initial flame to be centered, but with combinations
of sinusoidal perturbation controlled by random
number generators. While increasing the central 
density, a few effects are observed. 1. An earlier 
DDT occurs. It varies from 
1.5 s at $\rho_c = 1 \times 10^9$ g cm$^{-3}$
to 0.8 s at $\rho_c = 5 \times 10^9$ g cm$^{-3}$. 
2. A lower $^{56}$Ni mass and also $^{56}$Ni/$M$(NSE)
ratio. This shows that the flame takes longer
time to reach the DDT density and there is 
more time for electron capture before the 
expansion of matter after detonation. 
Our models are consistent
with theirs in the following ways. First, from Table 
\ref{table:Models}, we observe an earlier DDT
from 1.35 s down to 0.67 s from the Models
050-1-c3-1, 100-1-c3-1, 300-1-c3-1 and 500-1-c3-1. 
Second, the $^{56}$Ni mass drops from 0.89 to 
0.59 $M_{\odot}$.

\subsubsection{Jackson et al.}

In \cite{Jackson2010}, the dependence on
metallicity (namely $^{22}$Ne) and the DDT
density is explored. They carried out a
series of 2D models similar to \cite{Krueger2012}.
We first review the metallicity effects. 
They carry out models with a metallicity from 
$Z = 0.5 Z_{\odot}$ to $2.5 Z_{\odot}$. 
The NSE isotopes drops from an average to 0.85 $M_{\odot}$
to 0.70 $M_{\odot}$. Our models agree with 
their trend that, by comparing with 
our Models 300-0-c3-1, 300-1-c3-1, 300-3-c3-1 
and 300-5-c3-1, the $^{56}$Ni yield drop
significantly from $6.96 \times 10^{-1}$ to 
$4.38 \times 10^{-1} M_{\odot}$, although there is a 
very mild increase in other NSE isotopes, such as
$^{54}$Fe, $^{57}$Fe and $^{58}$Ni. The drop 
of NSE isotopes are still dominated by the change
of $^{56}$Ni.

Then we compare the DDT density effects. 
They observed that when the DDT density 
is decreased from $10^{7.5}$ to $10^{7.1}$
g cm$^{-3}$, the amount of NSE matter decreases
from $\approx 1.2$ to $ 1.0 M_{\odot}$,
showing that more matter is burnt to NQSE
matter, such as Si. 
This is consistent with our results. 
By comparing Models 300-1-c3-1, Test A1 
and Test A2, a decreases in the $^{56}$Ni mass from 0.627 
to 0.450 $M_{\odot}$ for the increase in the 
Karlorvitz number, which is equivalent to 
the decease in the DDT density.
At the same time, the $^{28}$Si mass 
increases from 0.235 to 0.324 $M_{\odot}$.
This suggests that the simplified 
nuclear burning scheme can still capture 
the essential nuclear reactions of
a much larger network.

\subsubsection{Seitenzahl et al.}

Then, we shall compare with some representative
three-dimensional models.
In \cite{Seitenzahl2011}, the first three-dimensional 
systematic study of nucleosynthesis is presented. 
The numerical modeling can also be traced back to 
that in \cite{Maeda2010}. The DDT criteria is different 
that Karlovitz number is not directly included in this work.
Instead, turbulent velocity threshold and the flame 
surface area are the criteria for the trigger for DDT. 
In this work, the nucleosynthesis is analysed based on the 
simplified energy scheme for hydrodynamics. They studied 
12 models with density from $10^9$ to $5.5 \times 10^{9}$ g cm$^{-3}$
in solar metallicity with off-center ignition kernels.  
Our benchmark model 3-1-c3-1 is the closet to their 0200 model at
$\rho_c = 2.9 \times 10^9$ g cm$^{-3}$ in the configuration. 
In terms of explosion energetic, their (our) model has a DDT time at 
0.802 (0.779) s. The energy released
by nuclear reaction at the transition time in their (our) model
is about $8.10 \times 10^{50}$ erg ($8.03 \times 10^{50}$) erg.
The DDT density is $2.27 \times 10^7$
($2.31 \times 10^7$) g cm$^{-3}$ in their (our) model.
In terms of chemical abundance, their (our) model produce 0.752 (0.63) $M_{\odot}$
$^{56}$Ni. 
We also observe a similar amount of unburnt $^{16}$O at
0.06 $M_{\odot}$. The difference in the $^{56}$Ni amount is
most likely related to the initial flame. Our centered flame
can have a stronger deflagration phase due to its larger
initial flame surface, which may further enhance the turbulence
generation. Therefore, more fuel is being burnt 
in the deflagration phase and starts electron capture. As
a result, the matter of lower electron fraction cannot produce
as much $^{56}$Ni as in theirs. 

In \cite{Seitenzahl2013}, they presented a first large-scale
study of nucleosynthesis in three-dimensional models. 
This study concerns about the effects of off-center
ignition kernels to the nucleosynthesis. The methodology
is still comparable with \cite{Seitenzahl2011} but
with an extra requirement on the timescale that the 
the DDT criteria should be satisfied for at least 
half an eddy turnover time. Our benchmark model can
be compared to their N100 model, which is also
selected to be the representative of typical 
SNe Ia model, which is characterized by an initial
flame of 100 flame bubbles with a size of 10 km 
and a mean radius of 60 km from center. 
They presented 14 models of different
ignition kernels at solar metallicity and one model
with metallicity from 0.01 $Z_{\odot}$ to solar
metallicity. In terms of spatial distribution, 
their N100 model shows that $^{56}$Ni is produced
from 0 to $9 \times 10^3$ km s$^{-1}$, which is significantly
higher than ours at $6 - 7 \times 10^3$ km s$^{-1}$. 
$^{58}$Ni has also a similar distribution as $^{56}$Ni 
but at a lower mass fraction below 0.1. $^{58}$Ni in our model
has also a lower mass fraction but has some  
peaks about 0.1 on the contrary. The IME distribution
is comparable to us that both models show 
signatures of IME starting from
 $3 - 4 \times 10^3$ km s$^{-1}$ and IME become the 
dominant isotopes at a velocity above 
6 - 7 $\times 10^3$ km s$^{-1}$. 
In terms of mass fraction, our model is also very close to
theirs qualitatively. For example, both models
show the iron-peak elements peaked at their isotopes
of highest electron fraction. A jump in the mass fraction
of $^{59}$Fe and $^{62}$Ni are also observed and their 
values but our models predict both isotopes with higher
mass fractions. 
One major difference is that our models predict a drop 
in $^{53}$Cr compared to other Cr isotopes, but its
production is boosted in their model.

We compare the central density 
dependence with the models in \cite{Seitenzahl2013}. 
In our models, the Models 050-1-c3-1
100-1-c3-1, 300-1-c3-1 and 500-1-c3-1 forms the series of 
models with the same configuration but different 
central densities from $5 \times 10^8$ g cm$^{-3}$
to $5 \times 10^{9}$ g cm$^{-3}$. In \cite{Seitenzahl2013},
the Models N100L, N100 and N100H are the series of 
models of similar properties, but for central density
from $1 \times 10^{9}$ g cm$^{-3}$ to $5.5 \times 10^{9}$ 
g cm$^{-3}$. In their models, an increase of central density
leads to the following trends: 
\par\noindent
1. The yields of intermediate mass
elements decrease. This can be observed from the variations
of $^{28}$Si ($^{32}$S) that it drops from $3.55 \times 10^{-1}$ 
($1.38 \times 10^{-1}$) to $2.12 \times 10^{-1}$
($8.55 \times 10^{-2}$) $M_{\odot}$. 
\par\noindent
2. More $^{56}$Ni is produced. This can be observed by 
the yield growing from $5.32 \times 10^{-1}$ to $6.94\times 10^{-1} M_{\odot}$.
\par\noindent
3. A significant increase of neutron-rich isotopes,
including $^{50}$Ti, $^{54}$Cr, $^{58}$Fe and so on. 
The yielded mass for $^{50}$Ti ($^{58}$Fe) increases
from $6.88 \times 10^{-10}$ ($1.27 \times 10^{-7}$)
to $1.02 \times 10^{-4}$ ($5.29 \times 10^{-3}$) $M_{\odot}$.

Then we compare their results with our models. 
\par\noindent
1. In our models when central density increases, 
yields of intermediate mass elements increases.
there is an increase of $^{28}$Si ($^{32}$S)
from $2.26 \times 10^{-1}$ ($1.24 \times 10^{-1}$) to 
$2.70 \times 10^{-1}$ ($1.8 \times 10^{-1}$) $M_{\odot}$. 
\par\noindent
2. The $^{56}$Ni mass decreases. 
It drops from $7.46 \times 10^{-1}$ to
$5.94 \times 10^{-1}$ $M_{\odot}$. 
\par\noindent
3. Masses of neutron-rich isotopes increase. 
The $^{50}$Ti ($^{54}$Cr) yield also increases from 
$6.22 \times 10^{-10}$ ($1.91 \times 10^{-7}$) to 
$5.21 \times 10^{-4}$ ($1.42 \times 10^{-2}$) $M_{\odot}$.
The difference in the first two trends 
shows that the center flame can behave differently
from off-center flame. Also, the dimensionality plays
a role.

We remind that the bubble configuration in 
our models corresponds to the ring in the 
three-dimensional realization. In lower dimensionality,
the smaller number of degree of freedom tends to 
enhance radial motion of the flame, thus 
boosting the burning. 
In our models, when central
density increases, the centred flame can propagate
faster because the thermalized core sustains
the flame to move outward. On the other hand, 
the off-centered flame may experience stronger
suppression after electron capture. The flame
may take longer time to reach the low density
region for triggering DDT, while burning 
more matter simultaneously.
On the other hand, the comparable neutron-rich
isotopes show that our calculation have a 
consistent electron capture rate from their works.

We compare the metallicity
dependence with the models in \cite{Seitenzahl2013}. 
In \cite{Seitenzahl2013}, the Models N100, N100-Z0.1, 
N100-Z0.01 form another series that studies the effects
of varying metallicity. We compare their results with our
Models 300-0-c3-1, 300-1-c3-1, 300-3-c3-1 and 
300-5-c3-1. In their models when metallicity increases, 
two trends can be observed.
1. $^{55}$Mn increases from $7.84 \times 10^{-3}$ to
$9.29 \times 10^{-3} M_{\odot}$. 2. Masses of isotopes
with electron fraction close 0.5 increase, 
such as $^{54}$Fe and $^{58}$Ni. An increase of yield
from $6.62 \times 10^{-2}$ ($5.01 \times 10^{-2}$) 
to $9.94 \times 10^{-2}$ ($6.90 \times 10^{-2}$) 
can be observed for $^{54}$Fe ($^{58}$Ni). Our models
also show similar trend that $^{55}$Mn grows
from $7.25 \times 10^{-3}$ to $9.55 \times 10^{-3}$,
while $^{54}$Fe ($^{58}$Ni) grows from 
$8.48 \times 10^{-2}$ ($4.29 \times 10^{-2}$) to 
$1.60 \times 10^{-1}$ ($6.35 \times 10^{-2}$) $M_{\odot}$.
This shows that our treatment of metallicity 
using $^{22}$Ne as a proxy is consistent with their 
results.

\subsection{Comments on the limitation of 2D models}

In this article we perform a number of two-dimensional
SNe Ia simulations with post-process nucleosynthesis, 
from which we extract the influence of model 
parameters on the chemical abundances. Our model
can represent WD where the fluid motion inside the 
WD preserves symmetry, such that the center remains
the most probable location for the first flame
to occur, and the initial flame can have sufficient time
to develop before being brought away from center 
by convection. However, in general the 3D SN Ia models
may provide higher flexibility to account for 
the diversity of WD environment prior to explosion. 
Here we briefly describe the short-coming of 
two-dimension models compared to the three-dimensional ones.

One of the shortcomings in 2D simulations is the 
boundary effects. Due to the 
assumed rotational symmetry, no bubble can be 
naturally constructed as the initial flame,
where flame bubble is often used as initial
configurations \citep{Roepke2005a} for 
SNe Ia simulation during the early phase of laminar
flame propagation, when hydrodynamics instabilities
are not strong enough to distort the flame structure.
As shown in \cite{Seitenzahl2011}, how many bubbles
at the beginning is one of the primary parameters
in controlling the explosion strength. 
One of the possible methods to model flame bubble in
a two-dimension model is by using the one bubble case, where 
a spherical flame is placed along the axis of 
rotational symmetry. However, we remark that this model
cannot be compared completely with the counterpart in 
three dimensional model. Due to the reflective
boundary along the symmetry axis, motion close to the 
boundaries tends to be enhanced. For example, a 
symmetry of an initially spherical-shaped bubble may quickly
be destroyed by the Rayleigh-Taylor instabilities.
Therefore, in the context of SNe Ia for the one-bubble 
configuration, where 
turbulence is believed to play a major role in 
flame propagation, this strong boundary flow may
provide unrealistic enhancement of turbulence production,
which over-estimates the flame propagation, as well as 
the transition time. These effects will entangle with 
the true hydrodynamics effects. This makes the extraction
of model parameter effects difficult.

Second, the symmetry boundary creates hot spots
after detonation. As shown in Figure \ref{fig:flame_std}
after the detonation is developed, spherical detonation
wave spreads rapidly and collide with the symmetry
axis. The reflected shock wave further interact with the
incoming shock, where intersection of shock wave compresses
the matter in-between and creates the hot spot. However,
the shock-boundary interaction is most likely to occur in
2D models. In a 3-dimension model, the use of randomized 
bubble configuration makes the shock waves colliding with
each other less likely. Such collisions create zones which 
experience a short period of time in comparatively 
high density and temperature. The temperature can be much higher than 
that after laminar deflagration or detonation
wave has passed through. Therefore, in nucleosynthesis it is
possible that such temperature fluctuation may lead to 
a boost in heavy isotope production, which might not be 
realized in 3-dimensional models.

\section{Summary}

In this present paper, we present two-dimensional
hydrodynamics simulations of 
near-Chandrasekhar mass white dwarf (WD) models for
SNe Ia using the turbulent 
deflagration model with deflagration-detonation transition. 
By calculating 41 models, we perform a parameter  
survey to study the effects of 
the initial central density (i.e., WD mass),
metallicity, flame shape and 
detonation transition criteria, and 
turbulent flame formula 
for a much wider parameter space than earlier studies.
The final isotopic abundances of $^{11}$C to $^{91}$Tc in these simulations
are obtained by post-process nucleosynthesis calculation. 
The parameter survey includes SNe Ia
models with the central density from $5 \times 10^8$ g cm$^{-3}$
to $5 \times 10^9$ g cm$^{-3}$
(WD masses of 1.30 - 1.38 $M_\odot$), a metallicity from
0 to 5 $Z_{\odot}$, C/O mass ratio from 0.3 - 1.0
and ignition kernels including centered and 
off-centered ignition kernels. 
The yield tables of 25 elements from carbon to 
zinc for a total of 41 models are computed. 
We examine the possible effects of 
$Y_{\rm e}$ mixing and find it is important to nucleosynthesis.
The results are compared with the solar composition
to derive constraints on each model parameters.
We also compare our models with some well-observed
SNe Ia including SN 2011fe, SN 2012cg and the supernova
remnant 3C 397. The possible supernova progenitors,
based on the abundance of $^{55}$Mn,  $^{57}$Fe
and $^{58}$Ni are suggested.
We have also carried out similar survey for the 
pure turbulent deflagration model and pure
carbon detonation model. The connection between
the pure turbulent deflagration model and the 
subluminous SNe Iax is discussed. 

We find that dependencies of the nucleosynthesis yields on the
metallicity and the central density (WD mass) are large
For comparisons with the observed abundance patterns of SNe Ia and their
remnants to constrain the explosion model and also for the application to the
galactic chemical evolution modeling, these dependencies on metallicity and WD mass
should be taken into account.
For this purpose, we present tables of the nucleosynthesis yields
of $^{12}$C to $^{70}$Zn as well as the major radioactive isotopes for 33 models.
For yields from core-collapse supernovae, see \cite{Nomoto2013}\footnote[1]{http://supernova.astron.s.u-tokyo.ac.jp/\~{}nomoto/reference}.

Our calculations may be applied to verify the 
validity of input physics from observational data. 
Recent observations of SN Ia remnants and
the luminosity evolution from the late-time light curves of SNe Ia
have shown possibilities to understand the supernova
physics through the abundance pattern of certain
representative isotopes. In future, the growing number
of this kind of objects may provide us the necessary
constraints on each of the model parameters.

\section{Acknowledgment}

This work has been supported by the World Premier International
Research Center Initiative (WPI Initiative), MEXT, Japan, and JSPS
KAKENHI Grant Numbers JP26400222, JP16H02168, JP17K05382, and
the Endowed Research Unit (Dark side of the Universe) by Hamamatsu Photonics K.K.. 
We thank F. X. Timmes for
his open-source micro-phsyics algorithm
including the Helmholtz equation of state 
subroutine, the torch nuclear reaction network
designed for an arbitrary choices of isotopes,
the seven-isotope nuclear reaction 
network and the galactic chemical evolution
code. We thank S. Blinnikov,
M. Ishigaki, C. Kobayashi and F.-K. Thielemann for their
insightful comments. We also thank 
T. Plewa for the discussion which helps 
us a lot in reviewing the methodology 
of SNe Ia modeling. 

\newpage

\appendix

\section{Review of our hydrodynamics code}
\label{sec:methods}

Here we briefly review the structure 
of the hydrodynamics code and then we present the 
new updates and changes incorporated in this article. 
For implementation details, code tests and applications
to pure turbulent deflagration and DDT models, refer \cite{Leung2015a}. 
The code has been used to standard SNe simulations 
and nucleosynthesis \citep{Leung2015b, Leung2016}. 

The code is an extension of the previous version of 
the hydrodynamics code which solves the two-dimensional Euler 
equations in cylindrical coordinate with sub-grid turbulence
and moving grid. Following the carbon deflagration and detonation,  
the explosion unbinds the WD. The WD quickly expands
and the matter cools down that all thermal nuclear 
reactions end. In order to understand the final 
nucleosynthesis, we keep track of the fluid 
motion until matter reaches homologous expansion. 
Using the moving-mesh algorithm \citep{Roepke2005a, Roepke2005b},  
the grid size is a time-dependent variables which 
varies with the WD. In particular, we make sure that
the matter of WD is accommodated inside the 
simulation box throughout the simulation. To do so, we 
assume that the grid boundary carries out an
homologous expansion, namely $\vec{v}_f (\tilde{r}) = v_f \tilde{r} / R$.
Here, $\tilde{r}$ and $R$ are the spherical distance 
from the origin and the radius of the star. $v_f$ is the 
magnitude of the expanding grid velocity, which is assumed 
to be equal or slightly larger than the surface velocity of 
the WD. The Euler equations are then rewritten as 
\begin{eqnarray}
\frac{\partial \rho}{\partial t} + \nabla \cdot [\rho (\vec{v} - \vec{v}_f] = -\rho \nabla \cdot \vec{v}_f,
\label{eq:euler_mass} \\
\frac{\partial \left( \rho \vec{v} \right)}{\partial t} + 
\nabla \cdot [\rho \vec{v} (\vec{v} - \vec{v}_f)] = -\nabla P - \rho \nabla \Phi -  
\rho \vec{v} \nabla \cdot \vec{v}_f,
\label{eq:euler_momentum} \\
\frac{\partial \tau}{\partial t} + \nabla \cdot [(\vec{v} - \vec{v}_f) \tau + p \vec{v}] = \rho \vec{v} \cdot \nabla \Phi + Q_{{\rm nuc}} - Q_{{\rm turb}} - Q_{\nu} - \tau \nabla \cdot \vec{v}_f,  
\label{eq:euler_energy} \\
\frac{\partial \left( \rho q \right)} {\partial t} 
+ \nabla \cdot \left[ \rho q (\vec{v} - \vec{v}_f) \right] = 
Q_{{\rm turb}} + \nabla \cdot (\rho \nu_{\rm turb} \nabla q) - \rho q \nabla \cdot \vec{v}_f, 
\label{eq:energy_eq2}
\end{eqnarray}
where $\rho$, $v_r$, $v_z$, $p_{{\rm NM}}$, $q$ and
$\tau$ are the mass density, velocities in 
the $r$ and $z$ directions, pressure and total energy 
density of the baryonic matter. The total energy
density includes both the thermal and kinetic components
$\tau = \rho \epsilon + \frac{1}{2} \rho v^2$, 
where $\epsilon$ is the specific internal energy density. The specific turbulence 
energy density $q = \tilde{v}^2$/2 corresponds to the 
velocity fluctuations $\tilde{v}$, where
the energy exchange between $\tau$ and $q$ is given by
\begin{eqnarray}
{Q}_{{\rm turb}} = -A \rho q \nabla \cdot v + \Sigma_{ij} \frac{\partial v_i}{\partial x_j}  
- \rho \epsilon_{{\rm dis}} + C_{{\rm Arch}} \rho g_{{\rm eff}}.
\label{eq:turb_eq}
\end{eqnarray}
The gravitational potential $\Phi$ is determined 
by the Poisson equation
\begin{equation}
\nabla^2 \Phi = 4 \pi G \rho. 
\end{equation}

We use the fifth-order Weighted Essentially 
non-oscillating scheme for the spatial discretization and 
the five-step third-order Non-strong-stability 
preserving Runge-Kutta scheme for the time
discretization. We use the successive over-relaxation
method with Gauss-Seidel iterations to solve the 
gravitational potential. The boundary is fixed to be
$\Phi(r,z) = G M / \sqrt{r^2 + z^2}$. To find the pressure 
and internal energy, we use the 
Helmholtz equation of state \citep{Timmes1999c} 
which provide these two
quantities as a function of density, temperature,
mean atomic mass $\bar{A}$ and mean atomic number
$\bar{Z}$. 

In this paper, in contrast to 
\cite{Leung2015a}, we try to represent the 
chemical composition of the matter by 7 isotopes.
They includes $^{4}$He, $^{12}$C, 
$^{16}$O, $^{20}$Ne, $^{24}$Mg, $^{28}$Si and
$^{56}$Ni \citep{Timmes2000a}. We choose to reduce the representative isotope number
because in the hydrodynamics the chemical abundance
is mainly responsible for providing the mean atomic mass
and mean atomic number, therefore a network which 
can describe C-burning to Ni and 
photodisintegration into $\alpha$-particle 
is sufficient to describe most energy generation and 
absorption processes. The detailed chemical 
abundance obtained from a much larger network to 
the post-processing nucleosynthesis (See section \ref{sec:nucleo}
for further details). 

We use a three-step burning to characterize the 
nuclear reactions \citep{Khokhlov1991a, Townsley2007}.
These three reactions capture the essential nuclear
bunring phase occurred in C-deflagration and detonation
They include the C-burning, NQSE-burning and 
NSE-burning. In terms of the isotopes we have, they are
\begin{eqnarray}
^{12}{\rm C} \rightarrow ~^{24}{\rm Mg}, \nonumber \\
^{16}{\rm O}, ~^{24}{\rm Mg} \rightarrow ~^{28}{\rm Si}, \nonumber \\
^{28}{\rm Si} \rightarrow ~^{56}{\rm Ni} + \alpha. 
\end{eqnarray}
The choices of these representative isotopes are
as follows. For the first step, C is the earliest 
and fastest isotope to be burned in the flame and 
detonation. It provides the first energy source 
for the coming O-burning and Si-burning. 
Si is used as the ash in the second reactions
because Si can sustain for a comparatively long time
compared to other intermediate mass elements,
such as Mg or S, before Fe-group elements emerge. 
It therefore provides a good approximation to the 
progress of NQSE burning. At last, Ni and $\alpha$
are the product of NSE burning as commonly used
in the literature \citep{Reinecke1999b, Reinecke2002a}.
It represents the end point of $\alpha$-chain network
plus the photo-disintegration effects at high 
temperature. 

\subsection{Nuclear Reactions and Flame Capturing Scheme}

The change of the NSE composition is assumed to take place
as long as the local temperature 
$T > 5 \times 10^9$ K.
The NSE composition is prepared in tabular form
with $X_{{\rm NSE}} = X_{{\rm NSE}}(\rho, T, Y_{\rm e})$.
The table is prepared similar to \cite{Seitenzahl2009},
where the corresponding binding energy and composition
are computed beforehand. 
After the hydrodynamics sub-step, the trial 
internal energy $\tilde{\epsilon}^{n+1}$ is obtained. 
Nuclear energy due to composition changes is 
included and the new temperature and internal
energy at step $n+1$ are solved iteratively such that they satisfy
\begin{equation}
\epsilon^{n+1} = \tilde{\epsilon}^{n+1} - \Delta E_{{\rm B}} (\rho^{n+1}, T^{n+1}).
\end{equation}
$\Delta E_{{\rm B}}$ is the change of binding energy,
where both the initial and final states are assumed to 
be in NSE, which are functions of the local density,
temperature and electron fraction.

To determine the energy release by carbon deflagration
and detonation, we use the level-set method \citep{Reinecke1999a}
to describe the front geometry. We implement the 
passive version of the level set method due to the known
numerical difficulty in reconstructing the exact 
thermodynamics state when there is numerical noise or error
\citep{Reinecke1999a}. This method introduces
the scalar $G$ whose zero-contour represents
the front surface and evolves as
\begin{equation}
\frac{\partial G_{\rm i}}{\partial t} + (\vec{v} + \vec{v}_{{\rm i}}) \cdot \nabla G_{\rm i} = 0.
\end{equation}
$\vec{v}$ and $\vec{v}_{{\rm i}}$
are the fluid velocity and the front
propagation speed, where $G_{\rm i}$ can represent 
both the deflagration or detonation front
by the notation ${\rm i} = $ flame or deton. 
We use operator splitting to solve the equation.
The advection is handled by the WENO scheme,
while the flame propagation is obtained by solving
$\frac{\partial G}{\partial t} + 
\vec{v}_{{\rm i}} \cdot \nabla G = 0$.
Notice that for ideal case where $| \nabla G| = 1$
exactly, 
\begin{equation}
\vec{v}_{\rm i} \cdot \nabla G = -{v}_{\rm i} \frac{\nabla G}{| \nabla G|} \cdot \nabla G = -{v}_{\rm i}.
\end{equation}
For deflagration, we use the laminar flame speed formula
reported in \citep{Timmes1992a} with the 
turbulent flame speed relation \citep{Pocheau1994,Schmidt2006b}
\begin{equation}
v_{{\rm flame}} = v_{{\rm lam}} \sqrt{1 + C_t \left( \frac{\tilde{v}}{v_{{\rm lam}}} \right)^2}
\label{eq:flameeq1}
\end{equation}
and
\begin{equation}
\vec{v}_{{\rm lam}} = 92 {\rm km~s^{-1}} \left( \frac{\rho}{2 \times 10^9 {\rm g~cm^{-3}}} \right) 
\left[ \frac{X(^{12}{\rm C})}{0.5} \right]^{0.889}.
\end{equation}
We use the standard value $C_t = 4/3$ in this article.
We notice that Eq. (\ref{eq:flameeq1}) is an empirical 
model based on renormalization and energy conservation,
which may not different from the actual turbulence-flame 
interaction. We notice that besides this formula, 
in the literature there are also other
turbulence flame speed formulae derived from some direct 
numerical experiments (See for example \cite{Hicks2015}),
where the empirical relation between 
the effective flame speed under the influence of turbulent
velocity fluctuation is studied by direct numerical 
simulations. They find the modified model
\begin{equation}
v_{{\rm flame}} = v_{{\rm lam}} \sqrt{1 + 0.614 \frac{\tilde{v}}{v_{{\rm lam}}}},
\label{eq:flameeq2}
\end{equation}
and the best-fit model
\begin{equation}
v_{{\rm flame}} = v_{{\rm lam}} 
\sqrt{1 + 0.654 \left( \frac{\tilde{v}}{v_{{\rm lam}}} \right)^{0.5985} }
\label{eq:flameeq3}
\end{equation}
can represent the turbulence-flame interaction
for their simulations.

In every timestep, after we have updated the propagation
part of the level-set function, we increase the 
internal energy according to the area swept
by the deflagration front. The corresponding
$^{12}$C is converted to $^{24}$Mg is converted correspondingly.
The carbon is assumed to be completely 
burnt independent of density.

Similar procedure is prepared for detonation. 
Another scalar field is used to represent the
geometry of the detonation front, where the 
advection and propagation are treated separately. 
The detonation velocity for matter with density
larger than $2 \times 10^7$ g cm$^{-3}$ is obtained 
by solving the detonation wave structure as described in 
\cite{Sharpe1999}. The detonation for the matter 
with a density below that is assumed to propagate in the form of 
Chapman-Jouget detonation. Similar to the deflagration, 
the area swept by the detonation front is assumed
to be completely burnt, which means all 
$^{12}$C is converted to $^{24}$Mg within that
region, and the corresponding binding energy
change is added to the internal energy.

\subsection{Criteria for Deflagration-Detonation Transition (DDT)}

DDT assumes that the deflagration develops into 
detonation \citep{Khokhlov1991c} when the flame enters 
the distributed regime, where the
heat conduction rate owing to turbulence diffusion 
becomes comparable with the consumption of fuel.
This process creates the detonation seed
by the Zeldovich gradient mechanism \citep{Khokhlov1997b}. 
It has been shown in numerical experiments that 
the detonation can be triggered through the
shock-flame interaction \citep{Khokhlov1997a}
which creates hot spots in shock-tube experiments,
and through the unsteady turbulent flame 
evolution \citep{Poludenko2011} in an unconfined media. 
However, there are arguments on the feasibility 
of this model based on arguments of whether 
turbulence can sustain the necessary
strong velocity fluctuation $(\sim 10^3$ km s$^{-1}$) 
\citep{Lisewski2000, Roepke2006}.
This model has been frequently applied to SN Ia 
explosion in both one-dimensional models \citep{Nomoto1984, 
Khokhlov1991a, Khokhlov1991b, Blondin2012},
and multi-dimensional models \citep{Gamezo2004, Gamezo2005,
Golombek2005, Blondin2011, Seitenzahl2013}. These models can show 
healthy explosions with a variety of nickel yield to explain
the observed SNe Ia diversity, while producing chemical 
abundance compatible with observations.

In our calculation, at the end of each step, we compare 
the eddy turnover scale with the flame width,
i.e. the {\sl Karlovitz Number Ka}. 
Here, we define the Karlovitz number as \citep{Niemeyer1995a}
the ratio of laminar flame width $l_{{\rm flame}}$
and the eddy size $l_{{\rm turb}}$. 
To determine the flame width, we use 
the deflagration wave structure
taken from \cite{Timmes1992a} as a function
of density. In general, the flame becomes
wider in its size when density decreases.
It can be as thin as $\sim 10^{-5}$ cm 
at high density $= 10^{10}$ g cm$^{-3}$, 
but can be as thick as $10^1$ cm at
low density $= 10^7$ g cm$^{-3}$ for 
matter with $X(^{12}$C$) = X(^{16}$O$) = 0.5$. 
For the eddy size, one use the Kolmogorov
scaling relation $v(l) = v(L) (l/L)^{1/3}$. 
The Gibb's length scale \citep{Niemeyer1995a} is given by 
\begin{equation}
\lambda_{{\rm Gibbs}} = \Delta \left( \frac{v^2_{{\rm lam}}}{2 q}\right). 
\end{equation}
Notice that the eddy size is not directly
used because below Gibb's length scale, 
anisotropy is always polished out by the 
flame propagation because of the faster
burning at cusp. 

To determine whether
detonation can start or not at the end of each step,
we check for grids which satisfy $Ka = 1$. 
A detonation seed of a ring with a radius
15 km is created around the grids which
fulfill this condition of $Ka = 1$. 
We remarked that, in the literature, there is not
yet any conclusive study that can pinpoint the 
exact detonation transition condition. 
Recent study of \cite{Poludenko2011} has 
found $Ka > 100$ is needed for a spontaneous
detonation for premixed H$_2$-air flame. 
In the hydrodynamics simulations,
we also perform models with different $Ka$ 
as the detonation transition criteria.

\subsection{Weak Interactions}

To study the effects of density, electron capture 
in NSE regions due to weak interaction cannot
be neglected. We include this process by introducing the
electron fraction $Y_{\rm e}$, which can be transported by 
the fluid flow, namely,
\begin{equation}
\frac{\partial \rho Y_{\rm e}}{\partial t} + \nabla \cdot (\rho Y_{\rm e} \vec{v}) = \dot{Y}_e.
\end{equation}
$\dot{Y}_e$ is the electron capture rate of matter in 
NSE as functions of density, temperature and
electron capture, which is derived from the table 
presented in \cite{Seitenzahl2009}. 
The change of local internal 
energy by the electron capture is given by:
\begin{equation}
\Delta \epsilon = (\mu_e + \mu_p - \mu_n) \dot{Y}_e - \dot{E}_{{\rm neut}}.
\end{equation}
Here, $\mu_i$ ($e$, $p$ or $n$) represents the chemical 
potential of electron, proton and neutron respectively. $\dot{E}_{{\rm neut}}$
is the energy loss by
the escaped neutrino during electron capture. 

We remark that this small network does not contain any isotope
which is mainly produced by NSE at $Y_{\rm e} < 0.50$. 
To incorporate the physics of electron capture into 
hydrodynamics, we define the effective atomic number
$\bar{Z}_{{\rm eff}} = Y_{\rm e} \bar{A}$, such that the pressure 
and internal energy are given by 
$p = p(\rho,T,\bar{A},\bar{Z}_{{\rm eff}})$ and 
$\epsilon = \epsilon(\rho,T,\bar{A}, \bar{Z}_{{\rm eff}})$ 
respectively. The reduced atomic number is to
mimic the reduction of electron fraction for the same amount of 
nucleons. 

\subsection{Post-processing nucleosynthesis and updates}
\label{sec:nucleo}

We use the tracer particle scheme \citep{Travaglio2004, Seitenzahl2010}
for the post-process nucleosynthesis. This scheme allocates some 
mass-less particles which follow the fluid motion during the 
hydrodynamics simulations. The thermodynamics history, including
their local densities, temperature, positions and velocities, 
of the tracer particles are recorded. The nuclear reaction 
history is then reconstructed. Following the suggestion
from \cite{Seitenzahl2010}, the number of tracers is
fixed to be $160^2$ to ensure the convergence of particles.
In the post-processing stage, we use the much larger 
495-isotope network which contains elements from 
hydrogen to technetium. In Table \ref{table:isotope_network}
we tabulate the isotopes included in the calculations.

We use the Torch subroutine \citep{Timmes1999a} to solve the system
of nuclear reaction equations. The subroutine solves the stiff system
of ordinary differential equations obtained from the chosen nuclear 
reaction network with the semi-implicit scheme.
The nuclear reaction rates are taken from \cite{Fowler1967, Thielemann1987}
and updated values from \cite{Rauscher2000}. 
The weak interaction takes the rates
from \cite{Fuller1982} with updated values 
from \cite{Langanke2001}. The screening function is taken from
\cite{Kitamura2000} which overwrites the default choice of  
the analytic formulae given in \cite{Alastuey1978}. 
In calculating the reaction rates, the faster rates among
($\alpha$,$p$), ($p$,$\gamma$) and ($\alpha$,$\gamma$) are chosen
as the preferred nuclear reactions.

We noticed that in \cite{Townsley2016}, they reported that
the post-process scheme requires a reconstruction for the thicken
flame in order to produce an accurate temperature-density history.
We remarked that such treatment is less important for the level-set 
formalism. First, in the 
advective-diffusive-reactive (ADR) flame formalism, the algorithm
estimates the injection the C-burning energy by the burning
variable $\phi_1$ where $\phi_1 \epsilon [0,1]$.
Cells with $\phi_1 \epsilon (0,1)$ stand for those 
actively burning carbon. Due to numerical diffusion and 
the inherent diffusion scheme, cells which are carrying out
carbon-burning are always dispersed into a few grids
in each direction. This phenomenon also appears even in high density,
where the real flame width can be as thin as $\sim$ cm. 
This creates a problem that the temperature growth
can be underestimated from the actual temperature 
evolution of the tracer particles and from the 
expected temperature evolution if the flame is treated 
as thin flame, which does not
experience the same smearing effect. 
To ameliorate the dispersion effect, 
in \cite{Townsley2016} the reconstruction process is
proposed. Instead of using the temperature of the 
Eulerian grid, the exact position of the 
deflagration is reconstructed by the composition.
The temperature of the tracer particle is interpolated 
by comparing the flame position and the particle 
position. On the other hand, 
in the level-set formalism, the flame geometry 
in a two-dimensional simulation is treated as an one-dimensional 
line provided by the zero-contour of the scalar field. The
injection of energy by the flame is always localized
and is not smeared. Thus the temperature in the 
Eulerian grid can consistent follow the average
temperature of that fluid element when the 
C-deflagration sweeps across. Notice that the 
temperature representation can be inaccurate
at very low density $< 10^7$ g cm$^{-3}$ where the
flame width extends to be comparable 
to the typical mesh size ($\sim$ km). In that case, the
ADR formalism provides a more accurate description
to the energy injection. But this feature will
be less important for the study of nucleosynthesis
since the important yields, in particular the 
iron-peak elements, are handled in the high density 
$\rho > 5 \times 10^7$ g cm$^{-3}$, where the 
thin flame treatment is a good approximation to
the actual energy injection.

\begin{table}

\begin{center}
\caption{Isotope network used for the 
post-process nucleosynthesis.}
\label{table:isotope_network}
\begin{tabular}{|c|c|c|c|}
\hline
Element & $Z$ & $A_{{\rm min}}$ & $A_{{\rm max}}$ \\ \hline
Hydrogen & 1 & 1 & 3 \\
Helium   & 2 & 3 & 4 \\
Lithium  & 3 & 6 & 7 \\
Beryllium & 4 & 7 & 9 \\
Boron & 5 & 8 & 11 \\
Carbon & 6 & 11 & 14 \\
Nitrogen & 7 & 12 & 15 \\
Oxygen & 8 & 14 & 19 \\
Fluorine & 9 & 17 & 21 \\
Neon & 10 & 17 & 24 \\
Sodium & 11 & 19 & 27 \\
Magnesium & 12 & 20 & 29 \\
Aluminum & 13 & 22 & 31 \\
Silicon & 14 & 23 & 34 \\ 
Phosphorous & 15 & 27 & 38 \\
Sulphur & 16 & 28 & 42 \\
Chlorine & 17 & 31 & 45 \\
Argon & 18 & 32 & 46 \\
Potassium & 19 & 35 & 49 \\
Calcium & 20 & 36 & 49 \\
Scandium & 21 & 40 & 51 \\
Titanium & 22 & 41 & 53 \\
Vanadium & 23 & 43 & 55 \\
Chromium & 24 & 44 & 59 \\
Manganese & 25 & 46 & 61 \\
Iron & 26 & 47 & 66 \\
Cobalt & 27 & 50 & 67 \\
Nickel & 28 & 51 & 68 \\
Copper & 29 & 55 & 69 \\
Zinc & 30 & 57 & 72 \\ \hline
Gallium & 31 & 59 & 75 \\
Germanium & 32 & 62 & 78 \\ 
Arsenic & 33 & 65 & 79 \\ 
Selenium & 34 & 67 & 83 \\ 
Bromine & 35 & 68 & 83 \\
Krypton & 36 & 69 & 87 \\ \hline
Rubidium & 37 & 73 & 85 \\ 
Strontium & 38 & 74 & 84 \\ 
Yttrium & 39 & 77 & 87 \\
Zirconium & 40 & 78 & 90 \\
Niobium & 41 & 82 & 90 \\
Molybdenum & 42 & 83 & 90 \\
Technetium & 43 & 89 & 91 \\ \hline
\end{tabular}
\end{center}
\end{table}

\section{Hydrodynamics of the Benchmark Model}
\label{sec:hydro}

\begin{figure}
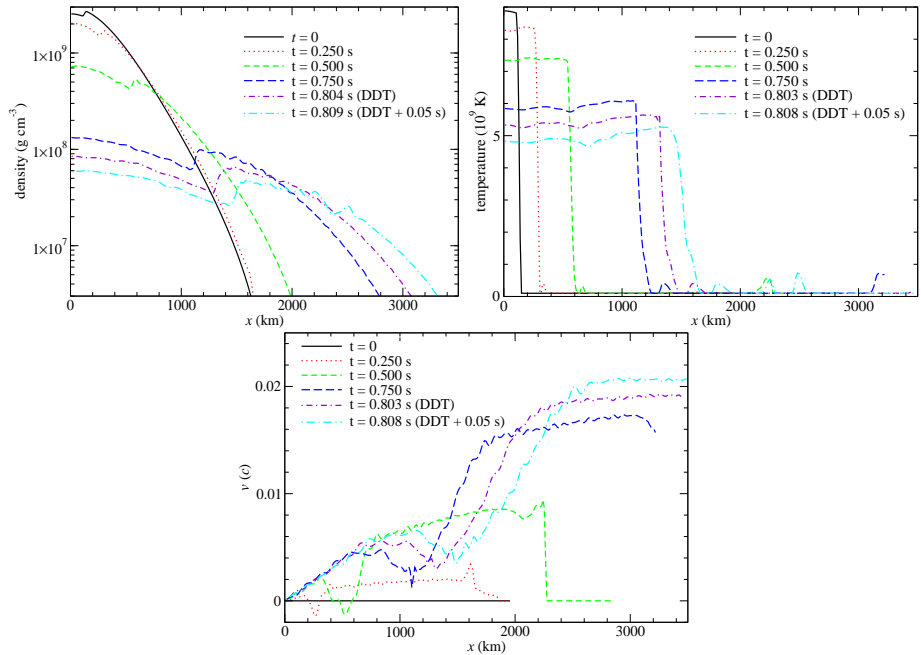

\centering
\includegraphics*[width=6cm,height=4.3cm]{fig25a.eps}
\includegraphics*[width=6cm,height=4.3cm]{fig25b.eps}
\includegraphics*[width=6cm,height=4.3cm]{fig25c.eps}
\caption{The density (left panel), 
temperature (middle panel) and velocity (right panel) profiles
of the benchmark model along the rotation axis 
at the beginning, 0.25, 0.50, 
0.75 s, at DDT and at 0.05 s after DDT.}
\label{fig:benchmark_axis}
\end{figure}

\begin{figure}
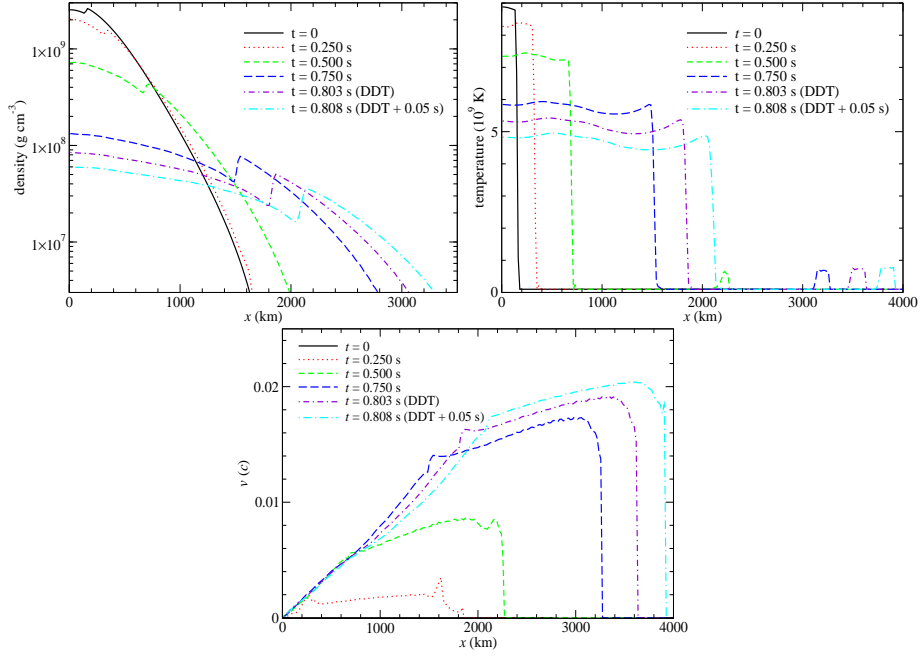

\centering
\includegraphics*[width=6cm,height=4.3cm]{fig26a.eps}
\includegraphics*[width=6cm,height=4.3cm]{fig26b.eps}
\includegraphics*[width=6cm,height=4.3cm]{fig26c.eps}
\caption{Similar to Figure \ref{fig:benchmark_axis}, but
along the diagonal direction.}
\label{fig:benchmark_diag}
\end{figure}

In the main text we have described the 
principle ideas about the benchmark model 
through its energy, luminosity and flame structure. 
In this section we further describe the hydrodynamics 
evolution of the benchmark models by its density, 
temperature and velocity at different time.
In Figures \ref{fig:benchmark_axis} and \ref{fig:benchmark_diag}
we plot the density, temperature and velocity profiles
along the rotation-axis and along diagonal direction 
respectively. Profiles at the beginning, at 0.25, 0.50, 0.75 s,
at DDT and at 0.05 s after DDT are chosen.

Along the rotation axis, the flame propagates
slower because the c3 flame has set this direction
to be a trough of the flame. At the beginning, the 
turbulent flame can efficiently burn the surrounding
matter of the core, making the core matter expand.
Within the first 0.75 s, the central density has dropped
by a factor of $\sim$ 10. The slow flame creates a 
small density contrast, which grows from a few percents
at a density of $10^9$ g cm$^{-3}$, to a few ten percents
at a density of $10^8$ g cm$^{-3}$. On the other
hand, due to the subsonic propagation, the temperature
profile appears to be very smooth within the 
ash region. A small temperature bump can be seen
just outside the deflagration front, which is because the 
isobaric condition is not perfectly implemented
at the beginning when the flame is implanted. 
At last, the velocity profile shows more 
interesting features along the rotation axis. 
At the beginning, there is a non-zero inflow 
of matter outside the flame front, which is related
to the Rayleigh-Taylor instabilities. The "finger" structure
allows the flame along the "finger" to propagate faster 
than the trough. As a result, the cold fuel is repelled from
the "finger" to the nearby trough which then hinders 
the propagation of flame. At the beginning, when the star
is mostly static, such effects can be most easily observed. 
Following the DDT, 
as noticed in Figure \ref{fig:flame_std}, the detonation 
wave creates disturbance along the axis because of the 
shock-boundary interaction. There are more wiggles in
the density profiles. From the velocity profile, it also shows
the inflow stops but the flame front along the axis
remains partially suppressed by the nearby flow.

Along the diagonal direction, the evolution becomes
much cleaner since this is along the "finger" structure of
the flame, whose propagation is more pronounced due to 
its geometry. The density evolution is almost comparable
to the typical one-dimensional models, where the 
flame creates the clean cut discontinuity. The 
density contrast grows with time. The temperature profiles
are also smooth compared to the previous plot. Again,
a small bump of temperature can be observed, but its
temperature is much below the ignition temperature 
of the fuel. At last, the velocity profiles show that
the homologous expansion quickly develops after DDT.
Before DDT, the ash shares a comparable velocity 
with some bumps near the surface similar to the temperature
profiles.

\section{Study of Minimum Temperature on the Flame Propagation}
\label{sec:tmin}

In the main text we have mentioned that the numerical scheme 
of how deflagration inputs energy is not completely quiet 
that the numerically resolved flame front can be 
different from the wave form solved analytically. 
Thus, just outside the flame front there can be disturbance
in the form of sound wave which mildly heats up the matter. 
However, it is unclear whether the non-zero sound wave
can affect the propagation of flame and also cause 
unrealistic heating of the matter, which may be 
misinterpreted by the post-processing nucleosythesis.
In this section we discuss the effects on the flame
propagation to the temperature profiles and 
luminosity. To analyze this effect, we present three
benchmark models but with different absolute minimum
temperature $T_{{\rm min}}$. This temperature corresponds
to the minimum temperature allowed in the simulation. 
When the EOS solvers convert the internal energy 
to the temperature for a given density and composition,
if the solved temperature is below $T_{{\rm min}}$, 
the temperature is reset to $T_{{\rm min}}$ with its
$\epsilon$ adjusted accordingly. To isolate this effect,
we pick $T_{{\rm min}} = 10^5$, $10^6$ and $10^7$ K 
respectively to check how large the difference can be.

\begin{figure}
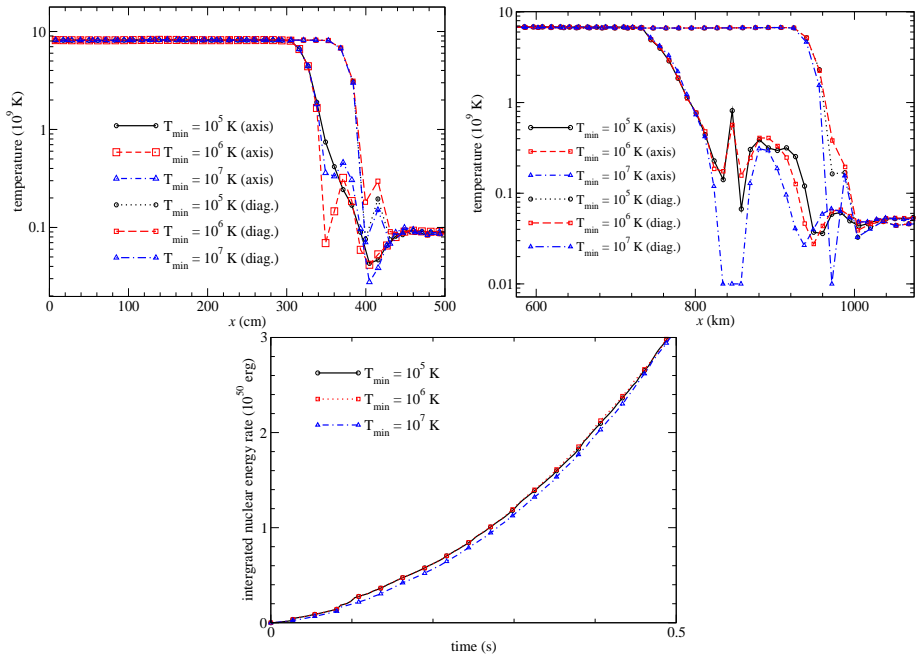

\centering
\includegraphics*[width=6cm,height=4.3cm]{fig27a.eps}
\includegraphics*[width=6cm,height=4.3cm]{fig27b.eps}
\includegraphics*[width=6cm,height=4.3cm]{fig27c.eps}
\caption{(left panel) Temperature profiles of the 
benchmark model along the rotation axis and diagonal 
direction at $t = 0.3$ s. Different choices 
of $T_{{\rm min}}$ from $10^5$ to $10^7$ K
are chosen. (middle panel) Similar to the
left panel, but at $t = 0.6$ s. (right panel)
The time-integrated nuclear energy production
for the three models. }
\label{fig:benchmark_Tmin}
\end{figure}

In Figure \ref{fig:benchmark_Tmin} we plot the temperature
profiles of the three models at 0.3 and 0.6 s, and the
corresponding time-integrated energy production
respectively. We take the directions along the rotation axis
and along the diagonal for sampling. At 0.3 s, we can 
see that along the axis 
for a high $T_{{\rm min}}$ (e.g. $10^7$ K), there is some 
temperature wave structure just outside the 
deflagration front. But the transition from fuel to ash
is smooth when a low $T_{{\rm min}}$ (e.g. $10^5$ K)
is chosen. However, the bump structure persists along
the diagonal. The difference in temperature profile
can extend from 30 - 50 km depending on the flame structure. 
The temperature profiles beyond are the same. 
At $t = 0.6$ s, where the flame front reaches a lower density,
the choice of $T_{{\rm min}}$ has more influences on
the temperature profile. Along the rotation axis, for
low $T_{{\rm min}}$ ($\leq 10^6$ K) the bump structure can be 
smoothly evolved where a spike at 820 km can be preserved.
For larger $T_{{\rm min}}$, the spike cannot be smoothly
produced when the temperature reaches below the limit. 
The size of this structure increases when
$T_{{\rm min}}$ becomes small. Again, the temperature 
profiles are independent to the choice of $T_{{\rm min}}$
behind the flame front (800 (900) km along the rotation
axis (diagonal)) and beyond the flame front ($\approx 1000$ km
for both directions).
Along the diagonal direction, similar observations 
appear that as $T_{{\rm min}} = 10^7$ K, the structure
beyond deflagration front cannot be fully captured. 
For lower temperature, the structure can be preserved 
when the temperature navigates around $T_{{\rm min}}$.

It can be seen that the choice of $T_{{\rm min}}$
can lead to some different temperature distributions 
just outside the flame front. However, we remark that,
despite all the differences, from the lower panel of 
Figure \ref{fig:benchmark_Tmin} we can see that the 
actual effects of the $T_{{\rm min}}$ choice 
are very small. At $t = 0.1$ s, the three models 
show a deviation from each other, where the 
one with $T_{{\rm min}} = 10^7$ K show a lower 
integrated nuclear energy about a few \%, 
meaning that its flame propagation is a bit slower
than the other case. But this effect is compensated at
around $t = 0.5$ s, when the deflagration becomes
large such that the small scale details become
less important. From this we observe that the choice of
$T_{{\rm min}}$ will affect the deflagration profile 
just outside the ash, but it has much smaller global 
effects compared to other numerical uncertainties.

\section{Effects of updated electron capture rates in 1-D models W7 and WDD2}

\label{sec:revisit}

In the present modeling, we use input physics as described in \S \ref{sec:input}.  In view of still existing uncertainties involved in
input physics, we examine how the updated electron capture rates
affect the nucleosynthesis yields in 1D Chandrasekhar mass models: PTD
model W7 \citep{Nomoto1984, Thielemann1986} and DDT model
WDD2 \citep{Iwamoto1999}.

These two models have been successfully 
describe the typical abundance distribution of SNe Ia and its coherence
with the solar abundance.
However, some problems have been notice in these two models. 
For example, stable Ni is overproduced to make
[Ni/Fe] $\sim$ 0.6 in W7, 
and Cr is overproduced in WDD2.

Due to the availability of new electron capture rates, 
where the rates for iron-peak elements are in general 
lower, it is interesting to see if the overproduction
problems in the W7 and WDD2 can be alleviated.
This is also a study of how the abundance ratios among iron-peak
elements depend on electron capture rates and possibly other 
nuclear reaction rates in view of still existing uncertainties of these rates
\citep[see also][]{Mori2016}.

Thus post-process nucleosynthesis in W7 and WDD2 are
re-calculated by using our updated nuclear reaction network.
In new W7 and WDD2, electron capture rates by
\cite{Langanke2001} are applied.
In Figure \ref{fig:benchmark_W7WDD2} we plot the scaled mass 
fraction for new W7 and WDD2 in the left and right panel,
respectively.

\begin{figure*}
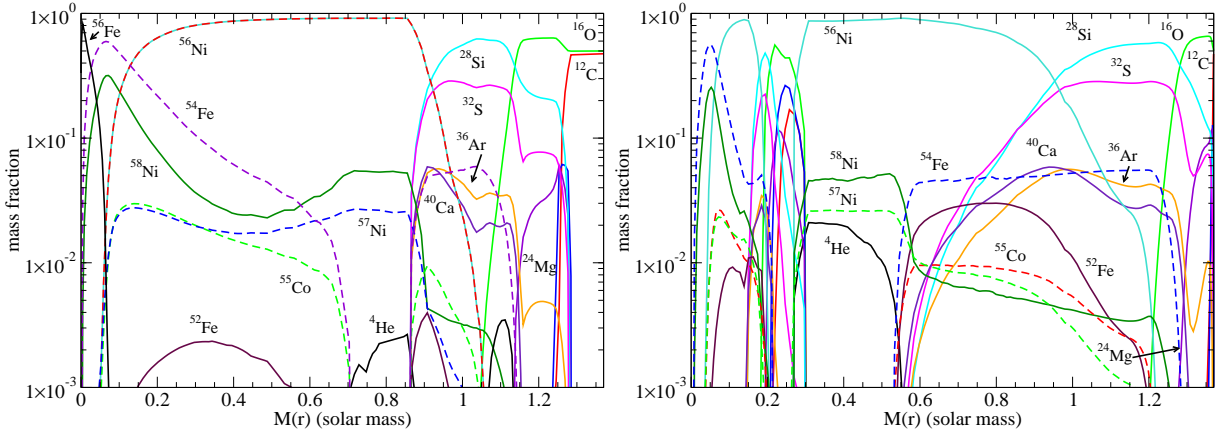

\centering
\includegraphics*[width=8cm,height=5.7cm]{fig28a.eps}
\includegraphics*[width=8cm,height=5.7cm]{fig28b.eps}
\caption{The mass fraction distribution of major isotopes 
for the W7 (left panel) and
WDD2 (right panel).}
\label{fig:benchmark_profile_W7WDD2}
\end{figure*}

\begin{figure*}
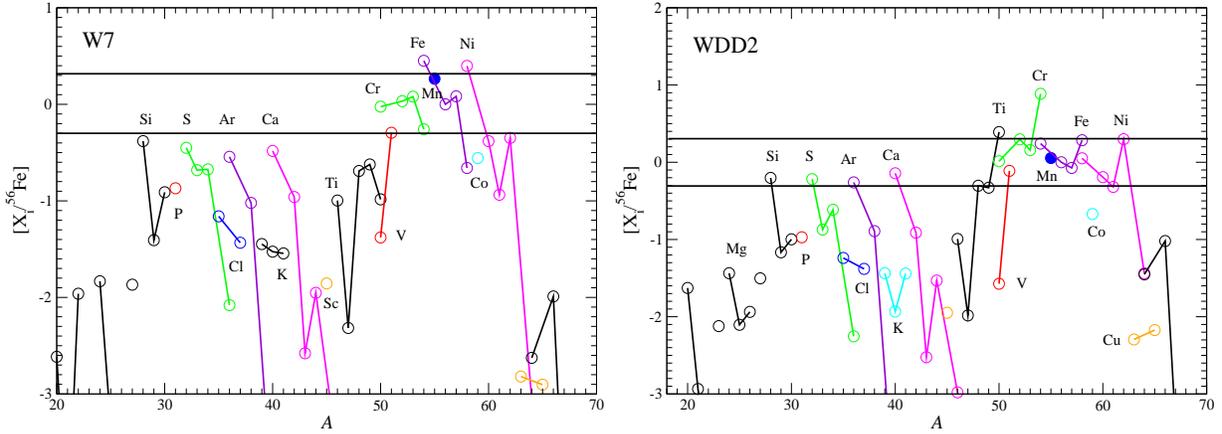

\centering
\includegraphics*[width=8cm,height=5.7cm]{fig29a_erratum.eps}
\includegraphics*[width=8cm,height=5.7cm]{fig29b_erratum.eps}
\caption{The scaled mass fraction [${\rm X_i}/^{56}$Fe] against 
atomic mass forW7 (left panel) and WDD2 (right panel).
($Remark$: The figure is replaced
due to the updated table.)}
\label{fig:benchmark_W7WDD2}
\end{figure*}

In Figure \ref{fig:benchmark_profile_W7WDD2} we
plot the chemical distribution of the major isotopes in 
the W7 and WDD2 against the mass coordinate.

In W7,
the turbulent deflagration produces a layered structure in the 
explosion ejecta. In the innermost part, electron capture
leads to the production of low-$Y_{\rm e}$ isotopes such as $^{56}$Fe and 
$^{54}$Fe. $^{58}$Ni is also abundantly produced
at a similar site as $^{54}$Fe. Then at $\sim 0.2 M_{\odot}$
$^{56}$Ni becomes the most abundant isotope, which extends 
up to $\sim 0.9 M_{\odot}$  (here $M_r$ is the Lagrangian mass coordinate).
At the same time, $^{55}$Co, which 
is the parent nuclei of $^{55}$Mn is also produced at
$M_r = 0.2 - 0.7 M_{\odot}$. Beyond 0.9 $M_{\odot}$, IMEs (including
Si, S, Ar and Ca) are the major isotopes in the middle layer.
At $M_r = 1.1 - 1.3 M_{\odot}$ matter is $^{16}$O-rich,
signifying that nuclear burning has become incomplete
as the flame reaches the low density zone. Above $M \sim 1.3 M_{\odot}$
no nuclear reaction occurs and the matter is pure $^{12}$C and $^{16}$O
(with $^{22}$Ne).

In WDD2, DDT produces two
distinct layers in the ejecta, the inner (outer) one making of product from
deflagration (detonation). In the innermost 0.3 $M_{\odot}$, where
matter is burnt by deflagration, the structure is similar to 
W7 with low-$Y_{\rm e}$ isotope in the core, outer core with $^{56}$Ni,
IMEs in the outer envelope, and $^{16}$O rich near the 
boundary of deflagration zone. Then in the detonation zone, $^{56}$Ni
is the dominant species which extends up to 0.9 $M_{\odot}$. 
Besides $^{56}$Ni, $^{57}$Ni and $^{58}$Ni are also produced from $M(r) = 0.3$
to 0.5 $M_{\odot}$. Then $^{54}$Fe and $^{55}$Co are 
produced in the ejecta. The matter becomes $^{28}$Si-rich at 
$M_r = 0.9 - 1.2 M_{\odot}$ and $^{16}$O-rich at $M_r > 1.2 M_{\odot}$.

In the yields of new W7, [$^{58}$Ni/$^{56}$Fe] $\sim$ 0.3, 
which is a factor of $\sim$ 2 smaller than old W7. This implies that the 
overproduction of $^{58}$Ni in old W7 
is due mainly to the less accurate electron capture rates.
The overproduction of $^{54}$Fe remains although it would not appear
in elemental abundances.

In the yields of both old and new WDD2, $^{50}$Ti and $^{54}$Cr are 
overproduced, which are synthesized in the low density detonated region 
rather than the electron capture effect
as in W7.
The overall masses of Ti and Cr remain 
compatible with the solar values because $^{50}$Ti
and $^{54}$Cr contribute to less than 10 \% of the total
mass of that element.

We also tabulate the 
mass yield of the radioactive isotopes 
from these models in Table \ref{table:Decay1} - \ref{table:Decay4}. 
The Models W7 and WDD2 are
re-computed by using our updated nuclear 
reaction network. \footnote{The electronic 
version of these yield tables are available in 
http://member.ipmu.jp/shingchi.leung/research$\_$gallery.htm$\#$Project1
and http://supernova.astron.s.u-tokyo.ac.jp/\~{}nomoto/yields}

\section{Study of parameters in numerical modelings}
\label{sec:numerical}

In this appendix, we present data for the models which 
explore the parameters not presented in the main text, 
they include the formula of the turbulent flame speed,
and the detonation transition criteria. In contrast
to the other model parameters presented in the main text,
which are relatively free parameters depending on the 
stellar evolution, its environment and its interaction
with its companion main-sequence stars, the model parameters
presented here are related to the input physics, which 
should be constrained by individual studies. However, due
to the resolution of resolving flame down to the Gibson
scale self-consistently in most SNe Ia simulations of the 
explosion phases, the exact value of these
parameters remain mostly unexplored. 
Here we show that the nucleosynthesis yield, 
can shed light on these parameters. In Table \ref{table:Models_app}
we tabulate the models we studied by varying the input
parameters which belongs to the theoretical uncertainties. 
The model 300-1-c3-1 is also listed as to contrast all the
variants with our benchmark model. 

\begin{table*}

\begin{center}
\caption{Model setup for the benchmark model: 
central densities of NM $\rho_{ c{\rm (NM)} }$
are in units of $10^{9}$ g cm$^{-3}$.  
Metallicity is in units of solar metallicity.
Masses of the baryonic matter $M_{{\rm NM}}$ and
and the final nickel-56 mass $M_{{\rm Ni}}$ 
are in units of solar mass. 
$R$ is the initial stellar radius. 
$E_{\rm nuc}$ and $E_{\rm tot}$ are the energy released by nuclear reactions
and final total energy, respectively, both in units of $10^{50}$ erg. 
$Y_{{\rm e(min)}}$ is the minimum value of electron fraction within
the simulation box at the end of simulation. $t_{{\rm DDT}}$ is 
the first detonation transition time in units of second. 
$M_{{\rm Fe}}$ is the mass of $^{56}$Fe at the end of 
simulation, after all short-live radioactive isotopes 
have decayed. In the last column, $Ka$ is the 
Karlovitz number for the detonation transition. 
$\alpha$ is the scale-down factor for the 
turbulent flame speed. Old screen stands for
using the default screening function for the
nuclear network and "No e-cap" stands for 
no electron capture is done in the hydrodynamics.}
\label{table:Models_app}
\begin{tabular}{|c|c|c|c|c|c|c|c|c|c|c|c|c|}

\hline
Model & $\rho_{c({\rm NM})}$ & Metallicity & flame shape & $X_{12C}$ & mass & 
$R$ & $Y_{{\rm e(min)}}$ & $E_{\rm nuc}$ & $E_{\rm tot}$ & $t_{{\rm DDT}}$ & $M_{{\rm Ni}}$ & Others \\ \hline
300-0-c3-1  & 3 & 1 & $c3$   & 0.49  & 1.38 & 1900 & 0.462 & 17.7 & 12.7 & 0.78 & 0.63 & \\ \hline
Test-A1  & 3 & 1 & $c3$   & 0.49  & 1.38 & 1900 & 0.462 & 16.3 & 11.3 & 0.80 & 0.52 & Ka $\times 2$ \\ 
Test-A2  & 3 & 1 & $c3$   & 0.49  & 1.38 & 1900 & 0.462 & 14.5 & 9.44 & 0.82 & 0.40 & Ka $\times 4$ \\ \hline
Test-B1  & 3 & 1 & $c3$   & 0.49  & 1.38 & 1900 & 0.462 & 18.7 & 13.6 & 0.89 & 0.76 & $\alpha = 0.50$ \\ 
Test-B2  & 3 & 1 & $c3$   & 0.49  & 1.38 & 1900 & 0.453 & 19.0 & 13.9 & 0.66 & 0.97 & $\alpha = 0.25$ \\ \hline

\end{tabular}
\end{center}
\end{table*}

\begin{figure}
\centering
\includegraphics*[width=8cm,height=5.7cm]{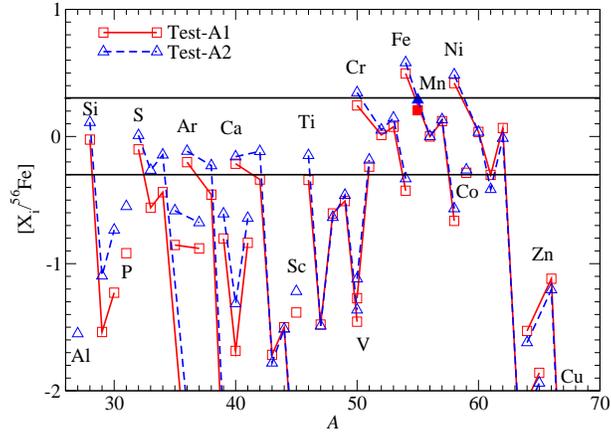}
\caption{Final mass fraction for Models
Test-A1 and Test-A2 after all radioactive 
decay of short-lived isotopes. ($Remark$: The figure is replaced
due to the updated table.)}
\label{fig:final_Ka}
\end{figure}

\subsection{Effects of detonation criteria}

Models 3-1-c3-1, Test-A1 and Test-A2 form a test which studies
the effects of the detonation criteria. Here, we modify the detonation
criteria $Ka$. In the main text, all models uses $Ka$ = 0.5. For Models Test-A1
and Test-A2, we study these parameters by increasing it to 1.0 and 2.0 
respectively. In order to know whether the flame can develop
into detonation or not in the distributed regime, resolving
the reaction zone is essential. However, even at density close
to the quenching of carbon-oxygen flame ($\sim 10^7$ g cm$^{-3}$),
the typical size of flame width has size below the 
simulated resolution. In the literature, one
picks the Karlovitz number, which represents the ratio 
between the typical flame width and the eddy turnover 
size, as to indicate whether to activate the trigger.
Certainly, the exact Karlovitz number depends on in general
extra information such as the exact velocity power spectrum, 
which cannot be known unless one consistently refines 
the resolution to keep track of these quantities. 
Therefore, there exists uncertainties in the exact trigger
time. To mimic this uncertainties, we vary the Karloritz number
to study its effects in nucleosynthesis. In Tables \ref{table:Isotopes_others} and \ref{table:Isotopes_othersb}
we tabulate the explosion energetic of these models. 
Since all models assume the same initial progenitor, 
there is no change in the initial global parameters. 
When the DDT criteria becomes higher, the deflagration
wave needs to reach lower densities in order to 
fulfill the condition, where the flame width increases
rapidly when density reaches $10^7$ g cm$^{-3}$. The later 
detonation time allows the matter to expand and have a
lower density before the detonation wave swept through. 
Thus, more matter undergoes incomplete burning, 
which reduces the energy production as well as the 
$^{56}$Ni synthesis. The $Y_{{\rm e(min)}}$ does not
vary much because the electron-capture zone lies
deep in the core, where deflagration takes part. 
In Figure \ref{fig:final_Ka} we plot the isotope abundance 
of the two models with respect to solar abundance. 
In general, most isotopes have a higher mass ratio
when $Ka$ increases. By comparing with Table 
\ref{table:Isotopes_othersb}, the iron-peaked elements
do not vary much when $Ka$ increases by a factor of 4.
This is consistent with the picture that most iron-peaked
elements, with those low $Y_{\rm e}$ isotopes inclusively, are
produced by deflagration. On the contrary, there is 
a rapid increase for most intermediate mass elements
up to $^{41}$K, which corresponds to the weaker 
detonation due to the expansion of matter. 
The study here suggests that the Karlovitz number is
crucial that it affects the global chemical abundance
by influencing the $^{56}$Ni production.

\subsection{Effects of turbulent flame models}

\begin{figure}
\centering
\includegraphics*[width=8cm,height=5.7cm]{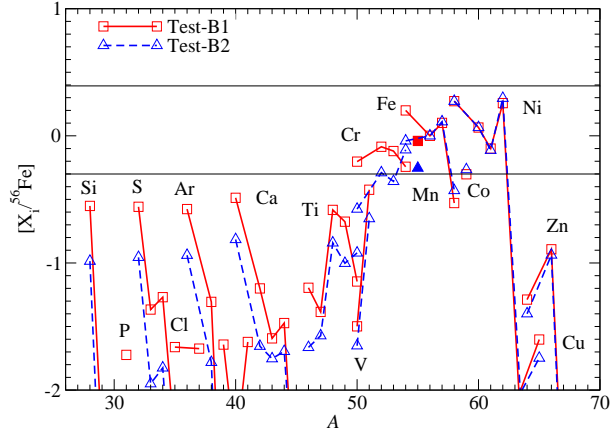}
\caption{Similar to Figure \ref{fig:final_Ka}, but for Models
Test-B1 and Test-B2. ($Remark$: The figure is replaced
due to the updated table.)}
\label{fig:final_alpha}
\end{figure}

Models 3-1-c3-1, Test-B1 and Test-B2 form another tests which
studies the effects of the turbulent flame formula to the 
global nucleosynthesis pattern. In the deflagration regime, 
two input physics are required, including the sub-grid
turbulence model and a formula relating the local velocity
fluctuations and the effective turbulent flame propagation.
In the literature, a number of turbulence models have been 
proposed to mimic the sub-grid development of velocity 
fluctuation induced by turbulent eddies and sub-grid
scale dissipation (See for example the one-equation
model \citep{Niemeyer1995a}, two-equation model 
\citep{Shih1994b}, the Rayleigh Stress-tensor model 
\citep{Shih1995} and the three-equation model 
\citep{Yoshizawa2012}). Similarly, a number of 
formula have been proposed to describe the process
(see for example the classical formula dervied from the 
Bunsen flame experiment \citep{Damkoehler1939}, the 
flame formula based on renormalizable scheme \citep{Pocheau1994},
and its variants based on empirical fitting \citep{Hicks2015}).
The variety of these models arise from the difficulties 
of resolving flame in a first-principle manner in a full star
SN Ia simulation, which requires resolution down to 
the Gibson's scale \citep{Niemeyer1995a}. Also, the
extremely high Reynold's number ($\sim 10^{16}$) in 
the scenario makes any direct modeling between 
turbulence and flame propagation formidable. 
In order to mimic the uncertainty in these models,
we add a scaling factor $\alpha$ to the flame formula, namely 
\begin{equation}
v_{{\rm flame}} = v_{{\rm lam}} \sqrt{1 + \alpha C \left( \frac{v'}{v_{{\rm lam}}} \right) }.
\end{equation}
This factor attempts to represent all uncertainties in the 
sub-grid turbulence generation, their inherent wall-proximity
relation and their corresponding dissipation rates. In this
test, we pick $\alpha = 0.25, 0.50, 1.00$. 
In Figure \ref{fig:final_alpha} we plot the isotope abundance
of the two tests, Test-B1 and Test-B2. 
It can be seen that the choice of $\alpha$ bring mild changes
to the mass ratio. The intermediate mass elements are in general
decreased when $\alpha$ decreases. Similar variations are found
for elements like Ti, V and Cr. No significant change is
found for iron-peak elements beyond $^{56}$Fe.

\newpage

\begin{table*}
\begin{center}
\caption{Nucleosynthesis yield for the 
Models presented in this articles. All models
in this table is based on the series with 
a $\rho_c = 10^9$ g cm$^{-3}$, $c3$ flame
and C/O ratio $ = 1$. The 
isotope masses are in units of solar mass.
($Remark$: The table is replaced due to typos while converting the 
raw data into the current table form. Changes are made for the 
isotopes including $^{22}$Ne, $^{26}$Mg, $^{26}$Al, $^{36}$S, $^{40}$K, 
$^{41}$K, $^{44}$Ca, $^{53}$Cr, $^{55}$Mn, $^{60}$Fe, $^{59}$Co and $^{63}$Cu.)}
\label{table:Isotopes}
\begin{tabular}{|c | c c c c c c c|}

\hline
Isotopes & $Z = 0$ & $Z = 0.1 Z_{\odot}$ & $Z = 0.5 Z_{\odot}$ & $Z = Z_{\odot}$ & 
		   $Z = 2 Z_{\odot}$ & $Z = 3 Z_{\odot}$ & $Z = 5 Z_{\odot}$ \\ \hline

 $^{12}$C & $1.58 \times 10^{-3}$ & $1.58 \times 10^{-3}$ & $1.32 \times 10^{-3}$ & $1.31 \times 10^{-3}$ & $1.29 \times 10^{-3}$ & $1.48 \times 10^{-3}$ & $1.45 \times 10^{-3}$ \\
 $^{13}$C & $7.17 \times 10^{-12}$ & $2.50 \times 10^{-12}$ & $3.79 \times 10^{-12}$ & $8.17 \times 10^{-12}$ & $2.44 \times 10^{-11}$ & $9.60 \times 10^{-12}$ & $5.68 \times 10^{-11}$ \\
 $^{14}$N & $2.3 \times 10^{-9}$ & $2.74 \times 10^{-10}$ & $3.34 \times 10^{-10}$ & $5.57 \times 10^{-10}$ & $1.14 \times 10^{-9}$ & $2.38 \times 10^{-10}$ & $8.34 \times 10^{-10}$ \\
 $^{15}$N & $7.64 \times 10^{-7}$ & $6.32 \times 10^{-9}$ & $9.8 \times 10^{-10}$ & $2.45 \times 10^{-10}$ & $6.56 \times 10^{-11}$ & $1.30 \times 10^{-11}$ & $5.92 \times 10^{-11}$ \\
 $^{16}$O & $4.19 \times 10^{-2}$ & $4.45 \times 10^{-2}$ & $5.38 \times 10^{-2}$ & $5.45 \times 10^{-2}$ & $5.49 \times 10^{-2}$ & $5.49 \times 10^{-2}$ & $6.55 \times 10^{-2}$ \\
 $^{17}$O & $7.89 \times 10^{-12}$ & $1.36 \times 10^{-12}$ & $4.27 \times 10^{-11}$ & $1.48 \times 10^{-10}$ & $4.95 \times 10^{-10}$ & $1.36 \times 10^{-10}$ & $4.7 \times 10^{-10}$ \\
 $^{18}$O & $1.91 \times 10^{-13}$ & $6.23 \times 10^{-14}$ & $1.47 \times 10^{-12}$ & $4.28 \times 10^{-12}$ & $1.5 \times 10^{-11}$ & $1.63 \times 10^{-12}$ & $5.78 \times 10^{-11}$ \\
 $^{19}$F & $3.67 \times 10^{-12}$ & $7.92 \times 10^{-13}$ & $1.46 \times 10^{-12}$ & $2.15 \times 10^{-12}$ & $7.92 \times 10^{-12}$ & $1.48 \times 10^{-12}$ & $8.35 \times 10^{-12}$ \\
 $^{20}$Ne & $2.18 \times 10^{-4}$ & $2.18 \times 10^{-4}$ & $5.48 \times 10^{-4}$ & $5.51 \times 10^{-4}$ & $5.51 \times 10^{-4}$ & $1.69 \times 10^{-4}$ & $4.36 \times 10^{-4}$ \\
 $^{21}$Ne & $3.33 \times 10^{-10}$ & $4.60 \times 10^{-10}$ & $4.73 \times 10^{-9}$ & $1.71 \times 10^{-8}$ & $6.20 \times 10^{-8}$ & $2.97 \times 10^{-8}$ & $3.54 \times 10^{-7}$ \\
 $^{22}$Ne & $1.52 \times 10^{-9}$ & $6.23 \times 10^{-6}$ & $2.59 \times 10^{-5}$ & $5.19 \times 10^{-5}$ & $1.3 \times 10^{-4}$ & $1.87 \times 10^{-4}$ & $3.11 \times 10^{-4}$ \\
 $^{23}$Na & $4.39 \times 10^{-7}$ & $3.81 \times 10^{-7}$ & $1.49 \times 10^{-6}$ & $2.9 \times 10^{-6}$ & $3.26 \times 10^{-6}$ & $1.34 \times 10^{-6}$ & $6.31 \times 10^{-6}$ \\
 $^{24}$Mg & $2.94 \times 10^{-3}$ & $2.61 \times 10^{-3}$ & $2.34 \times 10^{-3}$ & $1.70 \times 10^{-3}$ & $1.15 \times 10^{-3}$ & $8.69 \times 10^{-4}$ & $1.7 \times 10^{-3}$ \\
 $^{25}$Mg & $1.0 \times 10^{-8}$ & $5.47 \times 10^{-7}$ & $2.13 \times 10^{-6}$ & $5.6 \times 10^{-6}$ & $1.25 \times 10^{-5}$ & $8.4 \times 10^{-6}$ & $3.71 \times 10^{-5}$ \\
 $^{26}$Mg & $2.23 \times 10^{-7}$ & $4.23 \times 10^{-7}$ & $3.80 \times 10^{-6}$ & $7.17 \times 10^{-6}$ & $1.39 \times 10^{-5}$ & $8.65 \times 10^{-6}$ & $6.46 \times 10^{-5}$ \\
 $^{26}$Al & $3.45 \times 10^{-29}$ & $3.45 \times 10^{-29}$ & $4.77 \times 10^{-28}$ & $3.45 \times 10^{-29}$ & $3.45 \times 10^{-29}$ & $1.94 \times 10^{-11}$ & $2.61 \times 10^{-11}$ \\
 $^{27}$Al & $1.4 \times 10^{-5}$ & $3.29 \times 10^{-5}$ & $1.12 \times 10^{-4}$ & $1.36 \times 10^{-4}$ & $1.49 \times 10^{-4}$ & $1.47 \times 10^{-4}$ & $2.62 \times 10^{-4}$ \\
 $^{28}$Si & $1.62 \times 10^{-1}$ & $1.69 \times 10^{-1}$ & $2.16 \times 10^{-1}$ & $2.20 \times 10^{-1}$ & $2.23 \times 10^{-1}$ & $2.18 \times 10^{-1}$ & $2.52 \times 10^{-1}$ \\
 $^{29}$Si & $3.59 \times 10^{-5}$ & $7.62 \times 10^{-5}$ & $1.78 \times 10^{-4}$ & $2.65 \times 10^{-4}$ & $4.75 \times 10^{-4}$ & $6.59 \times 10^{-4}$ & $1.88 \times 10^{-3}$ \\
 $^{30}$Si & $4.14 \times 10^{-5}$ & $1.38 \times 10^{-5}$ & $1.85 \times 10^{-4}$ & $4.33 \times 10^{-4}$ & $1.3 \times 10^{-3}$ & $1.91 \times 10^{-3}$ & $5.83 \times 10^{-3}$ \\
 $^{31}$P & $5.10 \times 10^{-5}$ & $2.59 \times 10^{-5}$ & $9.91 \times 10^{-5}$ & $1.66 \times 10^{-4}$ & $2.99 \times 10^{-4}$ & $4.16 \times 10^{-4}$ & $8.76 \times 10^{-4}$ \\
 $^{32}$S & $1.11 \times 10^{-1}$ & $1.9 \times 10^{-1}$ & $1.25 \times 10^{-1}$ & $1.20 \times 10^{-1}$ & $1.9 \times 10^{-1}$ & $9.92 \times 10^{-2}$ & $9.45 \times 10^{-2}$ \\
 $^{33}$S & $1.87 \times 10^{-5}$ & $6.25 \times 10^{-5}$ & $1.63 \times 10^{-4}$ & $2.30 \times 10^{-4}$ & $3.44 \times 10^{-4}$ & $3.94 \times 10^{-4}$ & $6.19 \times 10^{-4}$ \\
 $^{34}$S & $1.25 \times 10^{-5}$ & $7.37 \times 10^{-5}$ & $7.59 \times 10^{-4}$ & $1.70 \times 10^{-3}$ & $3.87 \times 10^{-3}$ & $6.3 \times 10^{-3}$ & $1.39 \times 10^{-2}$ \\
 $^{36}$S & $3.79 \times 10^{-13}$ & $5.48 \times 10^{-10}$ & $1.4 \times 10^{-8}$ & $4.49 \times 10^{-8}$ & $3.36 \times 10^{-7}$ & $1.20 \times 10^{-6}$ & $1.1 \times 10^{-5}$ \\
 $^{35}$Cl & $1.13 \times 10^{-5}$ & $1.42 \times 10^{-5}$ & $6.34 \times 10^{-5}$ & $1.7 \times 10^{-4}$ & $1.76 \times 10^{-4}$ & $1.87 \times 10^{-4}$ & $2.84 \times 10^{-4}$ \\
 $^{37}$Cl & $2.29 \times 10^{-6}$ & $8.77 \times 10^{-6}$ & $2.50 \times 10^{-5}$ & $3.55 \times 10^{-5}$ & $4.98 \times 10^{-5}$ & $4.91 \times 10^{-5}$ & $6.66 \times 10^{-5}$ \\
 $^{36}$Ar & $2.61 \times 10^{-2}$ & $2.44 \times 10^{-2}$ & $2.49 \times 10^{-2}$ & $2.27 \times 10^{-2}$ & $1.91 \times 10^{-2}$ & $1.70 \times 10^{-2}$ & $1.41 \times 10^{-2}$ \\
 $^{38}$Ar & $1.31 \times 10^{-6}$ & $5.61 \times 10^{-5}$ & $5.71 \times 10^{-4}$ & $1.27 \times 10^{-3}$ & $2.72 \times 10^{-3}$ & $3.62 \times 10^{-3}$ & $7.52 \times 10^{-3}$ \\
 $^{40}$Ar & $2.83 \times 10^{-16}$ & $3.63 \times 10^{-12}$ & $1.27 \times 10^{-10}$ & $8.72 \times 10^{-10}$ & $9.57 \times 10^{-9}$ & $3.44 \times 10^{-8}$ & $2.27 \times 10^{-7}$ \\
 $^{39}$K & $3.52 \times 10^{-6}$ & $1.50 \times 10^{-5}$ & $7.41 \times 10^{-5}$ & $1.14 \times 10^{-4}$ & $1.64 \times 10^{-4}$ & $1.56 \times 10^{-4}$ & $1.88 \times 10^{-4}$ \\
 $^{40}$K & $3.63 \times 10^{-13}$ & $1.25 \times 10^{-9}$ & $9.99 \times 10^{-9}$ & $2.60 \times 10^{-8}$ & $6.31 \times 10^{-8}$ & $6.62 \times 10^{-8}$ & $1.5 \times 10^{-7}$ \\
 $^{41}$K & $5.46 \times 10^{-7}$ & $2.41 \times 10^{-6}$ & $6.34 \times 10^{-6}$ & $8.75 \times 10^{-6}$ & $1.11 \times 10^{-5}$ & $9.59 \times 10^{-6}$ & $1.7 \times 10^{-5}$ \\
 $^{40}$Ca & $2.64 \times 10^{-2}$ & $2.38 \times 10^{-2}$ & $2.21 \times 10^{-2}$ & $1.96 \times 10^{-2}$ & $1.62 \times 10^{-2}$ & $1.48 \times 10^{-2}$ & $1.16 \times 10^{-2}$ \\
 $^{42}$Ca & $1.6 \times 10^{-8}$ & $1.55 \times 10^{-6}$ & $1.88 \times 10^{-5}$ & $4.30 \times 10^{-5}$ & $8.70 \times 10^{-5}$ & $1.3 \times 10^{-4}$ & $1.81 \times 10^{-4}$ \\
 $^{43}$Ca & $4.8 \times 10^{-7}$ & $5.58 \times 10^{-7}$ & $1.77 \times 10^{-6}$ & $1.39 \times 10^{-6}$ & $9.50 \times 10^{-7}$ & $7.28 \times 10^{-7}$ & $7.90 \times 10^{-7}$ \\
 $^{44}$Ca & $4.60 \times 10^{-5}$ & $4.41 \times 10^{-5}$ & $3.71 \times 10^{-5}$ & $3.34 \times 10^{-5}$ & $2.86 \times 10^{-5}$ & $2.52 \times 10^{-5}$ & $2.13 \times 10^{-5}$ \\
 $^{46}$Ca & $1.92 \times 10^{-21}$ & $1.41 \times 10^{-15}$ & $1.4 \times 10^{-12}$ & $2.63 \times 10^{-11}$ & $4.9 \times 10^{-10}$ & $1.2 \times 10^{-9}$ & $1.86 \times 10^{-9}$ \\
 $^{48}$Ca & $2.62 \times 10^{-25}$ & $1.7 \times 10^{-22}$ & $8.55 \times 10^{-19}$ & $1.18 \times 10^{-16}$ & $1.81 \times 10^{-14}$ & $1.29 \times 10^{-13}$ & $1.93 \times 10^{-12}$ \\
 $^{45}$Sc & $6.51 \times 10^{-8}$ & $1.64 \times 10^{-7}$ & $3.80 \times 10^{-7}$ & $4.47 \times 10^{-7}$ & $4.89 \times 10^{-7}$ & $4.87 \times 10^{-7}$ & $5.88 \times 10^{-7}$ \\
 $^{46}$Ti & $3.73 \times 10^{-6}$ & $2.54 \times 10^{-6}$ & $1.5 \times 10^{-5}$ & $2.24 \times 10^{-5}$ & $4.38 \times 10^{-5}$ & $5.9 \times 10^{-5}$ & $7.97 \times 10^{-5}$ \\
 $^{47}$Ti & $1.94 \times 10^{-6}$ & $2.31 \times 10^{-6}$ & $4.70 \times 10^{-6}$ & $5.42 \times 10^{-6}$ & $5.51 \times 10^{-6}$ & $5.4 \times 10^{-6}$ & $6.47 \times 10^{-6}$ \\
 $^{48}$Ti & $6.56 \times 10^{-4}$ & $6.5 \times 10^{-4}$ & $5.1 \times 10^{-4}$ & $4.43 \times 10^{-4}$ & $3.69 \times 10^{-4}$ & $3.33 \times 10^{-4}$ & $2.52 \times 10^{-4}$ \\
 $^{49}$Ti & $3.94 \times 10^{-6}$ & $1.75 \times 10^{-5}$ & $2.53 \times 10^{-5}$ & $2.93 \times 10^{-5}$ & $3.14 \times 10^{-5}$ & $3.71 \times 10^{-5}$ & $3.87 \times 10^{-5}$ \\
 $^{50}$Ti & $3.40 \times 10^{-14}$ & $1.47 \times 10^{-13}$ & $6.21 \times 10^{-11}$ & $6.4 \times 10^{-10}$ & $2.15 \times 10^{-9}$ & $3.49 \times 10^{-9}$ & $1.90 \times 10^{-8}$ \\
 $^{50}$V & $1.5 \times 10^{-11}$ & $2.10 \times 10^{-11}$ & $8.0 \times 10^{-10}$ & $3.56 \times 10^{-9}$ & $9.96 \times 10^{-9}$ & $2.41 \times 10^{-8}$ & $1.10 \times 10^{-7}$ \\
 $^{51}$V & $2.58 \times 10^{-5}$ & $2.96 \times 10^{-5}$ & $5.83 \times 10^{-5}$ & $7.89 \times 10^{-5}$ & $1.3 \times 10^{-4}$ & $1.40 \times 10^{-4}$ & $2.4 \times 10^{-4}$ \\

\end{tabular}
\end{center}
\end{table*}  
  
\begin{table*}
\begin{center}
\caption{($cont'd$ of Table \ref{table:Isotopes}.
}
\label{table:Isotopesb}
\begin{tabular}{|c | c c c c c c c|}

\hline
Isotopes & $Z = 0$ & $Z = 0.1 Z_{\odot}$ & $Z = 0.5 Z_{\odot}$ & $Z = Z_{\odot}$ & 
		   $Z = 2 Z_{\odot}$ & $Z = 3 Z_{\odot}$ & $Z = 5 Z_{\odot}$ \\ \hline  
  
 $^{50}$Cr & $5.28 \times 10^{-5}$ & $6.12 \times 10^{-5}$ & $1.77 \times 10^{-4}$ & $3.50 \times 10^{-4}$ & $7.38 \times 10^{-4}$ & $1.6 \times 10^{-3}$ & $1.60 \times 10^{-3}$ \\
 $^{52}$Cr & $8.49 \times 10^{-3}$ & $7.98 \times 10^{-3}$ & $6.55 \times 10^{-3}$ & $5.91 \times 10^{-3}$ & $5.33 \times 10^{-3}$ & $5.76 \times 10^{-3}$ & $7.30 \times 10^{-3}$ \\
 $^{53}$Cr & $2.34 \times 10^{-4}$ & $4.3 \times 10^{-4}$ & $5.35 \times 10^{-4}$ & $6.30 \times 10^{-4}$ & $7.89 \times 10^{-4}$ & $1.3 \times 10^{-3}$ & $1.37 \times 10^{-3}$ \\
 $^{54}$Cr & $3.32 \times 10^{-8}$ & $3.39 \times 10^{-8}$ & $5.27 \times 10^{-8}$ & $1.19 \times 10^{-7}$ & $5.73 \times 10^{-7}$ & $1.71 \times 10^{-6}$ & $7.3 \times 10^{-6}$ \\
 $^{55}$Mn & $3.25 \times 10^{-3}$ & $3.79 \times 10^{-3}$ & $4.97 \times 10^{-3}$ & $5.89 \times 10^{-3}$ & $7.56 \times 10^{-3}$ & $9.70 \times 10^{-3}$ & $1.27 \times 10^{-2}$ \\
 $^{54}$Fe & $3.12 \times 10^{-2}$ & $3.28 \times 10^{-2}$ & $4.18 \times 10^{-2}$ & $5.28 \times 10^{-2}$ & $7.41 \times 10^{-2}$ & $9.67 \times 10^{-2}$ & $1.39 \times 10^{-1}$ \\
 $^{56}$Fe & $8.46 \times 10^{-1}$ & $8.39 \times 10^{-1}$ & $7.51 \times 10^{-1}$ & $7.25 \times 10^{-1}$ & $6.80 \times 10^{-1}$ & $6.41 \times 10^{-1}$ & $5.5 \times 10^{-1}$ \\
 $^{57}$Fe & $1.61 \times 10^{-2}$ & $1.72 \times 10^{-2}$ & $1.88 \times 10^{-2}$ & $2.17 \times 10^{-2}$ & $2.60 \times 10^{-2}$ & $2.91 \times 10^{-2}$ & $2.92 \times 10^{-2}$ \\
 $^{58}$Fe & $1.39 \times 10^{-7}$ & $1.41 \times 10^{-7}$ & $1.57 \times 10^{-7}$ & $1.86 \times 10^{-7}$ & $2.97 \times 10^{-7}$ & $4.66 \times 10^{-7}$ & $1.0 \times 10^{-6}$ \\
 $^{60}$Fe & $2.59 \times 10^{-21}$ & $4.73 \times 10^{-21}$ & $2.11 \times 10^{-19}$ & $2.39 \times 10^{-18}$ & $2.3 \times 10^{-18}$ & $6.67 \times 10^{-18}$ & $5.81 \times 10^{-17}$ \\
 $^{59}$Co & $1.85 \times 10^{-4}$ & $1.60 \times 10^{-4}$ & $4.55 \times 10^{-4}$ & $6.32 \times 10^{-4}$ & $8.15 \times 10^{-4}$ & $8.48 \times 10^{-4}$ & $8.12 \times 10^{-4}$ \\
 $^{58}$Ni & $2.14 \times 10^{-2}$ & $2.20 \times 10^{-2}$ & $3.14 \times 10^{-2}$ & $4.60 \times 10^{-2}$ & $7.48 \times 10^{-2}$ & $1.2 \times 10^{-1}$ & $1.39 \times 10^{-1}$ \\
 $^{60}$Ni & $1.25 \times 10^{-2}$ & $1.29 \times 10^{-2}$ & $1.2 \times 10^{-2}$ & $9.16 \times 10^{-3}$ & $7.53 \times 10^{-3}$ & $5.73 \times 10^{-3}$ & $4.14 \times 10^{-3}$ \\
 $^{61}$Ni & $3.21 \times 10^{-4}$ & $3.47 \times 10^{-4}$ & $3.48 \times 10^{-4}$ & $3.85 \times 10^{-4}$ & $4.25 \times 10^{-4}$ & $4.4 \times 10^{-4}$ & $3.79 \times 10^{-4}$ \\
 $^{62}$Ni & $1.4 \times 10^{-4}$ & $1.88 \times 10^{-4}$ & $1.25 \times 10^{-3}$ & $2.34 \times 10^{-3}$ & $4.19 \times 10^{-3}$ & $5.19 \times 10^{-3}$ & $6.95 \times 10^{-3}$ \\
 $^{64}$Ni & $2.45 \times 10^{-14}$ & $1.59 \times 10^{-13}$ & $1.70 \times 10^{-12}$ & $3.67 \times 10^{-10}$ & $4.89 \times 10^{-14}$ & $8.6 \times 10^{-14}$ & $1.92 \times 10^{-13}$ \\
 $^{63}$Cu & $4.90 \times 10^{-7}$ & $2.81 \times 10^{-6}$ & $1.59 \times 10^{-6}$ & $2.24 \times 10^{-6}$ & $4.2 \times 10^{-6}$ & $5.78 \times 10^{-6}$ & $1.2 \times 10^{-5}$ \\
 $^{65}$Cu & $2.73 \times 10^{-6}$ & $3.9 \times 10^{-6}$ & $3.29 \times 10^{-6}$ & $3.97 \times 10^{-6}$ & $4.88 \times 10^{-6}$ & $4.74 \times 10^{-6}$ & $5.51 \times 10^{-6}$ \\
 $^{64}$Zn & $2.48 \times 10^{-4}$ & $3.21 \times 10^{-4}$ & $4.52 \times 10^{-5}$ & $2.90 \times 10^{-5}$ & $1.99 \times 10^{-5}$ & $1.46 \times 10^{-5}$ & $1.7 \times 10^{-5}$ \\
 $^{66}$Zn & $3.46 \times 10^{-6}$ & $7.91 \times 10^{-6}$ & $2.25 \times 10^{-5}$ & $4.18 \times 10^{-5}$ & $7.37 \times 10^{-5}$ & $9.23 \times 10^{-5}$ & $1.31 \times 10^{-4}$ \\
 $^{67}$Zn & $3.26 \times 10^{-8}$ & $4.36 \times 10^{-8}$ & $1.58 \times 10^{-8}$ & $5.19 \times 10^{-8}$ & $1.56 \times 10^{-7}$ & $2.85 \times 10^{-7}$ & $6.10 \times 10^{-7}$ \\
 $^{68}$Zn & $3.6 \times 10^{-6}$ & $1.92 \times 10^{-6}$ & $1.59 \times 10^{-7}$ & $6.90 \times 10^{-8}$ & $3.57 \times 10^{-8}$ & $4.32 \times 10^{-8}$ & $9.41 \times 10^{-8}$ \\
 $^{70}$Zn & $3.51 \times 10^{-23}$ & $5.98 \times 10^{-18}$ & $9.98 \times 10^{-16}$ & $2.94 \times 10^{-15}$ & $2.66 \times 10^{-20}$ & $5.80 \times 10^{-22}$ & $7.72 \times 10^{-24}$ \\ \hline

\end{tabular}
\end{center}
\end{table*}

\begin{table*}

\begin{center}
\caption{($cont'd$) Nucleosynthesis yield for the 
Models presented in this articles. All models
in this table is based on the series with 
a $\rho_c = 3 \times 10^9$ g cm$^{-3}$, $c3$ flame
and C/O ratio $ = 1$. The 
isotope masses are in units of solar mass.
($Remark$: The table is replaced due to typos while converting the 
raw data into the current table form. Changes are made for the 
isotopes including $^{22}$Ne, $^{26}$Mg, $^{26}$Al, $^{36}$S, $^{40}$K, 
$^{41}$K, $^{44}$Ca, $^{53}$Cr, $^{55}$Mn, $^{60}$Fe, $^{59}$Co and $^{63}$Cu.)}
\label{table:Isotopes2}
\begin{tabular}{|c|c c c c c c c|}

\hline
Isotopes & $Z = 0$ & $Z = 0.1 Z_{\odot}$ & $Z = 0.5 Z_{\odot}$ & $Z = Z_{\odot}$ & 
		   $Z = 2 Z_{\odot}$ & $Z = 3 Z_{\odot}$ & $Z = 5 Z_{\odot}$ \\ \hline

 $^{12}$C & $5.93 \times 10^{-4}$ & $5.89 \times 10^{-4}$ & $1.8 \times 10^{-3}$ & $1.7 \times 10^{-3}$ & $1.5 \times 10^{-3}$ & $1.4 \times 10^{-3}$ & $1.3 \times 10^{-3}$ \\
 $^{13}$C & $1.11 \times 10^{-11}$ & $2.56 \times 10^{-12}$ & $1.91 \times 10^{-12}$ & $2.54 \times 10^{-12}$ & $4.48 \times 10^{-12}$ & $2.65 \times 10^{-11}$ & $1.37 \times 10^{-10}$ \\
 $^{14}$N & $5.20 \times 10^{-9}$ & $5.60 \times 10^{-10}$ & $1.70 \times 10^{-10}$ & $1.40 \times 10^{-10}$ & $1.40 \times 10^{-10}$ & $4.32 \times 10^{-10}$ & $7.3 \times 10^{-10}$ \\
 $^{15}$N & $9.22 \times 10^{-7}$ & $6.56 \times 10^{-9}$ & $2.53 \times 10^{-10}$ & $9.40 \times 10^{-11}$ & $3.39 \times 10^{-11}$ & $1.97 \times 10^{-11}$ & $1.95 \times 10^{-11}$ \\
 $^{16}$O & $4.22 \times 10^{-2}$ & $4.64 \times 10^{-2}$ & $5.62 \times 10^{-2}$ & $5.69 \times 10^{-2}$ & $5.73 \times 10^{-2}$ & $6.62 \times 10^{-2}$ & $7.26 \times 10^{-2}$ \\
 $^{17}$O & $3.29 \times 10^{-11}$ & $6.61 \times 10^{-12}$ & $4.68 \times 10^{-12}$ & $1.9 \times 10^{-11}$ & $2.71 \times 10^{-11}$ & $2.61 \times 10^{-10}$ & $5.38 \times 10^{-10}$ \\
 $^{18}$O & $4.94 \times 10^{-13}$ & $1.66 \times 10^{-13}$ & $1.17 \times 10^{-13}$ & $2.29 \times 10^{-13}$ & $4.27 \times 10^{-13}$ & $3.92 \times 10^{-12}$ & $1.73 \times 10^{-11}$ \\
 $^{19}$F & $1.4 \times 10^{-11}$ & $2.17 \times 10^{-12}$ & $4.51 \times 10^{-14}$ & $1.38 \times 10^{-13}$ & $3.68 \times 10^{-13}$ & $3.48 \times 10^{-12}$ & $6.69 \times 10^{-12}$ \\
 $^{20}$Ne & $6.32 \times 10^{-4}$ & $6.36 \times 10^{-4}$ & $1.40 \times 10^{-4}$ & $1.38 \times 10^{-4}$ & $1.33 \times 10^{-4}$ & $6.78 \times 10^{-4}$ & $4.18 \times 10^{-4}$ \\
 $^{21}$Ne & $6.95 \times 10^{-10}$ & $1.0 \times 10^{-9}$ & $1.18 \times 10^{-9}$ & $3.6 \times 10^{-9}$ & $8.35 \times 10^{-9}$ & $9.93 \times 10^{-8}$ & $2.14 \times 10^{-7}$ \\
 $^{22}$Ne & $5.91 \times 10^{-9}$ & $2.14 \times 10^{-6}$ & $2.14 \times 10^{-5}$ & $4.28 \times 10^{-5}$ & $8.56 \times 10^{-5}$ & $1.28 \times 10^{-4}$ & $2.14 \times 10^{-4}$ \\
 $^{23}$Na & $1.23 \times 10^{-6}$ & $1.17 \times 10^{-6}$ & $5.88 \times 10^{-7}$ & $8.9 \times 10^{-7}$ & $1.19 \times 10^{-6}$ & $3.76 \times 10^{-6}$ & $6.51 \times 10^{-6}$ \\
 $^{24}$Mg & $2.62 \times 10^{-3}$ & $2.30 \times 10^{-3}$ & $1.56 \times 10^{-3}$ & $1.10 \times 10^{-3}$ & $7.39 \times 10^{-4}$ & $1.0 \times 10^{-3}$ & $9.25 \times 10^{-4}$ \\
 $^{25}$Mg & $3.4 \times 10^{-8}$ & $4.53 \times 10^{-7}$ & $1.52 \times 10^{-6}$ & $2.36 \times 10^{-6}$ & $3.86 \times 10^{-6}$ & $2.46 \times 10^{-5}$ & $3.13 \times 10^{-5}$ \\
 $^{26}$Mg & $7.12 \times 10^{-7}$ & $1.6 \times 10^{-6}$ & $1.44 \times 10^{-6}$ & $2.56 \times 10^{-6}$ & $4.61 \times 10^{-6}$ & $2.22 \times 10^{-5}$ & $5.49 \times 10^{-5}$ \\
 $^{26}$Al & $6.23 \times 10^{-28}$ & $6.22 \times 10^{-28}$ & $4.55 \times 10^{-28}$ & $3.56 \times 10^{-29}$ & $2.20 \times 10^{-11}$ & $6.62 \times 10^{-11}$ & $2.7 \times 10^{-11}$ \\
 $^{27}$Al & $1.29 \times 10^{-5}$ & $2.88 \times 10^{-5}$ & $7.66 \times 10^{-5}$ & $9.14 \times 10^{-5}$ & $9.85 \times 10^{-5}$ & $1.62 \times 10^{-4}$ & $2.16 \times 10^{-4}$ \\
 $^{28}$Si & $2.4 \times 10^{-1}$ & $2.13 \times 10^{-1}$ & $2.30 \times 10^{-1}$ & $2.35 \times 10^{-1}$ & $2.39 \times 10^{-1}$ & $2.31 \times 10^{-1}$ & $2.47 \times 10^{-1}$ \\
 $^{29}$Si & $3.47 \times 10^{-5}$ & $9.92 \times 10^{-5}$ & $1.87 \times 10^{-4}$ & $2.58 \times 10^{-4}$ & $4.35 \times 10^{-4}$ & $8.48 \times 10^{-4}$ & $1.78 \times 10^{-3}$ \\
 $^{30}$Si & $2.96 \times 10^{-5}$ & $1.53 \times 10^{-5}$ & $1.51 \times 10^{-4}$ & $3.51 \times 10^{-4}$ & $8.58 \times 10^{-4}$ & $1.98 \times 10^{-3}$ & $5.75 \times 10^{-3}$ \\
 $^{31}$P & $6.0 \times 10^{-5}$ & $3.26 \times 10^{-5}$ & $1.16 \times 10^{-4}$ & $1.92 \times 10^{-4}$ & $3.44 \times 10^{-4}$ & $5.28 \times 10^{-4}$ & $9.62 \times 10^{-4}$ \\
 $^{32}$S & $1.32 \times 10^{-1}$ & $1.29 \times 10^{-1}$ & $1.28 \times 10^{-1}$ & $1.23 \times 10^{-1}$ & $1.11 \times 10^{-1}$ & $9.59 \times 10^{-2}$ & $8.98 \times 10^{-2}$ \\
 $^{33}$S & $1.84 \times 10^{-5}$ & $8.14 \times 10^{-5}$ & $2.2 \times 10^{-4}$ & $2.85 \times 10^{-4}$ & $4.28 \times 10^{-4}$ & $5.31 \times 10^{-4}$ & $7.18 \times 10^{-4}$ \\
 $^{34}$S & $8.43 \times 10^{-6}$ & $9.28 \times 10^{-5}$ & $9.41 \times 10^{-4}$ & $2.9 \times 10^{-3}$ & $4.76 \times 10^{-3}$ & $8.15 \times 10^{-3}$ & $1.71 \times 10^{-2}$ \\
 $^{36}$S & $1.68 \times 10^{-12}$ & $6.44 \times 10^{-10}$ & $8.15 \times 10^{-9}$ & $3.35 \times 10^{-8}$ & $2.60 \times 10^{-7}$ & $1.49 \times 10^{-6}$ & $7.61 \times 10^{-6}$ \\
 $^{35}$Cl & $1.35 \times 10^{-5}$ & $2.11 \times 10^{-5}$ & $9.26 \times 10^{-5}$ & $1.53 \times 10^{-4}$ & $2.49 \times 10^{-4}$ & $2.82 \times 10^{-4}$ & $3.43 \times 10^{-4}$ \\
 $^{37}$Cl & $2.82 \times 10^{-6}$ & $1.37 \times 10^{-5}$ & $3.53 \times 10^{-5}$ & $5.7 \times 10^{-5}$ & $7.13 \times 10^{-5}$ & $7.22 \times 10^{-5}$ & $8.6 \times 10^{-5}$ \\
 $^{36}$Ar & $2.96 \times 10^{-2}$ & $2.71 \times 10^{-2}$ & $2.46 \times 10^{-2}$ & $2.22 \times 10^{-2}$ & $1.84 \times 10^{-2}$ & $1.51 \times 10^{-2}$ & $1.29 \times 10^{-2}$ \\
 $^{38}$Ar & $1.9 \times 10^{-6}$ & $8.8 \times 10^{-5}$ & $8.17 \times 10^{-4}$ & $1.82 \times 10^{-3}$ & $3.94 \times 10^{-3}$ & $5.47 \times 10^{-3}$ & $9.66 \times 10^{-3}$ \\
 $^{40}$Ar & $4.61 \times 10^{-13}$ & $4.75 \times 10^{-12}$ & $1.23 \times 10^{-10}$ & $8.26 \times 10^{-10}$ & $8.46 \times 10^{-9}$ & $3.65 \times 10^{-8}$ & $1.76 \times 10^{-7}$ \\
 $^{39}$K & $4.57 \times 10^{-6}$ & $2.58 \times 10^{-5}$ & $1.11 \times 10^{-4}$ & $1.76 \times 10^{-4}$ & $2.55 \times 10^{-4}$ & $2.32 \times 10^{-4}$ & $2.48 \times 10^{-4}$ \\
 $^{40}$K & $6.80 \times 10^{-13}$ & $1.87 \times 10^{-9}$ & $1.57 \times 10^{-8}$ & $3.83 \times 10^{-8}$ & $8.84 \times 10^{-8}$ & $1.5 \times 10^{-7}$ & $1.20 \times 10^{-7}$ \\
 $^{41}$K & $7.74 \times 10^{-7}$ & $3.69 \times 10^{-6}$ & $9.47 \times 10^{-6}$ & $1.34 \times 10^{-5}$ & $1.71 \times 10^{-5}$ & $1.49 \times 10^{-5}$ & $1.38 \times 10^{-5}$ \\
 $^{40}$Ca & $2.79 \times 10^{-2}$ & $2.43 \times 10^{-2}$ & $2.5 \times 10^{-2}$ & $1.79 \times 10^{-2}$ & $1.44 \times 10^{-2}$ & $1.19 \times 10^{-2}$ & $1.1 \times 10^{-2}$ \\
 $^{42}$Ca & $1.67 \times 10^{-8}$ & $2.31 \times 10^{-6}$ & $2.84 \times 10^{-5}$ & $6.55 \times 10^{-5}$ & $1.35 \times 10^{-4}$ & $1.72 \times 10^{-4}$ & $2.51 \times 10^{-4}$ \\
 $^{43}$Ca & $3.37 \times 10^{-7}$ & $4.0 \times 10^{-7}$ & $1.37 \times 10^{-6}$ & $1.7 \times 10^{-6}$ & $7.99 \times 10^{-7}$ & $8.0 \times 10^{-7}$ & $8.66 \times 10^{-7}$ \\
 $^{44}$Ca & $3.98 \times 10^{-5}$ & $3.61 \times 10^{-5}$ & $2.95 \times 10^{-5}$ & $2.64 \times 10^{-5}$ & $2.23 \times 10^{-5}$ & $2.2 \times 10^{-5}$ & $1.71 \times 10^{-5}$ \\
 $^{46}$Ca & $6.23 \times 10^{-12}$ & $6.28 \times 10^{-12}$ & $7.38 \times 10^{-12}$ & $2.70 \times 10^{-11}$ & $3.15 \times 10^{-10}$ & $9.55 \times 10^{-10}$ & $1.62 \times 10^{-9}$ \\
 $^{48}$Ca & $2.68 \times 10^{-14}$ & $2.71 \times 10^{-14}$ & $3.7 \times 10^{-14}$ & $3.28 \times 10^{-14}$ & $5.34 \times 10^{-14}$ & $1.98 \times 10^{-13}$ & $1.49 \times 10^{-12}$ \\
 $^{45}$Sc & $9.65 \times 10^{-8}$ & $2.99 \times 10^{-7}$ & $4.87 \times 10^{-7}$ & $6.5 \times 10^{-7}$ & $7.9 \times 10^{-7}$ & $5.79 \times 10^{-7}$ & $6.59 \times 10^{-7}$ \\
 $^{46}$Ti & $2.20 \times 10^{-6}$ & $2.50 \times 10^{-6}$ & $1.54 \times 10^{-5}$ & $3.34 \times 10^{-5}$ & $6.66 \times 10^{-5}$ & $7.61 \times 10^{-5}$ & $1.1 \times 10^{-4}$ \\
 $^{47}$Ti & $1.36 \times 10^{-6}$ & $1.63 \times 10^{-6}$ & $3.29 \times 10^{-6}$ & $3.84 \times 10^{-6}$ & $4.37 \times 10^{-6}$ & $5.17 \times 10^{-6}$ & $6.71 \times 10^{-6}$ \\
 $^{48}$Ti & $5.90 \times 10^{-4}$ & $5.18 \times 10^{-4}$ & $3.96 \times 10^{-4}$ & $3.41 \times 10^{-4}$ & $2.78 \times 10^{-4}$ & $2.49 \times 10^{-4}$ & $2.2 \times 10^{-4}$ \\
 $^{49}$Ti & $6.13 \times 10^{-6}$ & $2.8 \times 10^{-5}$ & $2.48 \times 10^{-5}$ & $2.82 \times 10^{-5}$ & $2.95 \times 10^{-5}$ & $3.0 \times 10^{-5}$ & $3.55 \times 10^{-5}$ \\
 $^{50}$Ti & $2.43 \times 10^{-6}$ & $2.44 \times 10^{-6}$ & $2.57 \times 10^{-6}$ & $2.66 \times 10^{-6}$ & $2.85 \times 10^{-6}$ & $3.22 \times 10^{-6}$ & $3.89 \times 10^{-6}$ \\
 $^{50}$V & $1.25 \times 10^{-8}$ & $1.26 \times 10^{-8}$ & $1.41 \times 10^{-8}$ & $1.71 \times 10^{-8}$ & $2.64 \times 10^{-8}$ & $4.96 \times 10^{-8}$ & $1.50 \times 10^{-7}$ \\
 $^{51}$V & $4.72 \times 10^{-5}$ & $4.98 \times 10^{-5}$ & $7.68 \times 10^{-5}$ & $9.50 \times 10^{-5}$ & $1.17 \times 10^{-4}$ & $1.56 \times 10^{-4}$ & $2.22 \times 10^{-4}$ \\

\end{tabular}
\end{center}
\end{table*}

\begin{table*}

\begin{center}
\caption{($cont'd$) of Table \ref{table:Isotopes2}.
}
\label{table:Isotopes2b}
\begin{tabular}{|c|c c c c c c c|}

\hline
Isotopes & $Z = 0$ & $Z = 0.1 Z_{\odot}$ & $Z = 0.5 Z_{\odot}$ & $Z = Z_{\odot}$ & 
		   $Z = 2 Z_{\odot}$ & $Z = 3 Z_{\odot}$ & $Z = 5 Z_{\odot}$ \\ \hline

 $^{50}$Cr & $1.58 \times 10^{-4}$ & $1.74 \times 10^{-4}$ & $3.1 \times 10^{-4}$ & $4.96 \times 10^{-4}$ & $9.25 \times 10^{-4}$ & $1.21 \times 10^{-3}$ & $1.64 \times 10^{-3}$ \\
 $^{52}$Cr & $1.3 \times 10^{-2}$ & $9.73 \times 10^{-3}$ & $8.60 \times 10^{-3}$ & $8.4 \times 10^{-3}$ & $7.64 \times 10^{-3}$ & $8.10 \times 10^{-3}$ & $9.94 \times 10^{-3}$ \\
 $^{53}$Cr & $6.62 \times 10^{-4}$ & $8.17 \times 10^{-4}$ & $9.19 \times 10^{-4}$ & $1.0 \times 10^{-3}$ & $1.15 \times 10^{-3}$ & $1.37 \times 10^{-3}$ & $1.75 \times 10^{-3}$ \\
 $^{54}$Cr & $6.81 \times 10^{-5}$ & $6.84 \times 10^{-5}$ & $7.15 \times 10^{-5}$ & $7.34 \times 10^{-5}$ & $7.80 \times 10^{-5}$ & $8.63 \times 10^{-5}$ & $1.4 \times 10^{-4}$ \\
 $^{55}$Mn & $7.87 \times 10^{-3}$ & $8.36 \times 10^{-3}$ & $9.50 \times 10^{-3}$ & $1.3 \times 10^{-2}$ & $1.20 \times 10^{-2}$ & $1.38 \times 10^{-2}$ & $1.70 \times 10^{-2}$ \\
 $^{54}$Fe & $8.48 \times 10^{-2}$ & $8.66 \times 10^{-2}$ & $9.55 \times 10^{-2}$ & $1.6 \times 10^{-1}$ & $1.26 \times 10^{-1}$ & $1.46 \times 10^{-1}$ & $1.86 \times 10^{-1}$ \\
 $^{56}$Fe & $7.40 \times 10^{-1}$ & $7.32 \times 10^{-1}$ & $6.94 \times 10^{-1}$ & $6.71 \times 10^{-1}$ & $6.32 \times 10^{-1}$ & $5.98 \times 10^{-1}$ & $4.92 \times 10^{-1}$ \\
 $^{57}$Fe & $1.58 \times 10^{-2}$ & $1.66 \times 10^{-2}$ & $1.85 \times 10^{-2}$ & $2.8 \times 10^{-2}$ & $2.42 \times 10^{-2}$ & $2.66 \times 10^{-2}$ & $2.71 \times 10^{-2}$ \\
 $^{58}$Fe & $4.26 \times 10^{-4}$ & $4.28 \times 10^{-4}$ & $4.43 \times 10^{-4}$ & $4.54 \times 10^{-4}$ & $4.78 \times 10^{-4}$ & $5.18 \times 10^{-4}$ & $5.87 \times 10^{-4}$ \\
 $^{60}$Fe & $9.26 \times 10^{-11}$ & $9.33 \times 10^{-11}$ & $1.7 \times 10^{-10}$ & $1.9 \times 10^{-10}$ & $1.24 \times 10^{-10}$ & $1.43 \times 10^{-10}$ & $1.84 \times 10^{-10}$ \\
 $^{59}$Co & $5.44 \times 10^{-4}$ & $5.30 \times 10^{-4}$ & $7.54 \times 10^{-4}$ & $8.86 \times 10^{-4}$ & $1.1 \times 10^{-3}$ & $1.8 \times 10^{-3}$ & $1.3 \times 10^{-3}$ \\
 $^{58}$Ni & $4.29 \times 10^{-2}$ & $4.36 \times 10^{-2}$ & $5.16 \times 10^{-2}$ & $6.35 \times 10^{-2}$ & $8.69 \times 10^{-2}$ & $1.8 \times 10^{-1}$ & $1.38 \times 10^{-1}$ \\
 $^{60}$Ni & $1.31 \times 10^{-2}$ & $1.33 \times 10^{-2}$ & $1.21 \times 10^{-2}$ & $1.12 \times 10^{-2}$ & $1.1 \times 10^{-2}$ & $9.58 \times 10^{-3}$ & $8.12 \times 10^{-3}$ \\
 $^{61}$Ni & $2.16 \times 10^{-4}$ & $2.32 \times 10^{-4}$ & $2.43 \times 10^{-4}$ & $2.66 \times 10^{-4}$ & $2.89 \times 10^{-4}$ & $3.17 \times 10^{-4}$ & $2.70 \times 10^{-4}$ \\
 $^{62}$Ni & $3.55 \times 10^{-4}$ & $4.11 \times 10^{-4}$ & $1.15 \times 10^{-3}$ & $1.88 \times 10^{-3}$ & $3.9 \times 10^{-3}$ & $4.31 \times 10^{-3}$ & $4.99 \times 10^{-3}$ \\
 $^{64}$Ni & $1.5 \times 10^{-7}$ & $1.6 \times 10^{-7}$ & $1.17 \times 10^{-7}$ & $1.21 \times 10^{-7}$ & $1.30 \times 10^{-7}$ & $1.46 \times 10^{-7}$ & $1.79 \times 10^{-7}$ \\
 $^{63}$Cu & $5.98 \times 10^{-7}$ & $1.98 \times 10^{-6}$ & $1.35 \times 10^{-6}$ & $1.77 \times 10^{-6}$ & $2.96 \times 10^{-6}$ & $4.66 \times 10^{-6}$ & $7.16 \times 10^{-6}$ \\
 $^{65}$Cu & $1.68 \times 10^{-6}$ & $1.89 \times 10^{-6}$ & $1.98 \times 10^{-6}$ & $2.36 \times 10^{-6}$ & $2.86 \times 10^{-6}$ & $3.48 \times 10^{-6}$ & $3.34 \times 10^{-6}$ \\
 $^{64}$Zn & $1.64 \times 10^{-4}$ & $2.10 \times 10^{-4}$ & $2.76 \times 10^{-5}$ & $1.81 \times 10^{-5}$ & $1.30 \times 10^{-5}$ & $1.16 \times 10^{-5}$ & $7.24 \times 10^{-6}$ \\
 $^{66}$Zn & $2.30 \times 10^{-6}$ & $5.8 \times 10^{-6}$ & $1.50 \times 10^{-5}$ & $2.72 \times 10^{-5}$ & $4.79 \times 10^{-5}$ & $7.22 \times 10^{-5}$ & $8.72 \times 10^{-5}$ \\
 $^{67}$Zn & $2.40 \times 10^{-8}$ & $3.21 \times 10^{-8}$ & $1.3 \times 10^{-8}$ & $3.42 \times 10^{-8}$ & $9.53 \times 10^{-8}$ & $1.84 \times 10^{-7}$ & $3.17 \times 10^{-7}$ \\
 $^{68}$Zn & $1.99 \times 10^{-6}$ & $1.24 \times 10^{-6}$ & $3.96 \times 10^{-8}$ & $2.52 \times 10^{-8}$ & $2.33 \times 10^{-8}$ & $3.34 \times 10^{-8}$ & $6.37 \times 10^{-8}$ \\
 $^{70}$Zn & $2.11 \times 10^{-15}$ & $2.13 \times 10^{-15}$ & $2.55 \times 10^{-15}$ & $2.68 \times 10^{-15}$ & $2.96 \times 10^{-15}$ & $3.51 \times 10^{-15}$ & $4.63 \times 10^{-15}$ \\ \hline

\end{tabular}
\end{center}
\end{table*}


\begin{table*}

\begin{center}
\caption{($cont'd$) Nucleosynthesis yield for the 
Models presented in this articles. All models
in this table is based on the series with 
a $\rho_c = 5 \times 10^9$ g cm$^{-3}$, $c3$ flame
and C/O ratio $ = 1$. The 
isotope masses are in units of solar mass.
($Remark$: The table is replaced due to typos while converting the 
raw data into the current table form. Changes are made for the 
isotopes including $^{22}$Ne, $^{26}$Mg, $^{26}$Al, $^{36}$S, $^{40}$K, 
$^{41}$K, $^{44}$Ca, $^{53}$Cr, $^{55}$Mn, $^{60}$Fe, $^{59}$Co and $^{63}$Cu.)}
\label{table:Isotopes3}
\begin{tabular}{|c|c c c c c c c|}

\hline
Isotopes & $Z = 0$ & $Z = 0.1 Z_{\odot}$ & $Z = 0.5 Z_{\odot}$ & $Z = Z_{\odot}$ & 
		   $Z = 2 Z_{\odot}$ & $Z = 3 Z_{\odot}$ & $Z = 5 Z_{\odot}$ \\ \hline
		  
$^{12}$C & $5.48 \times 10^{-4}$ & $5.44 \times 10^{-4}$ & $5.88 \times 10^{-4}$ & $5.82 \times 10^{-4}$ & $5.71 \times 10^{-4}$ & $5.47 \times 10^{-4}$ & $5.36 \times 10^{-4}$ \\
 $^{13}$C & $1.3 \times 10^{-11}$ & $1.54 \times 10^{-12}$ & $3.24 \times 10^{-12}$ & $6.45 \times 10^{-12}$ & $1.62 \times 10^{-11}$ & $5.54 \times 10^{-11}$ & $8.29 \times 10^{-11}$ \\
 $^{14}$N & $2.15 \times 10^{-9}$ & $4.34 \times 10^{-10}$ & $2.94 \times 10^{-10}$ & $3.70 \times 10^{-10}$ & $5.57 \times 10^{-10}$ & $9.60 \times 10^{-10}$ & $8.28 \times 10^{-10}$ \\
 $^{15}$N & $7.59 \times 10^{-7}$ & $5.4 \times 10^{-9}$ & $4.87 \times 10^{-10}$ & $1.32 \times 10^{-10}$ & $4.38 \times 10^{-11}$ & $2.96 \times 10^{-11}$ & $1.60 \times 10^{-11}$ \\
 $^{16}$O & $3.46 \times 10^{-2}$ & $3.81 \times 10^{-2}$ & $4.85 \times 10^{-2}$ & $4.90 \times 10^{-2}$ & $4.94 \times 10^{-2}$ & $6.23 \times 10^{-2}$ & $7.54 \times 10^{-2}$ \\
 $^{17}$O & $6.65 \times 10^{-12}$ & $1.19 \times 10^{-12}$ & $1.39 \times 10^{-11}$ & $3.90 \times 10^{-11}$ & $1.15 \times 10^{-10}$ & $6.34 \times 10^{-10}$ & $4.72 \times 10^{-10}$ \\
 $^{18}$O & $1.87 \times 10^{-13}$ & $5.79 \times 10^{-14}$ & $9.92 \times 10^{-13}$ & $2.21 \times 10^{-12}$ & $4.48 \times 10^{-12}$ & $1.48 \times 10^{-11}$ & $1.14 \times 10^{-11}$ \\
 $^{19}$F & $4.2 \times 10^{-12}$ & $6.98 \times 10^{-13}$ & $2.60 \times 10^{-13}$ & $8.34 \times 10^{-13}$ & $2.66 \times 10^{-12}$ & $1.67 \times 10^{-11}$ & $6.27 \times 10^{-12}$ \\
 $^{20}$Ne & $1.63 \times 10^{-4}$ & $1.63 \times 10^{-4}$ & $6.58 \times 10^{-4}$ & $6.56 \times 10^{-4}$ & $6.46 \times 10^{-4}$ & $6.82 \times 10^{-4}$ & $2.81 \times 10^{-4}$ \\
 $^{21}$Ne & $2.36 \times 10^{-10}$ & $2.46 \times 10^{-10}$ & $5.19 \times 10^{-9}$ & $1.65 \times 10^{-8}$ & $5.64 \times 10^{-8}$ & $2.30 \times 10^{-7}$ & $1.43 \times 10^{-7}$ \\
 $^{22}$Ne & $2.69 \times 10^{-9}$ & $2.15 \times 10^{-6}$ & $1.7 \times 10^{-5}$ & $2.15 \times 10^{-5}$ & $4.31 \times 10^{-5}$ & $6.46 \times 10^{-5}$ & $1.7 \times 10^{-4}$ \\
 $^{23}$Na & $8.44 \times 10^{-8}$ & $9.50 \times 10^{-8}$ & $1.36 \times 10^{-6}$ & $2.0 \times 10^{-6}$ & $3.15 \times 10^{-6}$ & $4.45 \times 10^{-6}$ & $4.26 \times 10^{-6}$ \\
 $^{24}$Mg & $2.5 \times 10^{-3}$ & $1.84 \times 10^{-3}$ & $1.69 \times 10^{-3}$ & $1.22 \times 10^{-3}$ & $8.41 \times 10^{-4}$ & $7.57 \times 10^{-4}$ & $8.39 \times 10^{-4}$ \\
 $^{25}$Mg & $3.86 \times 10^{-9}$ & $4.2 \times 10^{-7}$ & $1.66 \times 10^{-6}$ & $4.65 \times 10^{-6}$ & $1.26 \times 10^{-5}$ & $2.35 \times 10^{-5}$ & $2.12 \times 10^{-5}$ \\
 $^{26}$Mg & $7.89 \times 10^{-8}$ & $1.67 \times 10^{-7}$ & $3.40 \times 10^{-6}$ & $6.59 \times 10^{-6}$ & $1.26 \times 10^{-5}$ & $2.89 \times 10^{-5}$ & $3.66 \times 10^{-5}$ \\
 $^{26}$Al & $4.78 \times 10^{-28}$ & $5.7 \times 10^{-28}$ & $3.58 \times 10^{-29}$ & $1.9 \times 10^{-10}$ & $7.69 \times 10^{-11}$ & $3.91 \times 10^{-11}$ & $1.49 \times 10^{-11}$ \\
 $^{27}$Al & $6.94 \times 10^{-6}$ & $2.6 \times 10^{-5}$ & $7.58 \times 10^{-5}$ & $9.18 \times 10^{-5}$ & $9.95 \times 10^{-5}$ & $1.18 \times 10^{-4}$ & $1.88 \times 10^{-4}$ \\
 $^{28}$Si & $1.72 \times 10^{-1}$ & $1.79 \times 10^{-1}$ & $2.3 \times 10^{-1}$ & $2.8 \times 10^{-1}$ & $2.13 \times 10^{-1}$ & $2.21 \times 10^{-1}$ & $2.25 \times 10^{-1}$ \\
 $^{29}$Si & $3.42 \times 10^{-5}$ & $8.43 \times 10^{-5}$ & $1.67 \times 10^{-4}$ & $2.43 \times 10^{-4}$ & $4.36 \times 10^{-4}$ & $7.29 \times 10^{-4}$ & $1.63 \times 10^{-3}$ \\
 $^{30}$Si & $3.70 \times 10^{-5}$ & $1.11 \times 10^{-5}$ & $1.37 \times 10^{-4}$ & $3.13 \times 10^{-4}$ & $7.52 \times 10^{-4}$ & $1.69 \times 10^{-3}$ & $5.64 \times 10^{-3}$ \\
 $^{31}$P & $5.32 \times 10^{-5}$ & $2.68 \times 10^{-5}$ & $9.94 \times 10^{-5}$ & $1.65 \times 10^{-4}$ & $3.1 \times 10^{-4}$ & $5.31 \times 10^{-4}$ & $9.58 \times 10^{-4}$ \\
 $^{32}$S & $1.14 \times 10^{-1}$ & $1.12 \times 10^{-1}$ & $1.14 \times 10^{-1}$ & $1.9 \times 10^{-1}$ & $9.81 \times 10^{-2}$ & $9.27 \times 10^{-2}$ & $8.0 \times 10^{-2}$ \\
 $^{33}$S & $1.56 \times 10^{-5}$ & $6.91 \times 10^{-5}$ & $1.74 \times 10^{-4}$ & $2.49 \times 10^{-4}$ & $3.80 \times 10^{-4}$ & $5.46 \times 10^{-4}$ & $7.15 \times 10^{-4}$ \\
 $^{34}$S & $1.11 \times 10^{-5}$ & $7.60 \times 10^{-5}$ & $8.14 \times 10^{-4}$ & $1.82 \times 10^{-3}$ & $4.20 \times 10^{-3}$ & $8.66 \times 10^{-3}$ & $1.72 \times 10^{-2}$ \\
 $^{36}$S & $1.81 \times 10^{-10}$ & $5.22 \times 10^{-10}$ & $7.34 \times 10^{-9}$ & $2.91 \times 10^{-8}$ & $2.55 \times 10^{-7}$ & $8.20 \times 10^{-7}$ & $6.67 \times 10^{-6}$ \\
 $^{35}$Cl & $1.20 \times 10^{-5}$ & $1.83 \times 10^{-5}$ & $8.17 \times 10^{-5}$ & $1.36 \times 10^{-4}$ & $2.23 \times 10^{-4}$ & $3.17 \times 10^{-4}$ & $3.48 \times 10^{-4}$ \\
 $^{37}$Cl & $2.43 \times 10^{-6}$ & $1.21 \times 10^{-5}$ & $3.25 \times 10^{-5}$ & $4.64 \times 10^{-5}$ & $6.50 \times 10^{-5}$ & $8.21 \times 10^{-5}$ & $8.17 \times 10^{-5}$ \\
 $^{36}$Ar & $2.64 \times 10^{-2}$ & $2.42 \times 10^{-2}$ & $2.20 \times 10^{-2}$ & $1.98 \times 10^{-2}$ & $1.62 \times 10^{-2}$ & $1.44 \times 10^{-2}$ & $1.14 \times 10^{-2}$ \\
 $^{38}$Ar & $1.25 \times 10^{-6}$ & $6.94 \times 10^{-5}$ & $7.36 \times 10^{-4}$ & $1.63 \times 10^{-3}$ & $3.49 \times 10^{-3}$ & $6.25 \times 10^{-3}$ & $9.30 \times 10^{-3}$ \\
 $^{40}$Ar & $2.28 \times 10^{-11}$ & $1.31 \times 10^{-11}$ & $1.19 \times 10^{-10}$ & $6.34 \times 10^{-10}$ & $6.16 \times 10^{-9}$ & $2.10 \times 10^{-8}$ & $1.56 \times 10^{-7}$ \\
 $^{39}$K & $3.83 \times 10^{-6}$ & $2.32 \times 10^{-5}$ & $1.4 \times 10^{-4}$ & $1.63 \times 10^{-4}$ & $2.33 \times 10^{-4}$ & $2.89 \times 10^{-4}$ & $2.44 \times 10^{-4}$ \\
 $^{40}$K & $3.34 \times 10^{-12}$ & $1.69 \times 10^{-9}$ & $1.28 \times 10^{-8}$ & $3.0 \times 10^{-8}$ & $6.48 \times 10^{-8}$ & $1.6 \times 10^{-7}$ & $1.40 \times 10^{-7}$ \\
 $^{41}$K & $7.34 \times 10^{-7}$ & $3.37 \times 10^{-6}$ & $8.93 \times 10^{-6}$ & $1.24 \times 10^{-5}$ & $1.55 \times 10^{-5}$ & $1.75 \times 10^{-5}$ & $1.36 \times 10^{-5}$ \\
 $^{40}$Ca & $2.55 \times 10^{-2}$ & $2.24 \times 10^{-2}$ & $1.84 \times 10^{-2}$ & $1.59 \times 10^{-2}$ & $1.26 \times 10^{-2}$ & $1.11 \times 10^{-2}$ & $8.93 \times 10^{-3}$ \\
 $^{42}$Ca & $2.7 \times 10^{-8}$ & $2.6 \times 10^{-6}$ & $2.57 \times 10^{-5}$ & $5.85 \times 10^{-5}$ & $1.19 \times 10^{-4}$ & $1.96 \times 10^{-4}$ & $2.42 \times 10^{-4}$ \\
 $^{43}$Ca & $3.56 \times 10^{-7}$ & $4.11 \times 10^{-7}$ & $1.35 \times 10^{-6}$ & $1.5 \times 10^{-6}$ & $7.69 \times 10^{-7}$ & $7.88 \times 10^{-7}$ & $9.1 \times 10^{-7}$ \\
 $^{44}$Ca & $3.66 \times 10^{-5}$ & $3.29 \times 10^{-5}$ & $2.76 \times 10^{-5}$ & $2.45 \times 10^{-5}$ & $2.4 \times 10^{-5}$ & $1.76 \times 10^{-5}$ & $1.30 \times 10^{-5}$ \\
 $^{46}$Ca & $3.37 \times 10^{-9}$ & $4.83 \times 10^{-10}$ & $3.40 \times 10^{-9}$ & $3.43 \times 10^{-9}$ & $3.65 \times 10^{-9}$ & $4.7 \times 10^{-9}$ & $5.5 \times 10^{-9}$ \\
 $^{48}$Ca & $4.89 \times 10^{-10}$ & $1.19 \times 10^{-11}$ & $4.99 \times 10^{-10}$ & $5.3 \times 10^{-10}$ & $5.11 \times 10^{-10}$ & $5.28 \times 10^{-10}$ & $5.60 \times 10^{-10}$ \\
 $^{45}$Sc & $8.52 \times 10^{-8}$ & $2.69 \times 10^{-7}$ & $4.97 \times 10^{-7}$ & $5.71 \times 10^{-7}$ & $5.97 \times 10^{-7}$ & $6.34 \times 10^{-7}$ & $6.48 \times 10^{-7}$ \\
 $^{46}$Ti & $2.10 \times 10^{-6}$ & $2.23 \times 10^{-6}$ & $1.50 \times 10^{-5}$ & $3.19 \times 10^{-5}$ & $6.29 \times 10^{-5}$ & $9.2 \times 10^{-5}$ & $9.17 \times 10^{-5}$ \\
 $^{47}$Ti & $1.12 \times 10^{-6}$ & $1.37 \times 10^{-6}$ & $3.12 \times 10^{-6}$ & $3.66 \times 10^{-6}$ & $4.15 \times 10^{-6}$ & $5.4 \times 10^{-6}$ & $5.90 \times 10^{-6}$ \\
 $^{48}$Ti & $5.60 \times 10^{-4}$ & $4.93 \times 10^{-4}$ & $3.56 \times 10^{-4}$ & $3.5 \times 10^{-4}$ & $2.50 \times 10^{-4}$ & $2.15 \times 10^{-4}$ & $1.81 \times 10^{-4}$ \\
 $^{49}$Ti & $6.70 \times 10^{-6}$ & $2.0 \times 10^{-5}$ & $2.49 \times 10^{-5}$ & $2.71 \times 10^{-5}$ & $2.70 \times 10^{-5}$ & $2.79 \times 10^{-5}$ & $3.28 \times 10^{-5}$ \\
 $^{50}$Ti & $5.19 \times 10^{-4}$ & $1.23 \times 10^{-4}$ & $5.23 \times 10^{-4}$ & $5.25 \times 10^{-4}$ & $5.28 \times 10^{-4}$ & $5.34 \times 10^{-4}$ & $5.46 \times 10^{-4}$ \\
 $^{50}$V & $6.14 \times 10^{-8}$ & $6.28 \times 10^{-8}$ & $6.35 \times 10^{-8}$ & $6.60 \times 10^{-8}$ & $7.51 \times 10^{-8}$ & $9.97 \times 10^{-8}$ & $1.98 \times 10^{-7}$ \\
 $^{51}$V & $2.39 \times 10^{-4}$ & $1.52 \times 10^{-4}$ & $2.68 \times 10^{-4}$ & $2.86 \times 10^{-4}$ & $3.6 \times 10^{-4}$ & $3.38 \times 10^{-4}$ & $4.4 \times 10^{-4}$ \\

\end{tabular}
\end{center}
\end{table*}  
  
\begin{table*}

\begin{center}
\caption{($cont'd$) of Table \ref{table:Isotopes3}.
}
\label{table:Isotopes3b}
\begin{tabular}{|c|c c c c c c c|}

\hline
Isotopes & $Z = 0$ & $Z = 0.1 Z_{\odot}$ & $Z = 0.5 Z_{\odot}$ & $Z = Z_{\odot}$ & 
		   $Z = 2 Z_{\odot}$ & $Z = 3 Z_{\odot}$ & $Z = 5 Z_{\odot}$ \\ \hline  
  
  $^{50}$Cr & $2.0 \times 10^{-4}$ & $2.14 \times 10^{-4}$ & $3.35 \times 10^{-4}$ & $5.22 \times 10^{-4}$ & $9.24 \times 10^{-4}$ & $1.13 \times 10^{-3}$ & $1.56 \times 10^{-3}$ \\
 $^{52}$Cr & $1.79 \times 10^{-2}$ & $1.81 \times 10^{-2}$ & $1.57 \times 10^{-2}$ & $1.52 \times 10^{-2}$ & $1.49 \times 10^{-2}$ & $1.52 \times 10^{-2}$ & $1.71 \times 10^{-2}$ \\
 $^{53}$Cr & $1.29 \times 10^{-3}$ & $1.31 \times 10^{-3}$ & $1.53 \times 10^{-3}$ & $1.60 \times 10^{-3}$ & $1.75 \times 10^{-3}$ & $1.94 \times 10^{-3}$ & $2.33 \times 10^{-3}$ \\
 $^{54}$Cr & $4.80 \times 10^{-3}$ & $1.83 \times 10^{-3}$ & $4.82 \times 10^{-3}$ & $4.84 \times 10^{-3}$ & $4.87 \times 10^{-3}$ & $4.92 \times 10^{-3}$ & $5.1 \times 10^{-3}$ \\
 $^{55}$Mn & $1.11 \times 10^{-2}$ & $1.12 \times 10^{-2}$ & $1.26 \times 10^{-2}$ & $1.34 \times 10^{-2}$ & $1.49 \times 10^{-2}$ & $1.66 \times 10^{-2}$ & $1.98 \times 10^{-2}$ \\
 $^{54}$Fe & $9.99 \times 10^{-2}$ & $1.1 \times 10^{-1}$ & $1.10 \times 10^{-1}$ & $1.20 \times 10^{-1}$ & $1.38 \times 10^{-1}$ & $1.55 \times 10^{-1}$ & $1.94 \times 10^{-1}$ \\
 $^{56}$Fe & $7.57 \times 10^{-1}$ & $7.60 \times 10^{-1}$ & $7.2 \times 10^{-1}$ & $6.81 \times 10^{-1}$ & $6.44 \times 10^{-1}$ & $5.87 \times 10^{-1}$ & $4.99 \times 10^{-1}$ \\
 $^{57}$Fe & $1.71 \times 10^{-2}$ & $1.79 \times 10^{-2}$ & $1.96 \times 10^{-2}$ & $2.17 \times 10^{-2}$ & $2.50 \times 10^{-2}$ & $2.61 \times 10^{-2}$ & $2.68 \times 10^{-2}$ \\
 $^{58}$Fe & $1.43 \times 10^{-2}$ & $8.19 \times 10^{-3}$ & $1.43 \times 10^{-2}$ & $1.43 \times 10^{-2}$ & $1.44 \times 10^{-2}$ & $1.45 \times 10^{-2}$ & $1.47 \times 10^{-2}$ \\
 $^{60}$Fe & $5.45 \times 10^{-8}$ & $8.39 \times 10^{-9}$ & $5.51 \times 10^{-8}$ & $5.67 \times 10^{-8}$ & $5.72 \times 10^{-8}$ & $5.82 \times 10^{-8}$ & $6.9 \times 10^{-8}$ \\
 $^{59}$Co & $9.25 \times 10^{-4}$ & $9.26 \times 10^{-4}$ & $1.12 \times 10^{-3}$ & $1.26 \times 10^{-3}$ & $1.38 \times 10^{-3}$ & $1.36 \times 10^{-3}$ & $1.37 \times 10^{-3}$ \\
 $^{58}$Ni & $4.82 \times 10^{-2}$ & $4.90 \times 10^{-2}$ & $5.64 \times 10^{-2}$ & $6.76 \times 10^{-2}$ & $8.97 \times 10^{-2}$ & $1.6 \times 10^{-1}$ & $1.33 \times 10^{-1}$ \\
 $^{60}$Ni & $1.47 \times 10^{-2}$ & $1.53 \times 10^{-2}$ & $1.39 \times 10^{-2}$ & $1.32 \times 10^{-2}$ & $1.21 \times 10^{-2}$ & $1.9 \times 10^{-2}$ & $1.0 \times 10^{-2}$ \\
 $^{61}$Ni & $2.40 \times 10^{-4}$ & $2.64 \times 10^{-4}$ & $2.79 \times 10^{-4}$ & $3.2 \times 10^{-4}$ & $3.24 \times 10^{-4}$ & $2.98 \times 10^{-4}$ & $2.69 \times 10^{-4}$ \\
 $^{62}$Ni & $4.22 \times 10^{-3}$ & $3.50 \times 10^{-3}$ & $5.0 \times 10^{-3}$ & $5.72 \times 10^{-3}$ & $6.91 \times 10^{-3}$ & $7.35 \times 10^{-3}$ & $8.24 \times 10^{-3}$ \\
 $^{64}$Ni & $2.50 \times 10^{-5}$ & $5.75 \times 10^{-6}$ & $2.51 \times 10^{-5}$ & $2.52 \times 10^{-5}$ & $2.54 \times 10^{-5}$ & $2.60 \times 10^{-5}$ & $2.67 \times 10^{-5}$ \\
 $^{63}$Cu & $3.21 \times 10^{-6}$ & $3.29 \times 10^{-6}$ & $4.8 \times 10^{-6}$ & $4.48 \times 10^{-6}$ & $5.64 \times 10^{-6}$ & $6.54 \times 10^{-6}$ & $8.79 \times 10^{-6}$ \\
 $^{65}$Cu & $1.82 \times 10^{-6}$ & $1.86 \times 10^{-6}$ & $2.37 \times 10^{-6}$ & $2.75 \times 10^{-6}$ & $3.28 \times 10^{-6}$ & $3.33 \times 10^{-6}$ & $3.3 \times 10^{-6}$ \\
 $^{64}$Zn & $1.51 \times 10^{-4}$ & $1.94 \times 10^{-4}$ & $2.89 \times 10^{-5}$ & $1.89 \times 10^{-5}$ & $1.37 \times 10^{-5}$ & $9.66 \times 10^{-6}$ & $6.70 \times 10^{-6}$ \\
 $^{66}$Zn & $2.3 \times 10^{-6}$ & $4.61 \times 10^{-6}$ & $1.56 \times 10^{-5}$ & $2.85 \times 10^{-5}$ & $5.7 \times 10^{-5}$ & $5.99 \times 10^{-5}$ & $7.90 \times 10^{-5}$ \\
 $^{67}$Zn & $2.22 \times 10^{-8}$ & $3.1 \times 10^{-8}$ & $1.17 \times 10^{-8}$ & $3.74 \times 10^{-8}$ & $1.4 \times 10^{-7}$ & $1.66 \times 10^{-7}$ & $2.59 \times 10^{-7}$ \\
 $^{68}$Zn & $1.85 \times 10^{-6}$ & $1.13 \times 10^{-6}$ & $6.7 \times 10^{-8}$ & $4.35 \times 10^{-8}$ & $4.11 \times 10^{-8}$ & $4.39 \times 10^{-8}$ & $7.45 \times 10^{-8}$ \\
 $^{70}$Zn & $3.76 \times 10^{-12}$ & $3.13 \times 10^{-13}$ & $3.81 \times 10^{-12}$ & $3.83 \times 10^{-12}$ & $3.87 \times 10^{-12}$ & $4.2 \times 10^{-12}$ & $4.25 \times 10^{-12}$ \\ \hline

\end{tabular}		 
\end{center} 
\end{table*}


\begin{table*}

\begin{center}
\caption{($cont'd$) Nucleosynthesis yield for the 
W7 and WDD2 models \citep{Nomoto1984} computed by
our updated nuclear reaction network. The 
isotope masses are in units of solar mass.
(Remark: The table is replaced due to inconsistency among previous 
publications and online tables. The yield is recomputed to restore 
the consistency among files.)}
\label{table:Isotopes4}
\begin{tabular}{|c | c c c c|}

\hline
Isotopes & W7 $Z = 0.1 Z_{{\odot}}$ & W7 $Z = 0.5 Z_{{\odot}}$ & W7 $Z = Z_{{\odot}}$ & 
           WDD2 $Z = Z_{{\odot}}$ \\ \hline

 $^{12}$C & $5.44 \times 10^{-2}$ & $5.32 \times 10^{-2}$ & $5.2 \times 10^{-2}$ & $1.0 \times 10^{-2}$ \\
 $^{13}$C & $1.37 \times 10^{-12}$ & $7.92 \times 10^{-12}$ & $1.81 \times 10^{-11}$ & $2.8 \times 10^{-7}$ \\
 $^{14}$N & $8.11 \times 10^{-10}$ & $7.19 \times 10^{-10}$ & $9.56 \times 10^{-10}$ & $2.0 \times 10^{-7}$ \\
 $^{15}$N & $1.41 \times 10^{-8}$ & $4.3 \times 10^{-10}$ & $1.35 \times 10^{-10}$ & $1.27 \times 10^{-8}$ \\
 $^{16}$O & $1.35 \times 10^{-1}$ & $1.65 \times 10^{-1}$ & $1.85 \times 10^{-1}$ & $9.94 \times 10^{-2}$ \\
 $^{17}$O & $1.27 \times 10^{-11}$ & $7.82 \times 10^{-11}$ & $1.75 \times 10^{-10}$ & $6.88 \times 10^{-8}$ \\
 $^{18}$O & $7.45 \times 10^{-13}$ & $3.82 \times 10^{-12}$ & $7.4 \times 10^{-12}$ & $3.46 \times 10^{-9}$ \\
 $^{19}$F & $3.89 \times 10^{-12}$ & $1.78 \times 10^{-12}$ & $4.54 \times 10^{-12}$ & $4.22 \times 10^{-10}$ \\
 $^{20}$Ne & $1.64 \times 10^{-3}$ & $1.65 \times 10^{-3}$ & $1.62 \times 10^{-3}$ & $1.54 \times 10^{-2}$ \\
 $^{21}$Ne & $3.16 \times 10^{-9}$ & $2.86 \times 10^{-8}$ & $6.97 \times 10^{-8}$ & $2.41 \times 10^{-6}$ \\
 $^{22}$Ne & $2.73 \times 10^{-4}$ & $1.53 \times 10^{-3}$ & $2.73 \times 10^{-3}$ & $1.38 \times 10^{-5}$ \\
 $^{23}$Na & $2.56 \times 10^{-6}$ & $4.69 \times 10^{-6}$ & $6.61 \times 10^{-6}$ & $1.47 \times 10^{-4}$ \\
 $^{24}$Mg & $9.74 \times 10^{-3}$ & $5.78 \times 10^{-3}$ & $4.26 \times 10^{-3}$ & $1.3 \times 10^{-2}$ \\
 $^{25}$Mg & $2.78 \times 10^{-6}$ & $1.3 \times 10^{-5}$ & $1.81 \times 10^{-5}$ & $2.98 \times 10^{-4}$ \\
 $^{26}$Mg & $2.96 \times 10^{-6}$ & $1.42 \times 10^{-5}$ & $2.35 \times 10^{-5}$ & $5.2 \times 10^{-4}$ \\
 $^{26}$Al & $5.4 \times 10^{-28}$ & $1.68 \times 10^{-12}$ & $2.60 \times 10^{-10}$ & $6.9 \times 10^{-9}$ \\
 $^{27}$Al & $1.75 \times 10^{-4}$ & $4.2 \times 10^{-4}$ & $4.50 \times 10^{-4}$ & $1.2 \times 10^{-3}$ \\
 $^{28}$Si & $1.43 \times 10^{-1}$ & $1.52 \times 10^{-1}$ & $1.55 \times 10^{-1}$ & $2.29 \times 10^{-1}$ \\
 $^{29}$Si & $1.91 \times 10^{-4}$ & $4.84 \times 10^{-4}$ & $7.73 \times 10^{-4}$ & $1.31 \times 10^{-3}$ \\
 $^{30}$Si & $7.10 \times 10^{-5}$ & $7.98 \times 10^{-4}$ & $1.64 \times 10^{-3}$ & $1.32 \times 10^{-3}$ \\
 $^{31}$P & $5.76 \times 10^{-5}$ & $2.48 \times 10^{-4}$ & $3.94 \times 10^{-4}$ & $3.4 \times 10^{-4}$ \\
 $^{32}$S & $8.29 \times 10^{-2}$ & $8.17 \times 10^{-2}$ & $7.80 \times 10^{-2}$ & $1.30 \times 10^{-1}$ \\
 $^{33}$S & $1.16 \times 10^{-4}$ & $2.89 \times 10^{-4}$ & $3.77 \times 10^{-4}$ & $2.38 \times 10^{-4}$ \\
 $^{34}$S & $1.17 \times 10^{-4}$ & $1.8 \times 10^{-3}$ & $2.20 \times 10^{-3}$ & $2.46 \times 10^{-3}$ \\
 $^{36}$S & $2.25 \times 10^{-9}$ & $6.11 \times 10^{-8}$ & $2.94 \times 10^{-7}$ & $1.93 \times 10^{-7}$ \\
 $^{35}$Cl & $1.73 \times 10^{-5}$ & $7.63 \times 10^{-5}$ & $1.25 \times 10^{-4}$ & $1.2 \times 10^{-4}$ \\
 $^{37}$Cl & $8.36 \times 10^{-6}$ & $1.71 \times 10^{-5}$ & $2.28 \times 10^{-5}$ & $2.53 \times 10^{-5}$ \\
 $^{36}$Ar & $1.77 \times 10^{-2}$ & $1.52 \times 10^{-2}$ & $1.33 \times 10^{-2}$ & $2.50 \times 10^{-2}$ \\
 $^{38}$Ar & $5.15 \times 10^{-5}$ & $4.36 \times 10^{-4}$ & $8.75 \times 10^{-4}$ & $1.15 \times 10^{-3}$ \\
 $^{40}$Ar & $1.55 \times 10^{-11}$ & $1.4 \times 10^{-9}$ & $7.39 \times 10^{-9}$ & $3.18 \times 10^{-9}$ \\
 $^{39}$K & $1.18 \times 10^{-5}$ & $4.41 \times 10^{-5}$ & $6.57 \times 10^{-5}$ & $6.59 \times 10^{-5}$ \\
 $^{40}$K & $2.0 \times 10^{-9}$ & $2.91 \times 10^{-8}$ & $8.27 \times 10^{-8}$ & $3.17 \times 10^{-8}$ \\
 $^{41}$K & $1.45 \times 10^{-6}$ & $3.9 \times 10^{-6}$ & $4.8 \times 10^{-6}$ & $5.7 \times 10^{-6}$ \\
 $^{40}$Ca & $1.74 \times 10^{-2}$ & $1.36 \times 10^{-2}$ & $1.15 \times 10^{-2}$ & $2.47 \times 10^{-2}$ \\
 $^{42}$Ca & $1.49 \times 10^{-6}$ & $1.43 \times 10^{-5}$ & $2.66 \times 10^{-5}$ & $2.92 \times 10^{-5}$ \\
 $^{43}$Ca & $4.70 \times 10^{-8}$ & $9.21 \times 10^{-8}$ & $1.49 \times 10^{-7}$ & $1.65 \times 10^{-7}$ \\
 $^{44}$Ca & $1.42 \times 10^{-5}$ & $1.8 \times 10^{-5}$ & $9.17 \times 10^{-6}$ & $2.36 \times 10^{-5}$ \\
 $^{46}$Ca & $4.35 \times 10^{-11}$ & $6.45 \times 10^{-11}$ & $3.26 \times 10^{-10}$ & $1.40 \times 10^{-9}$ \\
 $^{48}$Ca & $2.9 \times 10^{-12}$ & $2.14 \times 10^{-12}$ & $2.19 \times 10^{-12}$ & $1.36 \times 10^{-9}$ \\
 $^{45}$Sc & $9.74 \times 10^{-8}$ & $1.97 \times 10^{-7}$ & $2.81 \times 10^{-7}$ & $2.22 \times 10^{-7}$ \\
 $^{46}$Ti & $8.69 \times 10^{-7}$ & $6.90 \times 10^{-6}$ & $1.27 \times 10^{-5}$ & $1.26 \times 10^{-5}$ \\
 $^{47}$Ti & $1.39 \times 10^{-7}$ & $3.72 \times 10^{-7}$ & $5.74 \times 10^{-7}$ & $1.21 \times 10^{-6}$ \\
 $^{48}$Ti & $3.66 \times 10^{-4}$ & $2.88 \times 10^{-4}$ & $2.51 \times 10^{-4}$ & $5.99 \times 10^{-4}$ \\
 $^{49}$Ti & $1.53 \times 10^{-5}$ & $2.5 \times 10^{-5}$ & $2.23 \times 10^{-5}$ & $4.29 \times 10^{-5}$ \\
 $^{50}$Ti & $9.33 \times 10^{-6}$ & $9.47 \times 10^{-6}$ & $9.60 \times 10^{-6}$ & $2.22 \times 10^{-4}$ \\
 $^{50}$V & $6.94 \times 10^{-9}$ & $1.16 \times 10^{-8}$ & $1.84 \times 10^{-8}$ & $1.15 \times 10^{-8}$ \\
 $^{51}$V & $5.84 \times 10^{-5}$ & $8.29 \times 10^{-5}$ & $9.47 \times 10^{-5}$ & $1.41 \times 10^{-4}$ \\

\end{tabular}
\end{center}
\end{table*}

\begin{table*}

\begin{center}
\caption{($cont'd$) of Table \ref{table:Isotopes4}.
}
\label{table:Isotopes4b}
\begin{tabular}{|c | c c c c|}

\hline
Isotopes & W7 $Z = 0.1 Z_{{\odot}}$ & W7 $Z = 0.5 Z_{{\odot}}$ & W7 $Z = Z_{{\odot}}$ & 
           WDD2 $Z = Z_{{\odot}}$ \\ \hline  
  
 $^{50}$Cr & $1.54 \times 10^{-4}$ & $2.56 \times 10^{-4}$ & $3.74 \times 10^{-4}$ & $3.99 \times 10^{-4}$ \\
 $^{52}$Cr & $1.3 \times 10^{-2}$ & $9.19 \times 10^{-3}$ & $8.59 \times 10^{-3}$ & $1.54 \times 10^{-2}$ \\
 $^{53}$Cr & $8.73 \times 10^{-4}$ & $1.2 \times 10^{-3}$ & $1.11 \times 10^{-3}$ & $1.30 \times 10^{-3}$ \\
 $^{54}$Cr & $1.26 \times 10^{-4}$ & $1.28 \times 10^{-4}$ & $1.30 \times 10^{-4}$ & $1.77 \times 10^{-3}$ \\
 $^{55}$Mn & $1.3 \times 10^{-2}$ & $1.22 \times 10^{-2}$ & $1.36 \times 10^{-2}$ & $8.21 \times 10^{-3}$ \\
 $^{54}$Fe & $8.82 \times 10^{-2}$ & $1.2 \times 10^{-1}$ & $1.15 \times 10^{-1}$ & $6.98 \times 10^{-2}$ \\
 $^{56}$Fe & $7.24 \times 10^{-1}$ & $6.93 \times 10^{-1}$ & $6.68 \times 10^{-1}$ & $6.54 \times 10^{-1}$ \\
 $^{57}$Fe & $1.51 \times 10^{-2}$ & $1.77 \times 10^{-2}$ & $1.96 \times 10^{-2}$ & $1.34 \times 10^{-2}$ \\
 $^{58}$Fe & $5.32 \times 10^{-4}$ & $5.39 \times 10^{-4}$ & $5.46 \times 10^{-4}$ & $4.70 \times 10^{-3}$ \\
 $^{60}$Fe & $6.97 \times 10^{-10}$ & $7.18 \times 10^{-10}$ & $7.33 \times 10^{-10}$ & $4.10 \times 10^{-8}$ \\
 $^{59}$Co & $4.29 \times 10^{-4}$ & $4.96 \times 10^{-4}$ & $5.20 \times 10^{-4}$ & $3.92 \times 10^{-4}$ \\
 $^{58}$Ni & $4.71 \times 10^{-2}$ & $5.79 \times 10^{-2}$ & $6.80 \times 10^{-2}$ & $3.0 \times 10^{-2}$ \\
 $^{60}$Ni & $5.39 \times 10^{-3}$ & $4.85 \times 10^{-3}$ & $4.51 \times 10^{-3}$ & $6.82 \times 10^{-3}$ \\
 $^{61}$Ni & $5.77 \times 10^{-5}$ & $6.6 \times 10^{-5}$ & $5.81 \times 10^{-5}$ & $2.35 \times 10^{-4}$ \\
 $^{62}$Ni & $3.83 \times 10^{-4}$ & $5.90 \times 10^{-4}$ & $7.3 \times 10^{-4}$ & $3.5 \times 10^{-3}$ \\
 $^{64}$Ni & $4.6 \times 10^{-7}$ & $4.12 \times 10^{-7}$ & $4.17 \times 10^{-7}$ & $1.70 \times 10^{-5}$ \\
 $^{63}$Cu & $3.57 \times 10^{-7}$ & $4.42 \times 10^{-7}$ & $5.15 \times 10^{-7}$ & $1.69 \times 10^{-6}$ \\
 $^{65}$Cu & $1.96 \times 10^{-7}$ & $2.10 \times 10^{-7}$ & $1.96 \times 10^{-7}$ & $1.3 \times 10^{-6}$ \\
 $^{64}$Zn & $4.20 \times 10^{-6}$ & $1.95 \times 10^{-6}$ & $1.34 \times 10^{-6}$ & $1.96 \times 10^{-5}$ \\
 $^{66}$Zn & $1.14 \times 10^{-6}$ & $2.70 \times 10^{-6}$ & $3.42 \times 10^{-6}$ & $3.12 \times 10^{-5}$ \\
 $^{67}$Zn & $4.24 \times 10^{-10}$ & $1.48 \times 10^{-9}$ & $2.28 \times 10^{-9}$ & $1.90 \times 10^{-8}$ \\
 $^{68}$Zn & $2.95 \times 10^{-9}$ & $1.45 \times 10^{-9}$ & $1.29 \times 10^{-9}$ & $1.61 \times 10^{-8}$ \\
 $^{70}$Zn & $2.85 \times 10^{-14}$ & $2.89 \times 10^{-14}$ & $2.93 \times 10^{-14}$ & $1.29 \times 10^{-11}$ \\ \hline
 
\end{tabular}
\end{center}
\end{table*}


\begin{table*}

\begin{center}
\caption{Similar to Table \ref{table:Isotopes}, but for
the mass of major radioactive isotopes. The 
isotope masses are in units of solar mass.}
\label{table:Decay1}
\begin{tabular}{|c | c c c c c c c|}

\hline
Isotopes & $Z = 0$ & $Z = 0.1 Z_{\odot}$ & $Z = 0.5 Z_{\odot}$ & $Z = Z_{\odot}$ & 
		   $Z = 2 Z_{\odot}$ & $Z = 3 Z_{\odot}$ & $Z = 5 Z_{\odot}$ \\ \hline		   
		   
 $^{22}$Na & $1.75 \times 10^{-10}$ & $6.29 \times 10^{-10}$ & $1.48 \times 10^{
 -9}$ & $1.69 \times 10^{-9}$ & $1.46 \times 10^{-9}$ & $3.27 \times 10^{-10}$ &
  $5.64 \times 10^{-10}$ \\
 $^{26}$Al & $7.87 \times 10^{-8}$ & $2.54 \times 10^{-7}$ & $1.6 \times 10^{-6}
 $ & $1.7 \times 10^{-6}$ & $7.33 \times 10^{-7}$ & $1.90 \times 10^{-7}$ & $2.5
 1 \times 10^{-7}$ \\
 $^{39}$Ar & $5.1 \times 10^{-15}$ & $8.25 \times 10^{-11}$ & $1.14 \times 10^{-
 9}$ & $3.86 \times 10^{-9}$ & $1.39 \times 10^{-8}$ & $2.0 \times 10^{-8}$ & $6
 .82 \times 10^{-8}$ \\
 $^{40}$K & $3.65 \times 10^{-13}$ & $1.25 \times 10^{-9}$ & $1.0 \times 10^{-8}
 $ & $2.61 \times 10^{-8}$ & $6.35 \times 10^{-8}$ & $6.66 \times 10^{-8}$ & $1.
 5 \times 10^{-7}$ \\
 $^{41}$Ca & $5.20 \times 10^{-7}$ & $2.22 \times 10^{-6}$ & $6.42 \times 10^{-6
 }$ & $8.85 \times 10^{-6}$ & $1.11 \times 10^{-5}$ & $9.58 \times 10^{-6}$ & $1
 .6 \times 10^{-5}$ \\
 $^{44}$Ti & $4.25 \times 10^{-5}$ & $4.10 \times 10^{-5}$ & $3.42 \times 10^{-5
 }$ & $3.10 \times 10^{-5}$ & $2.64 \times 10^{-5}$ & $2.32 \times 10^{-5}$ & $1
 .93 \times 10^{-5}$ \\
 $^{48}$V & $9.66 \times 10^{-10}$ & $9.28 \times 10^{-9}$ & $3.51 \times 10^{-8
 }$ & $7.5 \times 10^{-8}$ & $1.46 \times 10^{-7}$ & $1.71 \times 10^{-7}$ & $2.
 39 \times 10^{-7}$ \\
 $^{49}$V & $1.9 \times 10^{-9}$ & $4.83 \times 10^{-9}$ & $5.81 \times 10^{-8}$
  & $1.68 \times 10^{-7}$ & $5.80 \times 10^{-7}$ & $1.22 \times 10^{-6}$ & $3.5
 6 \times 10^{-6}$ \\
 $^{53}$Mn & $1.27 \times 10^{-5}$ & $1.29 \times 10^{-5}$ & $1.79 \times 10^{-5
 }$ & $3.12 \times 10^{-5}$ & $1.15 \times 10^{-4}$ & $2.51 \times 10^{-4}$ & $5
 .98 \times 10^{-4}$ \\
 $^{60}$Fe & $4.32 \times 10^{-20}$ & $7.76 \times 10^{-20}$ & $3.3 \times 10^{-
 18}$ & $3.44 \times 10^{-17}$ & $3.0 \times 10^{-17}$ & $9.36 \times 10^{-17}$ 
 & $8.16 \times 10^{-16}$ \\
 $^{56}$Co & $3.5 \times 10^{-5}$ & $3.17 \times 10^{-5}$ & $3.68 \times 10^{-5}
 $ & $4.33 \times 10^{-5}$ & $5.55 \times 10^{-5}$ & $7.9 \times 10^{-5}$ & $1.2
  \times 10^{-4}$ \\
 $^{57}$Co & $1.6 \times 10^{-4}$ & $1.7 \times 10^{-4}$ & $1.13 \times 10^{-4}$
  & $1.22 \times 10^{-4}$ & $1.54 \times 10^{-4}$ & $1.85 \times 10^{-4}$ & $2.4
 7 \times 10^{-4}$ \\
 $^{60}$Co & $2.37 \times 10^{-13}$ & $2.48 \times 10^{-13}$ & $3.33 \times 10^{
 -13}$ & $8.53 \times 10^{-13}$ & $1.74 \times 10^{-12}$ & $5.13 \times 10^{-12}
 $ & $2.57 \times 10^{-11}$ \\
 $^{56}$Ni & $8.45 \times 10^{-1}$ & $8.38 \times 10^{-1}$ & $7.50 \times 10^{-1
 }$ & $7.24 \times 10^{-1}$ & $6.78 \times 10^{-1}$ & $6.38 \times 10^{-1}$ & $4
 .96 \times 10^{-1}$ \\
 $^{57}$Ni & $1.60 \times 10^{-2}$ & $1.71 \times 10^{-2}$ & $1.87 \times 10^{-2
 }$ & $2.15 \times 10^{-2}$ & $2.58 \times 10^{-2}$ & $2.89 \times 10^{-2}$ & $2
 .89 \times 10^{-2}$ \\
 $^{59}$Ni & $5.24 \times 10^{-5}$ & $5.28 \times 10^{-5}$ & $5.56 \times 10^{-5
 }$ & $5.94 \times 10^{-5}$ & $7.4 \times 10^{-5}$ & $8.39 \times 10^{-5}$ & $1.
 12 \times 10^{-4}$ \\
 $^{63}$Ni & $4.5 \times 10^{-15}$ & $6.96 \times 10^{-15}$ & $1.40 \times 10^{-
 14}$ & $1.19 \times 10^{-11}$ & $1.77 \times 10^{-14}$ & $4.78 \times 10^{-14}$
  & $3.19 \times 10^{-13}$ \\ \hline

\end{tabular}
\end{center}
\end{table*}


\begin{table*}

\begin{center}
\caption{$(cont'd)$ Similar to Table \ref{table:Isotopes2}, but for
the mass of major radioactive isotopes. The 
isotope masses are in units of solar mass.}
\label{table:Decay2}
\begin{tabular}{|c| c c c c c c c|}
\hline
Isotopes & $Z = 0$ & $Z = 0.1 Z_{\odot}$ & $Z = 0.5 Z_{\odot}$ & $Z = Z_{\odot}$ & 
		   $Z = 2 Z_{\odot}$ & $Z = 3 Z_{\odot}$ & $Z = 5 Z_{\odot}$ \\ \hline

  $^{22}$Na & $6.64 \times 10^{-10}$ & $1.6 \times 10^{-9}$ & $4.15 \times 10^{-1
 0}$ & $3.99 \times 10^{-10}$ & $3.12 \times 10^{-10}$ & $1.9 \times 10^{-9}$ & 
 $6.31 \times 10^{-10}$ \\
 $^{26}$Al & $3.47 \times 10^{-7}$ & $7.5 \times 10^{-7}$ & $3.83 \times 10^{-7}
 $ & $3.45 \times 10^{-7}$ & $2.24 \times 10^{-7}$ & $6.18 \times 10^{-7}$ & $2.
 27 \times 10^{-7}$ \\
 $^{39}$Ar & $1.19 \times 10^{-13}$ & $1.14 \times 10^{-10}$ & $1.65 \times 10^{
 -9}$ & $5.0 \times 10^{-9}$ & $1.56 \times 10^{-8}$ & $4.1 \times 10^{-8}$ & $5
 .51 \times 10^{-8}$ \\
 $^{40}$K & $6.84 \times 10^{-13}$ & $1.88 \times 10^{-9}$ & $1.58 \times 10^{-8
 }$ & $3.85 \times 10^{-8}$ & $8.89 \times 10^{-8}$ & $1.5 \times 10^{-7}$ & $1.
 21 \times 10^{-7}$ \\
 $^{41}$Ca & $7.81 \times 10^{-7}$ & $3.74 \times 10^{-6}$ & $9.58 \times 10^{-6
 }$ & $1.34 \times 10^{-5}$ & $1.70 \times 10^{-5}$ & $1.49 \times 10^{-5}$ & $1
 .37 \times 10^{-5}$ \\
 $^{44}$Ti & $3.73 \times 10^{-5}$ & $3.38 \times 10^{-5}$ & $2.76 \times 10^{-5
 }$ & $2.46 \times 10^{-5}$ & $2.6 \times 10^{-5}$ & $1.86 \times 10^{-5}$ & $1.
 54 \times 10^{-5}$ \\
 $^{48}$V & $2.1 \times 10^{-9}$ & $1.25 \times 10^{-8}$ & $5.15 \times 10^{-8}$
  & $1.10 \times 10^{-7}$ & $2.32 \times 10^{-7}$ & $2.76 \times 10^{-7}$ & $3.2
 2 \times 10^{-7}$ \\
 $^{49}$V & $7.52 \times 10^{-8}$ & $8.16 \times 10^{-8}$ & $1.58 \times 10^{-7}
 $ & $3.20 \times 10^{-7}$ & $9.25 \times 10^{-7}$ & $1.97 \times 10^{-6}$ & $4.
 46 \times 10^{-6}$ \\
 $^{53}$Mn & $3.52 \times 10^{-4}$ & $3.52 \times 10^{-4}$ & $3.63 \times 10^{-4
 }$ & $3.81 \times 10^{-4}$ & $4.91 \times 10^{-4}$ & $6.65 \times 10^{-4}$ & $9
 .91 \times 10^{-4}$ \\
 $^{60}$Fe & $1.33 \times 10^{-9}$ & $1.34 \times 10^{-9}$ & $1.54 \times 10^{-9
 }$ & $1.60 \times 10^{-9}$ & $1.75 \times 10^{-9}$ & $2.1 \times 10^{-9}$ & $2.
 58 \times 10^{-9}$ \\
 $^{56}$Co & $8.56 \times 10^{-5}$ & $8.70 \times 10^{-5}$ & $9.25 \times 10^{-5
 }$ & $9.96 \times 10^{-5}$ & $1.12 \times 10^{-4}$ & $1.24 \times 10^{-4}$ & $1
 .58 \times 10^{-4}$ \\
 $^{57}$Co & $1.17 \times 10^{-3}$ & $1.17 \times 10^{-3}$ & $1.17 \times 10^{-3
 }$ & $1.19 \times 10^{-3}$ & $1.24 \times 10^{-3}$ & $1.27 \times 10^{-3}$ & $1
 .38 \times 10^{-3}$ \\
 $^{60}$Co & $7.1 \times 10^{-8}$ & $7.4 \times 10^{-8}$ & $7.54 \times 10^{-8}$
  & $7.70 \times 10^{-8}$ & $8.5 \times 10^{-8}$ & $8.35 \times 10^{-8}$ & $9.48
  \times 10^{-8}$ \\
 $^{56}$Ni & $6.96 \times 10^{-1}$ & $6.89 \times 10^{-1}$ & $6.50 \times 10^{-1
 }$ & $6.27 \times 10^{-1}$ & $5.87 \times 10^{-1}$ & $5.51 \times 10^{-1}$ & $4
 .38 \times 10^{-1}$ \\
 $^{57}$Ni & $1.45 \times 10^{-2}$ & $1.53 \times 10^{-2}$ & $1.72 \times 10^{-2
 }$ & $1.95 \times 10^{-2}$ & $2.29 \times 10^{-2}$ & $2.52 \times 10^{-2}$ & $2
 .56 \times 10^{-2}$ \\
 $^{59}$Ni & $3.90 \times 10^{-4}$ & $3.91 \times 10^{-4}$ & $3.99 \times 10^{-4
 }$ & $4.5 \times 10^{-4}$ & $4.21 \times 10^{-4}$ & $4.36 \times 10^{-4}$ & $4.
 79 \times 10^{-4}$ \\
 $^{63}$Ni & $5.29 \times 10^{-8}$ & $5.32 \times 10^{-8}$ & $5.46 \times 10^{-8
 }$ & $5.61 \times 10^{-8}$ & $5.93 \times 10^{-8}$ & $6.27 \times 10^{-8}$ & $7
 .30 \times 10^{-8}$ \\ \hline

\end{tabular}
\end{center}
\end{table*}


\begin{table*}

\begin{center}
\caption{$(cont'd)$ Similar to Table \ref{table:Isotopes3}, but for
the mass of major radioactive isotopes. The 
isotope masses are in units of solar mass.}
\label{table:Decay3}
\begin{tabular}{|c| c c c c c c c|}
\hline
Isotopes & $Z = 0$ & $Z = 0.1 Z_{\odot}$ & $Z = 0.5 Z_{\odot}$ & $Z = Z_{\odot}$ & 
		   $Z = 2 Z_{\odot}$ & $Z = 3 Z_{\odot}$ & $Z = 5 Z_{\odot}$ \\ \hline		   
		   
 $^{22}$Na & $5.3 \times 10^{-11}$ & $1.58 \times 10^{-10}$ & $1.61 \times 10^{-
 9}$ & $1.93 \times 10^{-9}$ & $1.65 \times 10^{-9}$ & $1.48 \times 10^{-9}$ & $
 4.74 \times 10^{-10}$ \\
 $^{26}$Al & $9.41 \times 10^{-9}$ & $8.9 \times 10^{-8}$ & $9.97 \times 10^{-7}
 $ & $1.16 \times 10^{-6}$ & $7.92 \times 10^{-7}$ & $4.14 \times 10^{-7}$ & $1.
 60 \times 10^{-7}$ \\
 $^{39}$Ar & $1.2 \times 10^{-12}$ & $9.83 \times 10^{-11}$ & $1.35 \times 10^{-
 9}$ & $4.18 \times 10^{-9}$ & $1.45 \times 10^{-8}$ & $3.16 \times 10^{-8}$ & $
 6.61 \times 10^{-8}$ \\
 $^{40}$K & $3.36 \times 10^{-12}$ & $1.70 \times 10^{-9}$ & $1.29 \times 10^{-8
 }$ & $3.2 \times 10^{-8}$ & $6.52 \times 10^{-8}$ & $1.6 \times 10^{-7}$ & $1.4
 0 \times 10^{-7}$ \\
 $^{41}$Ca & $7.39 \times 10^{-7}$ & $3.41 \times 10^{-6}$ & $9.2 \times 10^{-6}
 $ & $1.23 \times 10^{-5}$ & $1.55 \times 10^{-5}$ & $1.75 \times 10^{-5}$ & $1.
 35 \times 10^{-5}$ \\
 $^{44}$Ti & $3.43 \times 10^{-5}$ & $3.8 \times 10^{-5}$ & $2.59 \times 10^{-5}
 $ & $2.28 \times 10^{-5}$ & $1.90 \times 10^{-5}$ & $1.61 \times 10^{-5}$ & $1.
 15 \times 10^{-5}$ \\
 $^{48}$V & $2.32 \times 10^{-9}$ & $1.24 \times 10^{-8}$ & $4.75 \times 10^{-8}
 $ & $1.1 \times 10^{-7}$ & $2.20 \times 10^{-7}$ & $3.20 \times 10^{-7}$ & $3.1
 2 \times 10^{-7}$ \\
 $^{49}$V & $1.28 \times 10^{-7}$ & $1.38 \times 10^{-7}$ & $2.3 \times 10^{-7}$
  & $3.47 \times 10^{-7}$ & $9.45 \times 10^{-7}$ & $2.6 \times 10^{-6}$ & $4.45
  \times 10^{-6}$ \\
 $^{53}$Mn & $5.17 \times 10^{-4}$ & $5.21 \times 10^{-4}$ & $5.29 \times 10^{-4
 }$ & $5.48 \times 10^{-4}$ & $6.56 \times 10^{-4}$ & $8.29 \times 10^{-4}$ & $1
 .14 \times 10^{-3}$ \\
 $^{60}$Fe & $7.82 \times 10^{-7}$ & $1.20 \times 10^{-7}$ & $7.94 \times 10^{-7
 }$ & $7.97 \times 10^{-7}$ & $8.4 \times 10^{-7}$ & $8.18 \times 10^{-7}$ & $8.
 56 \times 10^{-7}$ \\
 $^{56}$Co & $1.1 \times 10^{-4}$ & $1.2 \times 10^{-4}$ & $1.9 \times 10^{-4}$ 
 & $1.17 \times 10^{-4}$ & $1.30 \times 10^{-4}$ & $1.42 \times 10^{-4}$ & $1.73
  \times 10^{-4}$ \\
 $^{57}$Co & $1.56 \times 10^{-3}$ & $1.56 \times 10^{-3}$ & $1.58 \times 10^{-3
 }$ & $1.60 \times 10^{-3}$ & $1.66 \times 10^{-3}$ & $1.72 \times 10^{-3}$ & $1
 .86 \times 10^{-3}$ \\
 $^{60}$Co & $1.47 \times 10^{-6}$ & $1.4 \times 10^{-6}$ & $1.49 \times 10^{-6}
 $ & $1.49 \times 10^{-6}$ & $1.50 \times 10^{-6}$ & $1.50 \times 10^{-6}$ & $1.
 53 \times 10^{-6}$ \\
 $^{56}$Ni & $6.75 \times 10^{-1}$ & $6.69 \times 10^{-1}$ & $6.20 \times 10^{-1
 }$ & $5.98 \times 10^{-1}$ & $5.60 \times 10^{-1}$ & $5.0 \times 10^{-1}$ & $4.
 7 \times 10^{-1}$ \\
 $^{57}$Ni & $1.45 \times 10^{-2}$ & $1.53 \times 10^{-2}$ & $1.70 \times 10^{-2
 }$ & $1.91 \times 10^{-2}$ & $2.24 \times 10^{-2}$ & $2.33 \times 10^{-2}$ & $2
 .39 \times 10^{-2}$ \\
 $^{59}$Ni & $5.21 \times 10^{-4}$ & $5.23 \times 10^{-4}$ & $5.31 \times 10^{-4
 }$ & $5.39 \times 10^{-4}$ & $5.57 \times 10^{-4}$ & $5.75 \times 10^{-4}$ & $6
 .30 \times 10^{-4}$ \\
 $^{63}$Ni & $2.36 \times 10^{-6}$ & $1.18 \times 10^{-6}$ & $2.41 \times 10^{-6
 }$ & $2.41 \times 10^{-6}$ & $2.43 \times 10^{-6}$ & $2.42 \times 10^{-6}$ & $2
 .48 \times 10^{-6}$ \\ \hline

\end{tabular}
\end{center}		   
\end{table*}


\begin{table*}

\begin{center}
\caption{$(cont'd)$ Similar to Table \ref{table:Isotopes4}, but for
the mass of major radioactive isotopes. The 
isotope masses are in units of solar mass.}
\label{table:Decay4}
\begin{tabular}{|c| c c c c|}
\hline
Isotopes & W7 $Z = 0.1 Z_{{\odot}}$ & W7 $Z = 0.5 Z_{{\odot}}$ & W7 $Z = Z_{{\odot}}$ & 
           WDD2 $Z = Z_{{\odot}}$ \\ \hline		
		   
  $^{22}$Na & $2.76 \times 10^{-9}$ & $5.0 \times 10^{-9}$ & $4.60 \times 10^{-9}
 $ & $1.24 \times 10^{-8}$ \\
 $^{26}$Al & $1.61 \times 10^{-6}$ & $2.66 \times 10^{-6}$ & $2.9 \times 10^{-6}
 $ & $4.98 \times 10^{-6}$ \\
 $^{39}$Ar & $2.54 \times 10^{-10}$ & $4.4 \times 10^{-9}$ & $1.82 \times 10^{-8
 }$ & $7.13 \times 10^{-9}$ \\
 $^{40}$K & $2.34 \times 10^{-9}$ & $2.53 \times 10^{-8}$ & $8.83 \times 10^{-8}
 $ & $4.1 \times 10^{-8}$ \\
 $^{41}$Ca & $1.56 \times 10^{-6}$ & $2.95 \times 10^{-6}$ & $3.98 \times 10^{-6
 }$ & $5.0 \times 10^{-6}$ \\
 $^{44}$Ti & $9.67 \times 10^{-6}$ & $7.10 \times 10^{-6}$ & $5.53 \times 10^{-6
 }$ & $2.21 \times 10^{-5}$ \\
 $^{48}$V & $8.39 \times 10^{-9}$ & $2.43 \times 10^{-8}$ & $3.96 \times 10^{-8}
 $ & $5.83 \times 10^{-8}$ \\
 $^{49}$V & $5.56 \times 10^{-8}$ & $1.65 \times 10^{-7}$ & $2.75 \times 10^{-7}
 $ & $1.23 \times 10^{-7}$ \\
 $^{53}$Mn & $2.32 \times 10^{-4}$ & $2.35 \times 10^{-4}$ & $2.50 \times 10^{-4
 }$ & $1.44 \times 10^{-4}$ \\
 $^{60}$Fe & $1.10 \times 10^{-8}$ & $1.10 \times 10^{-8}$ & $1.10 \times 10^{-8
 }$ & $5.73 \times 10^{-7}$ \\
 $^{56}$Co & $1.2 \times 10^{-4}$ & $1.4 \times 10^{-4}$ & $1.6 \times 10^{-4}$ 
 & $4.72 \times 10^{-5}$ \\
 $^{57}$Co & $7.74 \times 10^{-4}$ & $7.76 \times 10^{-4}$ & $7.83 \times 10^{-4
 }$ & $3.48 \times 10^{-4}$ \\
 $^{60}$Co & $8.8 \times 10^{-8}$ & $8.8 \times 10^{-8}$ & $8.8 \times 10^{-8}$ 
 & $3.52 \times 10^{-7}$ \\
 $^{56}$Ni & $6.59 \times 10^{-1}$ & $6.51 \times 10^{-1}$ & $6.45 \times 10^{-1
 }$ & $6.32 \times 10^{-1}$ \\
 $^{57}$Ni & $1.77 \times 10^{-2}$ & $1.78 \times 10^{-2}$ & $1.78 \times 10^{-2
 }$ & $1.28 \times 10^{-2}$ \\
 $^{59}$Ni & $2.72 \times 10^{-4}$ & $2.73 \times 10^{-4}$ & $2.74 \times 10^{-4
 }$ & $1.1 \times 10^{-4}$ \\
 $^{63}$Ni & $8.98 \times 10^{-8}$ & $8.98 \times 10^{-8}$ & $8.98 \times 10^{-8
 }$ & $8.19 \times 10^{-7}$ \\ \hline
		   
\end{tabular}
\end{center}
\end{table*}


\begin{table*}

\begin{center}
\caption{Nucleosynthesis yield for the pure turbulent deflagration
models presented in this articles. The 
isotope masses are in units of solar mass.
($Remark$: The table is replaced due to typos while converting the 
raw data into the current table form. Changes are made for the 
isotopes including $^{22}$Ne, $^{26}$Mg, $^{26}$Al, $^{36}$S, $^{40}$K, 
$^{41}$K, $^{44}$Ca, $^{53}$Cr, $^{55}$Mn, $^{60}$Fe, $^{59}$Co and $^{63}$Cu.)}
\label{table:IsotopesPTD}
\begin{tabular}{|c | c c c c|}
\hline
Isotopes & 050-1-c3-1P & 100-1-c3-1P & 300-1-c3-1P & 500-1-c3-1P \\ \hline

$^{12}$C & $4.72 \times 10^{-1}$ & $4.34 \times 10^{-1}$ & $3.65 \times 10^{-1}$ & $3.21 \times 10^{-1}$ \\
 $^{13}$C & $3.38 \times 10^{-11}$ & $2.11 \times 10^{-11}$ & $1.34 \times 10^{-11}$ & $2.56 \times 10^{-12}$ \\
 $^{14}$N & $3.80 \times 10^{-9}$ & $1.67 \times 10^{-9}$ & $1.15 \times 10^{-9}$ & $2.68 \times 10^{-10}$ \\
 $^{15}$N & $7.15 \times 10^{-10}$ & $6.51 \times 10^{-10}$ & $5.5 \times 10^{-10}$ & $6.53 \times 10^{-11}$ \\
 $^{16}$O & $4.98 \times 10^{-1}$ & $4.63 \times 10^{-1}$ & $3.94 \times 10^{-1}$ & $3.47 \times 10^{-1}$ \\
 $^{17}$O & $1.51 \times 10^{-9}$ & $5.28 \times 10^{-10}$ & $3.40 \times 10^{-10}$ & $8.94 \times 10^{-11}$ \\
 $^{18}$O & $4.61 \times 10^{-11}$ & $1.43 \times 10^{-11}$ & $1.10 \times 10^{-11}$ & $1.36 \times 10^{-12}$ \\
 $^{19}$F & $9.3 \times 10^{-12}$ & $8.40 \times 10^{-12}$ & $7.24 \times 10^{-12}$ & $1.62 \times 10^{-12}$ \\
 $^{20}$Ne & $1.29 \times 10^{-3}$ & $2.35 \times 10^{-3}$ & $2.20 \times 10^{-3}$ & $8.47 \times 10^{-4}$ \\
 $^{21}$Ne & $4.6 \times 10^{-8}$ & $7.44 \times 10^{-8}$ & $6.70 \times 10^{-8}$ & $2.26 \times 10^{-8}$ \\
 $^{22}$Ne & $1.92 \times 10^{-2}$ & $1.77 \times 10^{-2}$ & $1.49 \times 10^{-2}$ & $1.31 \times 10^{-2}$ \\
 $^{23}$Na & $3.84 \times 10^{-6}$ & $6.4 \times 10^{-6}$ & $6.27 \times 10^{-6}$ & $1.66 \times 10^{-6}$ \\
 $^{24}$Mg & $2.36 \times 10^{-3}$ & $3.9 \times 10^{-3}$ & $3.22 \times 10^{-3}$ & $2.64 \times 10^{-3}$ \\
 $^{25}$Mg & $1.6 \times 10^{-5}$ & $1.48 \times 10^{-5}$ & $1.55 \times 10^{-5}$ & $7.40 \times 10^{-6}$ \\
 $^{26}$Mg & $1.39 \times 10^{-5}$ & $2.24 \times 10^{-5}$ & $2.21 \times 10^{-5}$ & $8.43 \times 10^{-6}$ \\
 $^{26}$Al & $3.37 \times 10^{-29}$ & $3.45 \times 10^{-29}$ & $3.56 \times 10^{-29}$ & $3.58 \times 10^{-29}$ \\
 $^{27}$Al & $1.72 \times 10^{-4}$ & $2.24 \times 10^{-4}$ & $2.36 \times 10^{-4}$ & $1.97 \times 10^{-4}$ \\
 $^{28}$Si & $3.57 \times 10^{-2}$ & $3.82 \times 10^{-2}$ & $3.46 \times 10^{-2}$ & $3.23 \times 10^{-2}$ \\
 $^{29}$Si & $2.27 \times 10^{-4}$ & $2.84 \times 10^{-4}$ & $2.86 \times 10^{-4}$ & $2.27 \times 10^{-4}$ \\
 $^{30}$Si & $3.72 \times 10^{-4}$ & $4.42 \times 10^{-4}$ & $4.61 \times 10^{-4}$ & $4.33 \times 10^{-4}$ \\
 $^{31}$P & $9.51 \times 10^{-5}$ & $1.9 \times 10^{-4}$ & $1.11 \times 10^{-4}$ & $1.0 \times 10^{-4}$ \\
 $^{32}$S & $1.50 \times 10^{-2}$ & $1.57 \times 10^{-2}$ & $1.37 \times 10^{-2}$ & $1.30 \times 10^{-2}$ \\
 $^{33}$S & $8.11 \times 10^{-5}$ & $8.77 \times 10^{-5}$ & $8.12 \times 10^{-5}$ & $7.87 \times 10^{-5}$ \\
 $^{34}$S & $6.96 \times 10^{-4}$ & $7.29 \times 10^{-4}$ & $6.47 \times 10^{-4}$ & $5.82 \times 10^{-4}$ \\
 $^{36}$S & $3.51 \times 10^{-8}$ & $3.91 \times 10^{-8}$ & $3.99 \times 10^{-8}$ & $4.18 \times 10^{-8}$ \\
 $^{35}$Cl & $3.69 \times 10^{-5}$ & $3.77 \times 10^{-5}$ & $3.76 \times 10^{-5}$ & $2.95 \times 10^{-5}$ \\
 $^{37}$Cl & $7.35 \times 10^{-6}$ & $7.27 \times 10^{-6}$ & $5.48 \times 10^{-6}$ & $5.48 \times 10^{-6}$ \\
 $^{36}$Ar & $2.26 \times 10^{-3}$ & $2.36 \times 10^{-3}$ & $1.98 \times 10^{-3}$ & $1.95 \times 10^{-3}$ \\
 $^{38}$Ar & $4.20 \times 10^{-4}$ & $3.94 \times 10^{-4}$ & $2.87 \times 10^{-4}$ & $2.53 \times 10^{-4}$ \\
 $^{40}$Ar & $4.43 \times 10^{-10}$ & $4.92 \times 10^{-10}$ & $5.51 \times 10^{-10}$ & $5.69 \times 10^{-10}$ \\
 $^{39}$K & $2.63 \times 10^{-5}$ & $2.50 \times 10^{-5}$ & $1.82 \times 10^{-5}$ & $1.78 \times 10^{-5}$ \\
 $^{40}$K & $9.18 \times 10^{-9}$ & $9.61 \times 10^{-9}$ & $1.16 \times 10^{-8}$ & $8.51 \times 10^{-9}$ \\
 $^{41}$K & $1.82 \times 10^{-6}$ & $1.71 \times 10^{-6}$ & $1.6 \times 10^{-6}$ & $1.6 \times 10^{-6}$ \\
 $^{40}$Ca & $1.71 \times 10^{-3}$ & $1.82 \times 10^{-3}$ & $1.50 \times 10^{-3}$ & $1.50 \times 10^{-3}$ \\
 $^{42}$Ca & $1.19 \times 10^{-5}$ & $1.12 \times 10^{-5}$ & $7.54 \times 10^{-6}$ & $7.2 \times 10^{-6}$ \\
 $^{43}$Ca & $1.95 \times 10^{-8}$ & $2.5 \times 10^{-8}$ & $2.5 \times 10^{-8}$ & $1.91 \times 10^{-8}$ \\
 $^{44}$Ca & $1.12 \times 10^{-6}$ & $1.21 \times 10^{-6}$ & $1.17 \times 10^{-6}$ & $1.25 \times 10^{-6}$ \\
 $^{46}$Ca & $6.72 \times 10^{-12}$ & $6.56 \times 10^{-12}$ & $1.69 \times 10^{-11}$ & $3.43 \times 10^{-9}$ \\
 $^{48}$Ca & $4.2 \times 10^{-17}$ & $3.32 \times 10^{-17}$ & $5.59 \times 10^{-14}$ & $5.8 \times 10^{-10}$ \\
 $^{45}$Sc & $4.20 \times 10^{-8}$ & $4.41 \times 10^{-8}$ & $3.52 \times 10^{-8}$ & $4.18 \times 10^{-8}$ \\
 $^{46}$Ti & $4.66 \times 10^{-6}$ & $4.56 \times 10^{-6}$ & $3.4 \times 10^{-6}$ & $3.36 \times 10^{-6}$ \\
 $^{47}$Ti & $9.65 \times 10^{-8}$ & $1.4 \times 10^{-7}$ & $9.88 \times 10^{-8}$ & $1.2 \times 10^{-7}$ \\
 $^{48}$Ti & $2.54 \times 10^{-5}$ & $3.11 \times 10^{-5}$ & $3.29 \times 10^{-5}$ & $3.84 \times 10^{-5}$ \\
 $^{49}$Ti & $1.94 \times 10^{-6}$ & $2.88 \times 10^{-6}$ & $4.14 \times 10^{-6}$ & $5.75 \times 10^{-6}$ \\
 $^{50}$Ti & $5.81 \times 10^{-11}$ & $6.9 \times 10^{-11}$ & $3.58 \times 10^{-6}$ & $5.25 \times 10^{-4}$ \\
 $^{50}$V & $4.3 \times 10^{-10}$ & $4.21 \times 10^{-10}$ & $1.53 \times 10^{-8}$ & $6.55 \times 10^{-8}$ \\
 $^{51}$V & $7.38 \times 10^{-6}$ & $1.41 \times 10^{-5}$ & $3.87 \times 10^{-5}$ & $2.33 \times 10^{-4}$ \\

\end{tabular}
\end{center}
\end{table*}

\begin{table*}

\begin{center}
\caption{$cont'd$ of Table \ref{table:IsotopesPTD}.
}
\label{table:IsotopesPTDb}
\begin{tabular}{|c | c c c c|}
\hline
Isotopes & 050-1-c3-1P & 100-1-c3-1P & 300-1-c3-1P & 500-1-c3-1P \\ \hline 
 
 $^{50}$Cr & $3.36 \times 10^{-5}$ & $7.1 \times 10^{-5}$ & $1.84 \times 10^{-4}$ & $2.26 \times 10^{-4}$ \\
 $^{52}$Cr & $7.53 \times 10^{-4}$ & $1.11 \times 10^{-3}$ & $4.46 \times 10^{-3}$ & $1.20 \times 10^{-2}$ \\
 $^{53}$Cr & $1.4 \times 10^{-4}$ & $2.10 \times 10^{-4}$ & $6.81 \times 10^{-4}$ & $1.32 \times 10^{-3}$ \\
 $^{54}$Cr & $7.87 \times 10^{-9}$ & $4.64 \times 10^{-8}$ & $9.16 \times 10^{-5}$ & $4.84 \times 10^{-3}$ \\
 $^{55}$Mn & $1.84 \times 10^{-3}$ & $3.86 \times 10^{-3}$ & $8.92 \times 10^{-3}$ & $1.23 \times 10^{-2}$ \\
 $^{54}$Fe & $1.22 \times 10^{-2}$ & $3.77 \times 10^{-2}$ & $9.27 \times 10^{-2}$ & $1.9 \times 10^{-1}$ \\
 $^{56}$Fe & $2.10 \times 10^{-1}$ & $2.65 \times 10^{-1}$ & $3.55 \times 10^{-1}$ & $4.2 \times 10^{-1}$ \\
 $^{57}$Fe & $6.27 \times 10^{-3}$ & $8.66 \times 10^{-3}$ & $1.19 \times 10^{-2}$ & $1.36 \times 10^{-2}$ \\
 $^{58}$Fe & $2.21 \times 10^{-9}$ & $1.62 \times 10^{-7}$ & $5.54 \times 10^{-4}$ & $1.43 \times 10^{-2}$ \\
 $^{60}$Fe & $1.67 \times 10^{-20}$ & $2.18 \times 10^{-20}$ & $1.52 \times 10^{-10}$ & $5.50 \times 10^{-8}$ \\
 $^{59}$Co & $1.2 \times 10^{-4}$ & $1.56 \times 10^{-4}$ & $5.82 \times 10^{-4}$ & $9.47 \times 10^{-4}$ \\
 $^{58}$Ni & $1.64 \times 10^{-2}$ & $3.0 \times 10^{-2}$ & $5.21 \times 10^{-2}$ & $5.79 \times 10^{-2}$ \\
 $^{60}$Ni & $6.52 \times 10^{-4}$ & $9.51 \times 10^{-4}$ & $6.17 \times 10^{-3}$ & $8.16 \times 10^{-3}$ \\
 $^{61}$Ni & $2.35 \times 10^{-5}$ & $2.59 \times 10^{-5}$ & $4.37 \times 10^{-5}$ & $8.26 \times 10^{-5}$ \\
 $^{62}$Ni & $2.10 \times 10^{-4}$ & $2.22 \times 10^{-4}$ & $5.73 \times 10^{-4}$ & $4.38 \times 10^{-3}$ \\
 $^{64}$Ni & $8.64 \times 10^{-17}$ & $3.36 \times 10^{-14}$ & $1.62 \times 10^{-7}$ & $2.56 \times 10^{-5}$ \\
 $^{63}$Cu & $1.24 \times 10^{-7}$ & $1.36 \times 10^{-7}$ & $3.99 \times 10^{-7}$ & $3.37 \times 10^{-6}$ \\
 $^{65}$Cu & $5.84 \times 10^{-8}$ & $6.52 \times 10^{-8}$ & $8.75 \times 10^{-8}$ & $4.22 \times 10^{-7}$ \\
 $^{64}$Zn & $4.36 \times 10^{-7}$ & $4.96 \times 10^{-7}$ & $6.73 \times 10^{-7}$ & $6.78 \times 10^{-7}$ \\
 $^{66}$Zn & $1.3 \times 10^{-6}$ & $1.10 \times 10^{-6}$ & $1.20 \times 10^{-6}$ & $1.22 \times 10^{-6}$ \\
 $^{67}$Zn & $5.54 \times 10^{-10}$ & $6.12 \times 10^{-10}$ & $7.14 \times 10^{-10}$ & $1.60 \times 10^{-9}$ \\
 $^{68}$Zn & $2.17 \times 10^{-10}$ & $2.41 \times 10^{-10}$ & $5.27 \times 10^{-10}$ & $1.53 \times 10^{-8}$ \\
 $^{70}$Zn & $9.84 \times 10^{-23}$ & $9.85 \times 10^{-24}$ & $4.16 \times 10^{-15}$ & $4.2 \times 10^{-12}$ \\ \hline

\end{tabular}
\end{center}
\end{table*}


\begin{table*}

\begin{center}
\caption{Nucleosynthesis yield for the 
Models presented in this articles. The 
isotope masses are in units of solar mass.
($Remark$: The table is replaced due to typos while converting the 
raw data into the current table form. Changes are made for the 
isotopes including $^{22}$Ne, $^{26}$Mg, $^{26}$Al, $^{36}$S, $^{40}$K, 
$^{41}$K, $^{44}$Ca, $^{53}$Cr, $^{55}$Mn, $^{60}$Fe, $^{59}$Co and $^{63}$Cu.)}
\label{table:Isotopes_others}
\begin{tabular}{|c | c c c c c|}

\hline
Isotopes & 300-1-c3-1 & Test-A1 & Test-A2 & Test-B1 & Test-B2  \\ \hline

$^{12}$C & $1.7 \times 10^{-3}$ & $2.21 \times 10^{-3}$ & $4.91 \times 10^{-3}$ & $5.34 \times 10^{-4}$ & $2.70 \times 10^{-6}$ \\
 $^{13}$C & $2.54 \times 10^{-12}$ & $8.18 \times 10^{-10}$ & $5.28 \times 10^{-10}$ & $2.40 \times 10^{-12}$ & $2.44 \times 10^{-12}$ \\
 $^{14}$N & $1.40 \times 10^{-10}$ & $2.20 \times 10^{-8}$ & $1.66 \times 10^{-8}$ & $5.47 \times 10^{-11}$ & $2.74 \times 10^{-11}$ \\
 $^{15}$N & $9.40 \times 10^{-11}$ & $3.86 \times 10^{-10}$ & $1.32 \times 10^{-9}$ & $4.78 \times 10^{-11}$ & $2.38 \times 10^{-11}$ \\
 $^{16}$O & $5.69 \times 10^{-2}$ & $8.95 \times 10^{-2}$ & $1.67 \times 10^{-1}$ & $2.38 \times 10^{-2}$ & $7.30 \times 10^{-3}$ \\
 $^{17}$O & $1.9 \times 10^{-11}$ & $1.24 \times 10^{-8}$ & $7.87 \times 10^{-9}$ & $5.82 \times 10^{-12}$ & $1.49 \times 10^{-13}$ \\
 $^{18}$O & $2.29 \times 10^{-13}$ & $1.36 \times 10^{-10}$ & $1.71 \times 10^{-10}$ & $7.94 \times 10^{-14}$ & $1.85 \times 10^{-15}$ \\
 $^{19}$F & $1.38 \times 10^{-13}$ & $1.45 \times 10^{-11}$ & $2.69 \times 10^{-11}$ & $3.35 \times 10^{-14}$ & $8.18 \times 10^{-17}$ \\
 $^{20}$Ne & $1.38 \times 10^{-4}$ & $1.22 \times 10^{-3}$ & $4.28 \times 10^{-3}$ & $1.16 \times 10^{-4}$ & $1.8 \times 10^{-6}$ \\
 $^{21}$Ne & $3.6 \times 10^{-9}$ & $9.37 \times 10^{-8}$ & $1.93 \times 10^{-7}$ & $1.54 \times 10^{-9}$ & $1.10 \times 10^{-11}$ \\
 $^{22}$Ne & $4.28 \times 10^{-5}$ & $7.49 \times 10^{-5}$ & $1.60 \times 10^{-4}$ & $2.14 \times 10^{-5}$ & $1.93 \times 10^{-11}$ \\
 $^{23}$Na & $8.9 \times 10^{-7}$ & $5.25 \times 10^{-6}$ & $1.62 \times 10^{-5}$ & $3.55 \times 10^{-7}$ & $9.98 \times 10^{-9}$ \\
 $^{24}$Mg & $1.10 \times 10^{-3}$ & $2.45 \times 10^{-3}$ & $7.76 \times 10^{-3}$ & $6.39 \times 10^{-4}$ & $2.30 \times 10^{-5}$ \\
 $^{25}$Mg & $2.36 \times 10^{-6}$ & $1.13 \times 10^{-5}$ & $3.86 \times 10^{-5}$ & $1.58 \times 10^{-6}$ & $8.60 \times 10^{-9}$ \\
 $^{26}$Mg & $2.56 \times 10^{-6}$ & $1.98 \times 10^{-5}$ & $5.50 \times 10^{-5}$ & $1.60 \times 10^{-6}$ & $2.52 \times 10^{-8}$ \\
 $^{26}$Al & $3.56 \times 10^{-29}$ & $3.56 \times 10^{-29}$ & $6.73 \times 10^{-10}$ & $3.56 \times 10^{-29}$ & $3.56 \times 10^{-29}$ \\
 $^{27}$Al & $9.14 \times 10^{-5}$ & $1.90 \times 10^{-4}$ & $6.27 \times 10^{-4}$ & $5.46 \times 10^{-5}$ & $1.58 \times 10^{-6}$ \\
 $^{28}$Si & $2.35 \times 10^{-1}$ & $2.95 \times 10^{-1}$ & $3.24 \times 10^{-1}$ & $1.35 \times 10^{-1}$ & $5.90 \times 10^{-2}$ \\
 $^{29}$Si & $2.58 \times 10^{-4}$ & $4.78 \times 10^{-4}$ & $1.6 \times 10^{-3}$ & $9.67 \times 10^{-5}$ & $1.98 \times 10^{-5}$ \\
 $^{30}$Si & $3.51 \times 10^{-4}$ & $6.63 \times 10^{-4}$ & $1.65 \times 10^{-3}$ & $1.50 \times 10^{-4}$ & $1.73 \times 10^{-5}$ \\
 $^{31}$P & $1.92 \times 10^{-4}$ & $2.94 \times 10^{-4}$ & $5.52 \times 10^{-4}$ & $7.13 \times 10^{-5}$ & $2.0 \times 10^{-5}$ \\
 $^{32}$S & $1.23 \times 10^{-1}$ & $1.45 \times 10^{-1}$ & $1.50 \times 10^{-1}$ & $7.86 \times 10^{-2}$ & $3.72 \times 10^{-2}$ \\
 $^{33}$S & $2.85 \times 10^{-4}$ & $4.17 \times 10^{-4}$ & $6.62 \times 10^{-4}$ & $1.0 \times 10^{-4}$ & $3.11 \times 10^{-5}$ \\
 $^{34}$S & $2.9 \times 10^{-3}$ & $3.17 \times 10^{-3}$ & $5.0 \times 10^{-3}$ & $7.22 \times 10^{-4}$ & $2.37 \times 10^{-4}$ \\
 $^{36}$S & $3.35 \times 10^{-8}$ & $6.82 \times 10^{-8}$ & $1.75 \times 10^{-7}$ & $1.44 \times 10^{-8}$ & $1.37 \times 10^{-9}$ \\
 $^{35}$Cl & $1.53 \times 10^{-4}$ & $2.12 \times 10^{-4}$ & $3.17 \times 10^{-4}$ & $5.10 \times 10^{-5}$ & $1.99 \times 10^{-5}$ \\
 $^{37}$Cl & $5.7 \times 10^{-5}$ & $6.84 \times 10^{-5}$ & $8.77 \times 10^{-5}$ & $1.70 \times 10^{-5}$ & $6.54 \times 10^{-6}$ \\
 $^{36}$Ar & $2.22 \times 10^{-2}$ & $2.44 \times 10^{-2}$ & $2.39 \times 10^{-2}$ & $1.59 \times 10^{-2}$ & $8.17 \times 10^{-3}$ \\
 $^{38}$Ar & $1.82 \times 10^{-3}$ & $2.67 \times 10^{-3}$ & $3.64 \times 10^{-3}$ & $5.87 \times 10^{-4}$ & $2.32 \times 10^{-4}$ \\
 $^{40}$Ar & $8.26 \times 10^{-10}$ & $1.30 \times 10^{-9}$ & $3.34 \times 10^{-9}$ & $3.20 \times 10^{-10}$ & $4.63 \times 10^{-11}$ \\
 $^{39}$K & $1.76 \times 10^{-4}$ & $2.42 \times 10^{-4}$ & $3.4 \times 10^{-4}$ & $5.42 \times 10^{-5}$ & $2.37 \times 10^{-5}$ \\
 $^{40}$K & $3.83 \times 10^{-8}$ & $4.78 \times 10^{-8}$ & $9.4 \times 10^{-8}$ & $1.16 \times 10^{-8}$ & $5.31 \times 10^{-9}$ \\
 $^{41}$K & $1.34 \times 10^{-5}$ & $1.74 \times 10^{-5}$ & $2.18 \times 10^{-5}$ & $4.42 \times 10^{-6}$ & $1.86 \times 10^{-6}$ \\
 $^{40}$Ca & $1.79 \times 10^{-2}$ & $1.79 \times 10^{-2}$ & $1.63 \times 10^{-2}$ & $1.47 \times 10^{-2}$ & $8.20 \times 10^{-3}$ \\
 $^{42}$Ca & $6.55 \times 10^{-5}$ & $9.33 \times 10^{-5}$ & $1.24 \times 10^{-4}$ & $1.98 \times 10^{-5}$ & $8.23 \times 10^{-6}$ \\
 $^{43}$Ca & $1.7 \times 10^{-6}$ & $9.7 \times 10^{-7}$ & $6.30 \times 10^{-7}$ & $1.87 \times 10^{-6}$ & $1.52 \times 10^{-6}$ \\
 $^{44}$Ca & $2.64 \times 10^{-5}$ & $2.17 \times 10^{-5}$ & $1.69 \times 10^{-5}$ & $3.57 \times 10^{-5}$ & $2.53 \times 10^{-5}$ \\
 $^{46}$Ca & $2.70 \times 10^{-11}$ & $4.36 \times 10^{-11}$ & $8.95 \times 10^{-11}$ & $3.95 \times 10^{-11}$ & $7.2 \times 10^{-11}$ \\
 $^{48}$Ca & $3.28 \times 10^{-14}$ & $3.50 \times 10^{-14}$ & $3.43 \times 10^{-14}$ & $3.48 \times 10^{-13}$ & $1.35 \times 10^{-12}$ \\
 $^{45}$Sc & $6.5 \times 10^{-7}$ & $6.97 \times 10^{-7}$ & $8.21 \times 10^{-7}$ & $2.60 \times 10^{-7}$ & $1.8 \times 10^{-7}$ \\
 $^{46}$Ti & $3.34 \times 10^{-5}$ & $4.82 \times 10^{-5}$ & $6.4 \times 10^{-5}$ & $1.4 \times 10^{-5}$ & $4.18 \times 10^{-6}$ \\
 $^{47}$Ti & $3.84 \times 10^{-6}$ & $3.30 \times 10^{-6}$ & $2.59 \times 10^{-6}$ & $6.34 \times 10^{-6}$ & $4.87 \times 10^{-6}$ \\
 $^{48}$Ti & $3.41 \times 10^{-4}$ & $2.55 \times 10^{-4}$ & $1.92 \times 10^{-4}$ & $4.16 \times 10^{-4}$ & $2.71 \times 10^{-4}$ \\
 $^{49}$Ti & $2.82 \times 10^{-5}$ & $2.43 \times 10^{-5}$ & $2.17 \times 10^{-5}$ & $2.55 \times 10^{-5}$ & $1.42 \times 10^{-5}$ \\
 $^{50}$Ti & $2.66 \times 10^{-6}$ & $2.70 \times 10^{-6}$ & $2.69 \times 10^{-6}$ & $8.56 \times 10^{-6}$ & $1.69 \times 10^{-5}$ \\
 $^{50}$V & $1.71 \times 10^{-8}$ & $1.97 \times 10^{-8}$ & $2.25 \times 10^{-8}$ & $1.81 \times 10^{-8}$ & $1.50 \times 10^{-8}$ \\
 $^{51}$V & $9.50 \times 10^{-5}$ & $9.3 \times 10^{-5}$ & $8.20 \times 10^{-5}$ & $9.9 \times 10^{-5}$ & $6.37 \times 10^{-5}$ \\

\end{tabular}
\end{center}
\end{table*}

\begin{table*}

\begin{center}
\caption{$cont'd$ of Table \ref{table:Isotopes_others}.
}
\label{table:Isotopes_othersb}
\begin{tabular}{|c | c c c c c|}

\hline
Isotopes & 300-1-c3-1 & Test-A1 & Test-A2 & Test-B1 & Test-B2  \\ \hline
  
 $^{50}$Cr & $4.96 \times 10^{-4}$ & $5.81 \times 10^{-4}$ & $5.89 \times 10^{-4}$ & $3.20 \times 10^{-4}$ & $1.60 \times 10^{-4}$ \\
 $^{52}$Cr & $8.4 \times 10^{-3}$ & $6.86 \times 10^{-3}$ & $6.3 \times 10^{-3}$ & $8.46 \times 10^{-3}$ & $6.25 \times 10^{-3}$ \\
 $^{53}$Cr & $1.0 \times 10^{-3}$ & $9.25 \times 10^{-4}$ & $8.72 \times 10^{-4}$ & $9.11 \times 10^{-4}$ & $6.21 \times 10^{-4}$ \\
 $^{54}$Cr & $7.34 \times 10^{-5}$ & $7.38 \times 10^{-5}$ & $7.39 \times 10^{-5}$ & $1.74 \times 10^{-4}$ & $2.80 \times 10^{-4}$ \\
 $^{55}$Mn & $1.3 \times 10^{-2}$ & $9.96 \times 10^{-3}$ & $9.65 \times 10^{-3}$ & $8.72 \times 10^{-3}$ & $6.30 \times 10^{-3}$ \\
 $^{54}$Fe & $1.6 \times 10^{-1}$ & $1.7 \times 10^{-1}$ & $1.4 \times 10^{-1}$ & $8.39 \times 10^{-2}$ & $5.70 \times 10^{-2}$ \\
 $^{56}$Fe & $6.71 \times 10^{-1}$ & $5.59 \times 10^{-1}$ & $4.49 \times 10^{-1}$ & $8.64 \times 10^{-1}$ & $10.21 \times 10^{-1}$ \\
 $^{57}$Fe & $2.8 \times 10^{-2}$ & $1.79 \times 10^{-2}$ & $1.48 \times 10^{-2}$ & $2.65 \times 10^{-2}$ & $3.20 \times 10^{-2}$ \\
 $^{58}$Fe & $4.54 \times 10^{-4}$ & $4.52 \times 10^{-4}$ & $4.53 \times 10^{-4}$ & $9.54 \times 10^{-4}$ & $1.41 \times 10^{-3}$ \\
 $^{60}$Fe & $1.9 \times 10^{-10}$ & $1.29 \times 10^{-10}$ & $1.22 \times 10^{-10}$ & $4.68 \times 10^{-10}$ & $1.11 \times 10^{-9}$ \\
 $^{59}$Co & $8.86 \times 10^{-4}$ & $8.14 \times 10^{-4}$ & $6.90 \times 10^{-4}$ & $1.21 \times 10^{-3}$ & $1.55 \times 10^{-3}$ \\
 $^{58}$Ni & $6.35 \times 10^{-2}$ & $5.98 \times 10^{-2}$ & $5.59 \times 10^{-2}$ & $6.59 \times 10^{-2}$ & $7.77 \times 10^{-2}$ \\
 $^{60}$Ni & $1.12 \times 10^{-2}$ & $9.86 \times 10^{-3}$ & $7.90 \times 10^{-3}$ & $1.63 \times 10^{-2}$ & $1.93 \times 10^{-2}$ \\
 $^{61}$Ni & $2.66 \times 10^{-4}$ & $2.9 \times 10^{-4}$ & $1.29 \times 10^{-4}$ & $5.16 \times 10^{-4}$ & $5.91 \times 10^{-4}$ \\
 $^{62}$Ni & $1.88 \times 10^{-3}$ & $1.53 \times 10^{-3}$ & $1.2 \times 10^{-3}$ & $3.67 \times 10^{-3}$ & $4.73 \times 10^{-3}$ \\
 $^{64}$Ni & $1.21 \times 10^{-7}$ & $1.20 \times 10^{-7}$ & $1.20 \times 10^{-7}$ & $4.18 \times 10^{-7}$ & $8.96 \times 10^{-7}$ \\
 $^{63}$Cu & $1.77 \times 10^{-6}$ & $1.43 \times 10^{-6}$ & $9.53 \times 10^{-7}$ & $3.32 \times 10^{-6}$ & $4.8 \times 10^{-6}$ \\
 $^{65}$Cu & $2.36 \times 10^{-6}$ & $1.81 \times 10^{-6}$ & $1.21 \times 10^{-6}$ & $5.7 \times 10^{-6}$ & $4.28 \times 10^{-6}$ \\
 $^{64}$Zn & $1.81 \times 10^{-5}$ & $1.40 \times 10^{-5}$ & $9.13 \times 10^{-6}$ & $3.78 \times 10^{-5}$ & $3.45 \times 10^{-5}$ \\
 $^{66}$Zn & $2.72 \times 10^{-5}$ & $2.12 \times 10^{-5}$ & $1.40 \times 10^{-5}$ & $5.53 \times 10^{-5}$ & $5.90 \times 10^{-5}$ \\
 $^{67}$Zn & $3.42 \times 10^{-8}$ & $2.65 \times 10^{-8}$ & $1.95 \times 10^{-8}$ & $7.25 \times 10^{-8}$ & $5.61 \times 10^{-8}$ \\
 $^{68}$Zn & $2.52 \times 10^{-8}$ & $2.0 \times 10^{-8}$ & $1.63 \times 10^{-8}$ & $7.29 \times 10^{-8}$ & $4.15 \times 10^{-8}$ \\
 $^{70}$Zn & $2.68 \times 10^{-15}$ & $4.62 \times 10^{-15}$ & $2.81 \times 10^{-15}$ & $1.75 \times 10^{-14}$ & $4.71 \times 10^{-14}$ \\ \hline

\end{tabular}  
\end{center}
\end{table*}

\begin{table*}
\begin{center}
\caption{Mass of major radioactive isotopes. The 
isotope masses are in units of solar mass.}
\label{table:DecayPTD}
\begin{tabular}{|c| c c c c |}
\hline
Isotopes & 050-1-c3-1P & 100-1-c3-1P & 300-1-c3-1P & 500-1-c3-1P \\ \hline

$^{22}$Na & $3.79 \times 10^{-9}$ & $7.49 \times 10^{-9}$ & $6.91 \times 10^{-9
 }$ & $2.34 \times 10^{-9}$ \\
 $^{26}$Al & $2.22 \times 10^{-6}$ & $3.64 \times 10^{-6}$ & $3.54 \times 10^{-6
 }$ & $1.51 \times 10^{-6}$ \\
 $^{39}$Ar & $1.75 \times 10^{-9}$ & $1.80 \times 10^{-9}$ & $2.19 \times 10^{-9
 }$ & $1.74 \times 10^{-9}$ \\
 $^{40}$K & $9.23 \times 10^{-9}$ & $9.67 \times 10^{-9}$ & $1.17 \times 10^{-8}
 $ & $8.56 \times 10^{-9}$ \\
 $^{41}$Ca & $1.57 \times 10^{-6}$ & $1.54 \times 10^{-6}$ & $1.1 \times 10^{-6}
 $ & $1.6 \times 10^{-6}$ \\
 $^{44}$Ti & $1.3 \times 10^{-6}$ & $1.19 \times 10^{-6}$ & $1.15 \times 10^{-6}
 $ & $1.21 \times 10^{-6}$ \\
 $^{48}$V & $1.15 \times 10^{-8}$ & $1.17 \times 10^{-8}$ & $8.56 \times 10^{-9}
 $ & $9.28 \times 10^{-9}$ \\
 $^{49}$V & $2.25 \times 10^{-8}$ & $2.38 \times 10^{-8}$ & $1.2 \times 10^{-7}$
  & $1.54 \times 10^{-7}$ \\
 $^{53}$Mn & $1.97 \times 10^{-6}$ & $1.54 \times 10^{-5}$ & $3.94 \times 10^{-4
 }$ & $5.49 \times 10^{-4}$ \\
 $^{60}$Fe & $2.63 \times 10^{-19}$ & $3.28 \times 10^{-19}$ & $2.41 \times 10^{
 -9}$ & $8.28 \times 10^{-7}$ \\
 $^{56}$Co & $1.4 \times 10^{-5}$ & $3.50 \times 10^{-5}$ & $9.38 \times 10^{-5}
 $ & $1.11 \times 10^{-4}$ \\
 $^{57}$Co & $2.30 \times 10^{-6}$ & $1.12 \times 10^{-4}$ & $1.23 \times 10^{-3
 }$ & $1.64 \times 10^{-3}$ \\
 $^{60}$Co & $1.3 \times 10^{-14}$ & $3.45 \times 10^{-13}$ & $9.65 \times 10^{-
 8}$ & $1.56 \times 10^{-6}$ \\
 $^{56}$Ni & $2.10 \times 10^{-1}$ & $2.64 \times 10^{-1}$ & $3.9 \times 10^{-1}
 $ & $3.19 \times 10^{-1}$ \\
 $^{57}$Ni & $6.27 \times 10^{-3}$ & $8.54 \times 10^{-3}$ & $1.5 \times 10^{-2}
 $ & $1.10 \times 10^{-2}$ \\
 $^{59}$Ni & $4.17 \times 10^{-6}$ & $5.72 \times 10^{-5}$ & $4.10 \times 10^{-4
 }$ & $5.42 \times 10^{-4}$ \\
 $^{63}$Ni & $5.89 \times 10^{-17}$ & $6.25 \times 10^{-15}$ & $7.3 \times 10^{-
 8}$ & $2.57 \times 10^{-6}$ \\ \hline

\end{tabular}
\end{center}
\end{table*}

\begin{table*}

\begin{center}
\caption{Mass of major radioactive isotopes. The 
isotope masses are in units of solar mass.}
\label{table:Decay_others}
\begin{tabular}{|c| c c c c c|}
\hline
Isotopes & 300-1-c3-1 & Test-A1 & Test-A2 & Test-B1 & Test-B2  \\ \hline

$^{22}$Na & $3.99 \times 10^{-10}$ & $4.19 \times 10^{-9}$ & $1.43 \times 10^{-
 8}$ & $3.82 \times 10^{-10}$ & $3.89 \times 10^{-11}$  \\
 $^{26}$Al & $3.45 \times 10^{-7}$ & $1.90 \times 10^{-6}$ & $6.96 \times 10^{-6
 }$ & $2.25 \times 10^{-7}$ & $2.33 \times 10^{-9}$  \\
  $^{39}$Ar & $5.0 \times 10^{-9}$ & $7.17 \times 10^{-9}$ & $1.49 \times 10^{-8}
 $ & $3.31 \times 10^{-9}$ & $8.28 \times 10^{-10}$  \\
 $^{40}$K & $3.85 \times 10^{-8}$ & $4.81 \times 10^{-8}$ & $9.9 \times 10^{-8}$
  & $2.38 \times 10^{-8}$ & $4.95 \times 10^{-9}$  \\
 $^{41}$Ca & $1.34 \times 10^{-5}$ & $1.76 \times 10^{-5}$ & $2.18 \times 10^{-5
 }$ & $1.7 \times 10^{-5}$ & $2.0 \times 10^{-6}$ \\
 $^{44}$Ti & $2.46 \times 10^{-5}$ & $2.4 \times 10^{-5}$ & $1.58 \times 10^{-5}
 $ & $2.58 \times 10^{-5}$ & $2.26 \times 10^{-5}$ \\
 $^{48}$V & $1.10 \times 10^{-7}$ & $1.38 \times 10^{-7}$ & $1.76 \times 10^{-7}
 $ & $8.32 \times 10^{-8}$ & $1.76 \times 10^{-8}$ \\
 $^{49}$V & $3.20 \times 10^{-7}$ & $3.88 \times 10^{-7}$ & $4.91 \times 10^{-7}
 $ & $2.46 \times 10^{-7}$ & $9.45 \times 10^{-8}$ \\
 $^{53}$Mn & $3.81 \times 10^{-4}$ & $3.89 \times 10^{-4}$ & $3.97 \times 10^{-4
 }$ & $3.53 \times 10^{-4}$ & $2.63 \times 10^{-4}$ \\
 $^{60}$Fe & $1.60 \times 10^{-9}$ & $1.83 \times 10^{-9}$ & $1.72 \times 10^{-9
 }$ & $5.87 \times 10^{-9}$ & $9.20 \times 10^{-9}$ \\
 $^{56}$Co & $9.96 \times 10^{-5}$ & $9.85 \times 10^{-5}$ & $9.90 \times 10^{-5
 }$ & $9.75 \times 10^{-5}$ & $9.18 \times 10^{-5}$ \\
 $^{57}$Co & $1.19 \times 10^{-3}$ & $1.18 \times 10^{-3}$ & $1.20 \times 10^{-3
 }$ & $1.16 \times 10^{-3}$ & $1.0 \times 10^{-3}$ \\
 $^{60}$Co & $7.70 \times 10^{-8}$ & $8.37 \times 10^{-8}$ & $8.8 \times 10^{-8}
 $ & $7.86 \times 10^{-8}$ & $8.23 \times 10^{-8}$  \\
 $^{56}$Ni & $6.27 \times 10^{-1}$ & $5.15 \times 10^{-1}$ & $4.5 \times 10^{-1}
 $ & $7.62 \times 10^{-1}$ & $9.68 \times 10^{-1}$  \\
 $^{57}$Ni & $1.95 \times 10^{-2}$ & $1.66 \times 10^{-2}$ & $1.35 \times 10^{-2
 }$ & $2.46 \times 10^{-2}$ & $3.18 \times 10^{-2}$ \\
 $^{59}$Ni & $4.5 \times 10^{-4}$ & $4.8 \times 10^{-4}$ & $4.6 \times 10^{-4}$ 
 & $4.14 \times 10^{-4}$ & $3.84 \times 10^{-4}$ \\
 $^{63}$Ni & $5.61 \times 10^{-8}$ & $5.83 \times 10^{-8}$ & $5.58 \times 10^{-8
 }$ & $8.0 \times 10^{-8}$ & $9.50 \times 10^{-8}$  \\ \hline
 
 \end{tabular}
 \end{center}
 \end{table*}


\bigskip

\begin{thebibliography}{}
\expandafter\ifx\csname natexlab\endcsname\relax\def\natexlab#1{#1}\fi

\bibitem[{Alastuey \& Jancovici(1978)}]{Alastuey1978}
Alastuey, A., \& Jancovici, B. 1978, Astrophys. J., 226, 1034

\bibitem[{Arnett(1969)}]{Arnett1969}
Arnett, D. 1969, Astrophys. \& Space Sci., 5, 180

\bibitem[{Arnett(1996)}]{Arnett1996}
Arnett, D. 1996, Nucleosynthesis and Supernovae (Princeton Univ. Press)

\bibitem[{Asplund {et~al.}(2009)Asplund, Grevesse, Sauval, \&
  Scott}]{Asplund2009}
Asplund, M., Grevesse, N., Sauval, A.~J., \& Scott, P. 2009, Annu. Rev. Astron.
  \& Astrophy., 47, 481

\bibitem[{Benvenuto {et~al.}(2013)Benvenuto, Bersten, \&
  Nomoto}]{Benvenuto2013}
Benvenuto, O.~G., Bersten, M.~C., \& Nomoto, K. 2013, Astrophys. J., 74

\bibitem[{Blondin {et~al.}(2013)Blondin, Dessart, Hillier, \&
  Khokhlov}]{Blondin2012}
Blondin, S., Dessart, L., Hillier, D.~J., \& Khokhlov, A.~M. 2013, Mon. Not. R.
  astr. Soc., 429, 2127

\bibitem[{Blondin {et~al.}(2011)Blondin, Kasen, Roepke, Kirshner, \&
  Mandel}]{Blondin2011}
Blondin, S., Kasen, D., Roepke, F.~K., Kirshner, R.~P., \& Mandel, K.~S. 2011,
  Mon. Not. R. astr. Soc., 417, 1280

 
\bibitem[\protect\citeauthoryear{Brachwitz et al.}{2000}]{Brachwitz2000}
Brachwitz, F., Dean, D.J., Hix, W..R. et al.,
ApJ \textbf{536}, 934 (2000)
  

\bibitem[{Branch \& Wheeler(2017)Branch \& Wheeler}]{Branch2017}
Branch, D. \& Wheeler, J.~C., 2017, Supernova Explosions (Springer)

\bibitem[{Damkoehler(1939)}]{Damkoehler1939}
Damkoehler, G. 1939, Jahrb. Deut. Luftfarhforsch., 113

\bibitem[{Dave {et~al.}(2017)Dave, Kashyap, Fisher, {et~al.}}]{Dave2017}
Dave, P., Kashyap, R., Fisher, R., {et~al.} 2017, Astrophys. J., 841, 58

\bibitem[{Dimitriadis {et~al.}(2017)Dimitriadis, Sullivan, Kerzendorf,
  {et~al.}}]{Dimitriadis2017}
Dimitriadis, G., Sullivan, M., Kerzendorf, W., {et~al.} 2017, Mon. Not. R.
  astr. Soc., 468, 3798

\bibitem[{Fink {et~al.}(2014)}]{Fink2014}
Fink, M., {et~al.} 2014, Mon. Not. R. astr. Soc., 438, 438

\bibitem[{Fowler {et~al.}(1967)Fowler, Caughlan, \& Zimmerman}]{Fowler1967}
Fowler, W.~A., Caughlan, G.~R., \& Zimmerman, B.~A. 1967, ARAA, 5, 525

\bibitem[{Fuller {et~al.}(1982)Fuller, Fowler, \& Newman}]{Fuller1982}
Fuller, G.~M., Fowler, W.~A., \& Newman, M.~J. 1982, Astrophys. J., 48, 279

\bibitem[{Gamezo {et~al.}(2004)Gamezo, Khokhlov, \& Oran}]{Gamezo2004}
Gamezo, V.~N., Khokhlov, A.~M., \& Oran, E.~S. 2004, Phys. Rev. Lett., 92,
  211102

\bibitem[{Gamezo {et~al.}(2005)Gamezo, Khokhlov, \& Oran}]{Gamezo2005}
---. 2005, Astrophys. J., 623, 337

\bibitem[{Graur {et~al.}(2016)}]{Graur2016}
Graur, O., {et~al.} 2016, Astrophys. J., 819, 36

\bibitem[{Hicks(2015)}]{Hicks2015}
Hicks, E.~P. 2015, Astrophys. J., 803, 72

\bibitem[{Hillebrandt \& Niemeyer(2000)}]{Hillebrandt2000}
Hillebrandt, W., \& Niemeyer, J.~C. 2000, Annu. Rev. Astron. Astrophys., 38,
  191

\bibitem[{Hinkel {et~al.}(2014)}]{Hinkel2014}
Hinkel, N.~R., {et~al.} 2014, Astrophys. J., 148, 54

\bibitem[{Iwamoto {et~al.}(1999)}]{Iwamoto1999}
Iwamoto, K., {et~al.} 1999, Astrophys. J. Suppl., 125, 439

\bibitem[{Jackson {et~al.}(2010)Jackson, Calder, Townsley,
  {et~al.}}]{Jackson2010}
Jackson, A.~P., Calder, A.~C., Townsley, D.~M., {et~al.} 2010, Astrophys. J.,
  720, 99

\bibitem[{{Jha}(2017)}]{Jha2017}
{Jha}, S.~W. 2017, in Handbook of Supernovae,
ed. A.~W. Alsabti \& P.~Murdin (Springer), 1, 375 (arXiv:1707.01110)

\bibitem[{Kato {et~al.}(2014)Kato, Saio, Hachisu, \& Nomoto}]{Kato2014}
Kato, M., Saio, H., Hachisu, I., \& Nomoto, K. 2014, Astrophys. J., 793, 136

\bibitem[{Khokhlov(1991{\natexlab{a}})}]{Khokhlov1991a}
Khokhlov, A.~M. 1991{\natexlab{a}}, Astron. Astrophys., 245, 114

\bibitem[{Khokhlov(1991{\natexlab{b}})}]{Khokhlov1991c}
---. 1991{\natexlab{b}}, Astrophys. J., 246, 383

\bibitem[{Khokhlov(1991{\natexlab{c}})}]{Khokhlov1991b}
---. 1991{\natexlab{c}}, Astron. Astrophys., 245, L25

\bibitem[{Khokhlov {et~al.}(1997{\natexlab{a}})Khokhlov, Oran, \&
  Wheeler}]{Khokhlov1997b}
Khokhlov, A.~M., Oran, E.~S., \& Wheeler, J.~C. 1997{\natexlab{a}}, Astrophys.
  J., 478, 678

\bibitem[{Khokhlov {et~al.}(1997{\natexlab{b}})Khokhlov, Oran, \&
  Wheeler}]{Khokhlov1997a}
---. 1997{\natexlab{b}}, Combustion and Flame, 108, 503

\bibitem[{Kitamura(2000)}]{Kitamura2000}
Kitamura, H. 2000, Astrophys. J., 539, 888

\bibitem[{Kobayashi {et~al.}(2015)Kobayashi, Nomoto, \&
  Hachisu}]{Kobayashi2015}
Kobayashi, C., Nomoto, K., \& Hachisu, I. 2015, Astrophys. J., 804, L24

\bibitem[{Krueger {et~al.}(2010)Krueger, p.~Jackson, Calder,
  {et~al.}}]{Krueger2010}
Krueger, B.~K., p.~Jackson, A., Calder, A.~C., {et~al.} 2010, Astrophys. J.,
  719, L5

\bibitem[{Krueger {et~al.}(2012)Krueger, p.~Jackson, Calder,
  {et~al.}}]{Krueger2012}
---. 2012, Astrophys. J., 757, 175

\bibitem[{Langanke \& Martinez-Pinedo(2001)}]{Langanke2001}
Langanke, K., \& Martinez-Pinedo, G. 2001, ADNDT, 79, 1

\bibitem[{Larsen {et~al.}(2014)}]{Larsen2014}
Larsen, S.~S., {et~al.} 2014, Mon. Not. R. astr. Soc., 565, A98


\bibitem[{Lesaffre {et~al.}(2006)Lesaffre, Han, Tout, Podsiadlowski, \&
  Martin}]{Lesaffre2006}
Lesaffre, P., Han, Z., Tout, C.~A., Podsiadlowski, P., \& Martin, R.~G. 2006,
  Mon. Not. R. astr. Soc., 368, 187

\bibitem[{Leung {et~al.}(2015{\natexlab{a}})Leung, Chu, \& Lin}]{Leung2015a}
Leung, S.-C., Chu, M.-C., \& Lin, L.-M. 2015{\natexlab{a}}, Mon. Not. R. astr.
  Soc., 454, 1238

\bibitem[{Leung {et~al.}(2015{\natexlab{b}})Leung, Chu, \& Lin}]{Leung2015b}
---. 2015{\natexlab{b}}, Astrophys. J., 812, 110

\bibitem[{Leung \& Nomoto(2017)}]{Leung2016}
Leung, S.-C., \& Nomoto, K. 2017, Proc. NIC2016, 14, 040506

\bibitem[{Lisewski {et~al.}(2000)Lisewski, Hillebrandt, \&
  Woosley}]{Lisewski2000}
Lisewski, A.~M., Hillebrandt, W., \& Woosley, S.~E. 2000, Astrophys. J., 538,
  831

\bibitem[{lolombek \& Niemeyer(2005)}]{Golombek2005}
lolombek, I., \& Niemeyer, J.~C. 2005, Astron. Astrophys., 438, 611

\bibitem[{Maeda {et~al.}(2010)Maeda, Roepke, Fink, {et~al.}}]{Maeda2010}
Maeda, K., Roepke, F.~K., Fink, M., {et~al.} 2010, Astrophys. J., 712, 624

\bibitem[{Martinez-Rodringuez {et~al.}(2016)Martinez-Rodringuez, Piro, Schwab,
  \& Badenes}]{MartinezRodriguez2016}
Martinez-Rodringuez, H., Piro, A.~L., Schwab, J., \& Badenes, C. 2016,
  Astrophys. J., 825, 57

\bibitem[{Matteucci(2001)}]{Matteucci2001}
Matteucci, F. 2001, The Chemical Evolution of the Galaxy (Springer)

\bibitem[{{Matteucci}(2012)}]{Matteucci2012}
Matteucci, F. 2012, Chemical Evolution of Galaxies (Springer)
  

\bibitem[{Mishenina {et~al.}(2015)Mishenina, Gorbaneva, Pignatari, Thielemann,
  \& Korotin}]{Mishenina2015}
Mishenina, T., Gorbaneva, T., Pignatari, M., Thielemann, F.-K., \& Korotin,
  S.~A. 2015, Mon. Not. R. astr. Soc., 454, 1585

\bibitem[{Mori {et~al.}(2016)Mori, Famiano, Kajino, {et~al.}}]{Mori2016}
Mori, K., Famiano, M.~A., Kajino, T., {et~al.} 2016, Astrophys. J., 833, 179

\bibitem[{Mueller \& Arnett(1982)}]{Mueller1982}
Mueller, E., \& Arnett, D. 1982, Astrophys. J., 261, L109

\bibitem[{Niemeyer {et~al.}(1995)Niemeyer, Hillebrandt, \&
  Woosley}]{Niemeyer1995a}
Niemeyer, J.~C., Hillebrandt, W., \& Woosley, S.~E. 1995, Astrophys. J., 452,
  769

\bibitem[{Nomoto {et~al.}(1976)Nomoto, Sugimoto, \& Neo}]{Nomoto1976}
Nomoto, K., Sugimoto, D., \& Neo, S. 1976, Ap \& SS, 39, 37

\bibitem[{Nomoto(1982{\natexlab{a}})}]{Nomoto1982a}
Nomoto, K. 1982{\natexlab{a}}, Astrophys. J., 253, 798

\bibitem[{Nomoto(1982{\natexlab{b}})}]{Nomoto1982b}
---. 1982{\natexlab{b}}, Astrophys. J., 257, 780

\bibitem[{Nomoto {et~al.}(1984)Nomoto, Thielemann, \& Yokoi}]{Nomoto1984}
Nomoto, K., Thielemann, F.-K., \& Yokoi, K. 1984, Astrophys. J., 286, 644

\bibitem[{Nomoto \& Iben(1985)}]{Nomoto1985}
Nomoto, K., \& Iben, I.~J. 1985, Astrophys. J., 297, 531

\bibitem[\protect\citeauthoryear{Nomoto et al.}{1994}]{Nomoto1994}
Nomoto, K., Yamaoka, H., Shigeyama, T., Kumagai, S., Tsujimoto, T. 1994, 
in Supernovae, NATO ASI Ser. C, Proc. of Session LIV held in Les Houche
1990, ed. S. Bludmann et al. (North-Holland) p.199
\footnote[1]{http://supernova.astron.s.u-tokyo.ac.jp/\~{}nomoto/reference}

\bibitem[{Nomoto {et~al.}(1997)}]{Nomoto1997}
Nomoto, K., Iwamoto, K., \& Kishimoto, N., 1997, Science 276, 1378

\bibitem[{Nomoto {et~al.}(2013)Nomoto, Kobayashi, \& Tominaga}]{Nomoto2013}
Nomoto, K., Kobayashi, C., \& Tominaga, N. 2013, Annu. Rev. Astron. and
  Astrophys., 51, 457 $^1$

\bibitem[{Nomoto \& Leung(2017)}]{Nomoto2017}
Nomoto, K., \& Leung, S.-C. 2017, in Handbook of Supernovae, 
ed. A.~W. Alsabti \& P.~Murdin (Springer), 2, 1275 $^1$

\bibitem[{Nomoto \& Leung(2018)}]{Nomoto2018}
---. 2018, Space Science Review, 214: 67 $^1$

\bibitem[{Nomoto {et~al.}(2007)Nomoto, Saio, Kato, \& Hachisu}]{Nomoto2007}
Nomoto, K., Saio, H., Kato, M., \& Hachisu, I. 2007, Astrophys. J., 663, 1269


\bibitem[{Nugent {et~al.}(2011)}]{Nugent2011}
Nugent, P.~E., {et~al.} 2011, Nature, 480, 344

\bibitem[{Ohlmann {et~al.}(2014)Ohlmann, Kormer, Fink, {et~al.}}]{Ohlmann2014}
Ohlmann, S.~T., Kormer, M., Fink, M., {et~al.} 2014, Astron. Astrophys., 572,
  A57

\bibitem[{Park {et~al.}(2013)Park, Badenes, Mori, {et~al.}}]{Park2013}
Park, S., Badenes, C., Mori, K., {et~al.} 2013, Astrophys. J., 767, L10

\bibitem[{Perlmutter {et~al.}(1999)}]{Perlmutter1999}
Perlmutter, S., {et~al.} 1999, Astrophys. J., 517, 565

\bibitem[{Piro \& Chang(2008)}]{Piro2008}
Piro, A.~L., \& Chang, P. 2008, Astrophys. J., 678

\bibitem[{Pocheau(1994)}]{Pocheau1994}
Pocheau, A. 1994, Phys. Rev. E, 49, 1109

\bibitem[{Poludenko {et~al.}(2011)Poludenko, Gardiner, \& Oran}]{Poludenko2011}
Poludenko, A.~Y., Gardiner, T.~Y., \& Oran, E.~S. 2011, Phys. Rev. Lett., 107,
  054501

\bibitem[{Rauscher \& Thielemann(2009)}]{Rauscher2000}
Rauscher, T., \& Thielemann, F.-K. 2009, ADNDT, 75, 1

\bibitem[{Reddy {et~al.}(2003)Reddy, Tomkin, Lambert, \& Prieto}]{Reddy2003}
Reddy, B.~E., Tomkin, J., Lambert, D.~L., \& Prieto, C.~A. 2003, Mon. Not. R.
  astr. Soc., 340, 304

\bibitem[{Reinecke {et~al.}(1999{\natexlab{a}})Reinecke, Hillebrandt, \&
  Niemeyer}]{Reinecke1999b}
Reinecke, M., Hillebrandt, W., \& Niemeyer, J.~C. 1999{\natexlab{a}}, Astron.
  Astrophys., 347, 739

\bibitem[{Reinecke {et~al.}(2002)Reinecke, Hillebrandt, \&
  Niemeyer}]{Reinecke2002a}
---. 2002, Astron. Astrophys., 386, 936

\bibitem[{Reinecke {et~al.}(1999{\natexlab{b}})Reinecke, Hillebrandt, Niemeyer,
  Klein, \& Gloebl}]{Reinecke1999a}
Reinecke, M., Hillebrandt, W., Niemeyer, J.~C., Klein, R., \& Gloebl, A.
  1999{\natexlab{b}}, Astron. Astrophys., 347, 724

\bibitem[{Riess {et~al.}(1998)}]{Riess1998}
Riess, A.~G., {et~al.} 1998, Astron. J., 116, 1009

\bibitem[{Roepke(2005)}]{Roepke2005b}
Roepke, F.~K. 2005, Astron. Astrophys., 432, 969

\bibitem[{Roepke {et~al.}(2006)Roepke, Gieseler, Reinecke, Travaglio, \&
  Hillebrandt}]{Roepke2006}
Roepke, F.~K., Gieseler, M., Reinecke, M., Travaglio, C., \& Hillebrandt, W.
  2006, Astron. Astrophys., 453, 203

\bibitem[{Roepke \& Hillebrandt(2005)}]{Roepke2005a}
Roepke, F.~K., \& Hillebrandt, W. 2005, Astron. Astrophys., 431, 635

\bibitem[{Roepke {et~al.}(2012)}]{Roepke2012}
Roepke, F.~K., {et~al.} 2012, Astrophys. J., 750, L19

\bibitem[{Sbordone {et~al.}(2015)}]{Sbordone2015}
Sbordone, L., {et~al.} 2015, Astron. Astrophys., 579, A104

\bibitem[{Schmidt {et~al.}(2006)Schmidt, Niemeyer, Hillebrndt, \&
  Roepke}]{Schmidt2006b}
Schmidt, W., Niemeyer, J.~C., Hillebrndt, W., \& Roepke, F.~K. 2006, Astron.
  Astrophys., 450, 283

\bibitem[{Seitenzahl {et~al.}(2011)Seitenzahl, Ciaraldi-Schoolmann, \&
  Roepke}]{Seitenzahl2011}
Seitenzahl, I.~R., Ciaraldi-Schoolmann, F., \& Roepke, F.~K. 2011, MNRAS, 414,
  2709

\bibitem[{Seitenzahl {et~al.}(2010)Seitenzahl, Ropeke, Fink, \&
  Pakmor}]{Seitenzahl2010}
Seitenzahl, I.~R., Ropeke, F.~K., Fink, M., \& Pakmor, R. 2010, Mon. Not. R.
  astr. Soc., 407, 2297

\bibitem[{Seitenzahl {et~al.}(2009)Seitenzahl, Townsley, Peng, \&
  Truran}]{Seitenzahl2009}
Seitenzahl, I.~R., Townsley, D.~M., Peng, F., \& Truran, J.~W. 2009, ADNDT, 95,
  96

\bibitem[{Seitenzahl {et~al.}(2013)}]{Seitenzahl2013}
Seitenzahl, I.~R., {et~al.} 2013, Mon. Not. R. astr. Soc., 429, 1156

\bibitem[{Shappee {et~al.}(2017)Shappee, Stanek, Kochanek, \&
  Garnavich}]{Shappee2016}
Shappee, B.~J., Stanek, K.~Z., Kochanek, C.~S., \& Garnavich, P.~M. 2017,
  Astrophys. J., 841, 48

\bibitem[{Sharpe(1999)}]{Sharpe1999}
Sharpe, G.~J. 1999, Mon. Not. R. astr. Soc., 310, 1039

\bibitem[{Shen {et~al.}(2017)Shen, Kasen, Miles, \& Townsley}]{Shen2017}
Shen, K.~J., Kasen, D., Miles, B.~J., \& Townsley, D.~M. 2017, arXiv:1706.01898

\bibitem[{Shigeyama {et~al.}(1992)Shigeyama, Nomoto, Yamaoka, \&
  F.-K.Thielemann}]{Shigeyama1992}
Shigeyama, T., Nomoto, K., Yamaoka, H., \& F.-K.Thielemann. 1992, Astrophys.
  J., 386, L13

\bibitem[{Shih {et~al.}(1994)Shih, Liou, Shabbir, Yang, \& Zhu}]{Shih1994b}
Shih, T.-H., Liou, W.~W., Shabbir, A., Yang, Z., \& Zhu, J. 1994, Comp. Fluids,
  24, 227

\bibitem[{Shih {et~al.}(1995)Shih, Zhu, \& Lumley}]{Shih1995}
Shih, T.-H., Zhu, J., \& Lumley, J.~L. 1995, Comput. Methods Appl. Mech.
  Engrg., 125, 287

\bibitem[{Sobeck {et~al.}(2006)}]{Sobeck2006}
Sobeck, J.~S., {et~al.} 2006, Astrophys. J., 131, 2949

\bibitem[{{Taubenberger}(2017)}]{Taubenberger2017}
{Taubenberger}, S. 2017, in Handbook of Supernovae,
ed. A.~W. Alsabti \& P.~Murdin (Springer), 1, 317 (arXiv:1703.00528)

\bibitem[{Thielemann {et~al.}(1987)Thielemann, Arnould, \&
  Turran}]{Thielemann1987}
Thielemann, F.-K., Arnould, M., \& Turran, J.~W. 1987, Advances in Nuclar
  Astrophysics (Editions Frontiere), 525

\bibitem[{{Thielemann} {et~al.}(1986){Thielemann}, {Nomoto}, \&
  {Yokoi}}]{Thielemann1986}
{Thielemann}, F.-K., {Nomoto}, K., \& {Yokoi}, K. 1986, \aap, 158, 17

\bibitem[{Timmes(1999)}]{Timmes1999a}
Timmes, F.~X. 1999, Astrophys. J., 124, 241

\bibitem[{Timmes {et~al.}(2000)Timmes, Hoofman, \& Woosley}]{Timmes2000a}
Timmes, F.~X., Hoofman, R.~D., \& Woosley, S.~E. 2000, Astrophys. J. Suppl.,
  129, 377

\bibitem[{Timmes \& Swesty(1999)}]{Timmes1999c}
Timmes, F.~X., \& Swesty, F.~D. 1999, Astrophys. J. Suppl., 126, 501

\bibitem[{Timmes \& Woosley(1992)}]{Timmes1992a}
Timmes, F.~X., \& Woosley, S.~E. 1992, Astrophys. J., 396, 649

\bibitem[{Townsley {et~al.}(2002)Townsley, Calder, Asida,
  {et~al.}}]{Townsley2007}
Townsley, D.~M., Calder, A.~M., Asida, S.~M., {et~al.} 2002, Astrophys. J.
  Suppl., 143, 201

\bibitem[{Townsley {et~al.}(2016)Townsley, Miles, Timmes, Calder, \&
  Brown}]{Townsley2016}
Townsley, D.~M., Miles, B.~J., Timmes, F.~X., Calder, A.~C., \& Brown, E.~F.
  2016, Astrophys. J. Suppl., 225, 3

\bibitem[{Travaglio {et~al.}(2004)Travaglio, Hillebrandt, Reinecke, Thielemann,
  {et~al.}}]{Travaglio2004}
Travaglio, C., Hillebrandt, W., Reinecke, M., Thielemann, F.-K., {et~al.} 2004,
  Astrophys. J., 425, 1029

\bibitem[{Umeda {et~al.}(1999)Umeda, Nomoto, Yamaoka, \& Wanajo}]{Umeda1999}
Umeda, H., Nomoto, K., Yamaoka, H., \& Wanajo, S. 1999, Astrophys. J., 513, 861

\bibitem[{Unno(1967)}]{Unno1967}
Unno, W. 1967, PASJ, 19, 140

\bibitem[{Woosley(1997)}]{Woosley1997}
Woosley, S.~E. 1997, Astrophys. J., 476, 801

\bibitem[{Woosley {et~al.}(2004)Woosley, Wunsch, \& Kuhlen}]{Woosley2004}
Woosley, S.~E., Wunsch, S., \& Kuhlen, M. 2004, Astrophys. J., 607, 921

\bibitem[{Wunsch \& Woosley(2006)}]{Wunsch2006}
Wunsch, S., \& Woosley, S.~E. 2006, Astrophys. J., 616, 1102

\bibitem[{Yamaguchi {et~al.}(2014)Yamaguchi, Eriksen, Badenes,
  {et~al.}}]{Yamaguchi2014}
Yamaguchi, H., Eriksen, K.~A., Badenes, C., {et~al.} 2014, Astrophys. J., 780,
  L136

\bibitem[{Yamaguchi {et~al.}(2015)}]{Yamaguchi2015}
Yamaguchi, H., {et~al.} 2015, Astrophys. J., 801, L31

\bibitem[{Yoshizawa {et~al.}(2012)Yoshizawa, Abe, Matsuo, Fujiwara, \&
  Mizobuchi}]{Yoshizawa2012}
Yoshizawa, A., Abe, H., Matsuo, Y., Fujiwara, H., \& Mizobuchi, Y. 2012, Phys.
  of Fluids, 24, 075109

\bibitem[{Zingale \& Dursi(2007)}]{Zingale2007}
Zingale, M., \& Dursi, L.~J. 2007, Astrophys. J., 656, 333

\bibitem[{Zingale {et~al.}(2011)Zingale, Nonaka, Almgren,
  {et~al.}}]{Zingale2011}
Zingale, M., Nonaka, A., Almgren, A.~S., {et~al.} 2011, Astrophys. J., 740, 8

\end{thebibliography}
\end{document}